\documentclass[12pt]{article}

\usepackage[usenames]{color}
\usepackage{setspace}

\usepackage{graphicx}
\usepackage{amsmath}
\usepackage{amssymb}
\usepackage{amsthm}

\usepackage{bm}
 
\newcommand{\submatrix}[4]{{#1}^{#4}_{#2\,#3}}
 
\newcommand{\VEV}[1]{\left\langle #1 \right\rangle}

\newcommand{\Z}[1]{{\mathbb Z}_#1}
\newcommand{\abs}[1]{\left| #1 \right|}

\newcommand{\bequ}{\begin{equation}}
\newcommand{\eequ}{\end{equation}}
\newcommand{\beqn}{\begin{eqnarray}}
\newcommand{\eeqn}{\end{eqnarray}}
\newcommand{\bctr}{\begin{center}}
\newcommand{\ectr}{\end{center}}
\newcommand{\bit}{\begin{itemize}}
\newcommand{\eit}{\end{itemize}}

\allowdisplaybreaks[4]

\usepackage{multirow}
\numberwithin{equation}{section}

\begin{document}

\begin{flushright}
\today
\end{flushright}

\begin{center}

{\LARGE\bf On representation matrices of boundary conditions in $SU(n)$
gauge theories compactified
on two-dimensional orbifolds}

\vskip 1.4cm

{\large  
$^{a}$Yoshiharu Kawamura\footnote{e-mail:haru@azusa.shinshu-u.ac.jp},
$^{a}$Eiji Kodaira\footnote{e-mail:20hs302b@shinshu-u.ac.jp},
$^{b}$Kentaro Kojima\footnote{e-mail:kojima@artsci.kyushu-u.ac.jp}\\
and\\
$^{c}$Toshifumi Yamashita\footnote{e-mail:tyamashi@aichi-med-u.ac.jp}
}
\\
\vskip 1.0cm
{\it $^a$Department of Physics, Shinshu University, Matsumoto 390-8621, Japan}\\
{\it $^b$ Faculty of Arts and Science, Kyushu University, Fukuoka 819-0395, Japan}\\
{\it $^c$Department of Physics, Aichi Medical University, Nagakute 480-1195, Japan}
\vskip 1.0cm

\begin{abstract}
We study the existence of diagonal representatives 
in each equivalence class of representation matrices 
of boundary conditions in $SU(n)$ or $U(n)$ gauge theories compactified
on the orbifolds $T^2/{\mathbb Z}_N$ ($N = 2, 3, 4, 6$).
We suppose that the theory has a global $G' = U(n)$ symmetry.
Using constraints, unitary transformations and gauge transformations,
we examine whether the representation matrices can simultaneously
become diagonal or not.
We show that at least one 
diagonal representative necessarily exists in each equivalence class 
on $T^2/{\mathbb Z}_2$ and $T^2/{\mathbb Z}_3$, 
but the representation matrices on $T^2/{\mathbb Z}_4$ and $T^2/{\mathbb Z}_6$
can contain not only diagonal matrices but also non-diagonal $2 \times 2$ ones and 
non-diagonal $3 \times 3$ and $2 \times 2$ ones, respectively,
as members of block-diagonal submatrices.
These non-diagonal matrices have discrete parameters, 
which means that the rank-reducing symmetry breaking can be caused 
by the discrete Wilson line phases.
\end{abstract}

\end{center}

\vskip 1.0 cm

%
\section{Introduction}
\label{sec:intro}
%

The standard model of particle physics has been established
as an effective theory around the weak scale,
but it possesses several riddles.
The origin of the gauge bosons and the Higgs boson has been a big mystery.
The standard model looks complicated at first glance,
and it seems to suggest a simple and beautiful theory beyond it.
Theories defined on a higher-dimensional space-time
are possible candidates.
The gauge bosons and the Higgs boson can be unified
as a higher-dimensional gauge multiplet~\cite{M}.
To realize the chiral fermions in the higher-dimensional theories, 
an orbifold is considered as an extra space. 
Such models are phenomenologically attractive, because the 
Higgs mass splitting between the doublet components and the triplet ones
in the grand unified theories (GUTs)~\cite{GUT,SUSYGUT-DG,SUSYGUT-S}
can be elegantly realized by orbifolding~\cite{K1,K2,H&N}.
In addition, the idea to unify the gauge and Higgs bosons can be 
applied to the electroweak symmetry breaking~\cite{GHU-KL&Y,GHU-CG&M,GHU-SS&S}, 
since the doublet components become able to be extracted from the 
original adjoint representation as zero modes.
In this scenario, 
the effective potential of the Higgs field is finite without supersymmetry 
(SUSY)~\cite{finiteness-K,finiteness-HI&L}, since 
the Higgs field is a pseudo Nambu-Goldstone mode~\cite{finiteness-AC&G}, 
with respect to spatially separated symmetry breakings. 
The finiteness is thanks to the non-locality of the breakings, 
and there are local divergences in subdiagrams~\cite{finiteness-M&Y,finiteness-HMT&Y,divcont}.  
These divergences
are removed by the lower-loop counter terms with no need to 
introduce additional counterterms~\cite{finiteness-M&Y,finiteness-HMT&Y},
which means that the effective potential is free from the divergences 
when it is written with the finite renormalized couplings.
This idea is also considered in GUT contexts to break the electroweak 
symmetry~\cite{GHU-HHK&Y,Lim:2007jv,Hosotani:2015hoa} 
as above, or to reduce the rank of 
the GUT symmetry~\cite{gGHU-E6}.
It is also applied to the breaking of the unified 
symmetry to show that the boundary conditions for the Higgs mass 
splitting assumed in refs.~\cite{K1,K2,H&N} can be naturally obtained 
as the minimum of the effective potential~\cite{gGHU,gGHU-KT&Y,gGHU-Y}, and its 
phenomenologies are studied~\cite{gGHU-pheno,gGHU-pheno-NSS&Y}.

The standard model particles are supposed to be composed of zero modes 
from bulk fields on a higher-dimensional space-time
and 4-dimensional fields localized on boundaries called brane fields.
Physical symmetries are determined
in cooperation of boundary conditions of fields and the dynamics of the Wilson line phases,
by the Hosotani mechanism~\cite{H1,H2,HHH&K}.
Hence, the study of the boundary conditions of fields 
as well as the dynamics is important.
Boundary conditions on orbifolds are specified by representation matrices and classified 
by equivalence relations of gauge symmetries~\cite{HH&K}.
We refer to the representation matrices as twist matrices.
In ref.~\cite{HH&K}, the classification of twist matrices has been carried out
on $S^1/\Z 2$, and it is shown that each equivalence class has at least
one diagonal representative.
Relying on this generality, a general form of the effective 
potential was derived~\cite{generalFormula}. 
For the orbifolds $T^2/\Z N$ ($N=2, 3, 4, 6$), it has been done, in a limited way,
for a class with diagonal representatives~\cite{KK&M,K&M,G&K},
and there is no definite answer 
whether every equivalence class has at least one diagonal representative
on the orbifolds.\footnote{
In refs.~\cite{HN&T,K&M}, the classification of the twist matrices
has been examined in an $SU(2)$ gauge theory compactified on $T^2/\Z 2$.
In ref.~\cite{K&N}, an attempt has been carried out to arrive at an answer 
by using matrix exponential representations,
but it remains incomplete because the conditions are not fully considered.
}

In this paper, we study the existence of diagonal representatives 
in each equivalence class of the twist matrices
in $SU(n)$ or $U(n)$ gauge theories compactified
on the orbifolds $T^2/\Z N$ ($N = 2, 3, 4, 6$).
We suppose that the theory has a global $G' = U(n)$ symmetry.
Using constraints, unitary transformations and gauge transformations,
we examine whether the twist matrices can simultaneously become diagonal or not.
We show that at least one diagonal representative 
necessarily exists in each equivalence class 
on $T^2/\Z 2$ and $T^2/\Z 3$, 
but the twist matrices on $T^2/\Z 4$ and $T^2/\Z 6$
can contain not only diagonal matrices but also non-diagonal $2 \times 2$ ones and 
non-diagonal $3 \times 3$ and $2 \times 2$ ones, respectively,
as members of block-diagonal submatrices.

The outline of this paper is as follows.
In the next section, we give a proof that there exists at least one 
diagonal representative
in all equivalence classes of the twist matrices on $S^1/\Z 2$,
using a notation which is useful to analyze twist matrices on $T^2/\Z N$.
We explain basic properties of $T^2/\Z N$ in section~\ref{sec:2Dorb}.
Using constraints and unitary transformations, 
we show that any twist matrices
can become block-diagonal forms with specific types of submatrices 
on $T^2/\Z 2$, $T^2/\Z 3$, $T^2/\Z 4$ and $T^2/\Z 6$ 
in section~\ref{sec:t2z2}, \ref{sec:t2z3}, \ref{sec:t2z4} and \ref{sec:t2z6}, respectively.
Performing gauge transformations on $T^2/\Z N$ for 
the twist matrices with the block-diagonal forms,
we examine whether the block-diagonal ones can become diagonal or not
in section~\ref{Sec:GaugeTr}.
We explore physical implications on our results
in section~\ref{Sec:PhysImp}.
In the last section, we give conclusions and discussions.
We present the details of block-diagonalization of twist matrices
on $T^2/\Z 2$, $T^2/\Z 3$, $T^2/\Z 4$ and $T^2/\Z 6$
in appendix~\ref{app:t2z2}, \ref{app:t2z3}, \ref{app:t2z4} and \ref{app:t2z6}, respectively.
We explain possible forms of a matrix under a certain condition
in appendix~\ref{ap:mmdag} and derivations of useful relations 
in our analyses in appendix~\ref{app:MM} and \ref{app:ap}.

%
\section{$S^1/\mathbb{Z}_2$}
\label{sec:s1z2}
%

Before working on two-dimensional (2D) orbifolds, we examine the
$S^1/{\mathbb Z}_2$ orbifold first, in a way slightly different from
the literature~\cite{HH&K}. Let $x^\mu$
$(\mu=0,1,2,3)$ and $y$ be the coordinates on the Minkowski space-time
$M^4$ and the orbifold $S^1/{\mathbb Z}_2$, respectively. The orbifold
$S^1/{\mathbb Z}_2$ is described by $y$ that satisfies the
identifications $y\sim y+2\pi R$ and $y\sim -y$, where $R$ is the
radius of $S^1$. Since the size of the extra dimension is irrelevant
to the following discussions, we take $2\pi R=1$ for simplicity.
Related to the identifications, we define the operations
${\cal T}:y\to y+1$ and ${\cal P}_0:y\to -y$.  Besides them, we can
also define a parity operation around $y=\pi R=1/2$ as
${\cal P}_1={\cal T}{\cal P}_0$.  Note that
${\cal P}_0^2={\cal P}_1^2={\cal I}$ holds, where ${\cal I}$ is the
identity operation.

Let us consider a gauge theory on $M^4\times S^1/{\mathbb Z}_2$, whose
bulk gauge group is $G=SU(n)$ or $U(n)$, and suppose that the Lagrangian of the
theory is invariant under global $G'=U(n)$ transformations for fields
for simplicity.
In general, the translation ${\cal T}$ and the parity operations
${\cal P}_0$ and ${\cal P}_1$ accompany non-trivial transformations
for fields in representation spaces of $G'$, which are described by
boundary conditions. We define that the transformations corresponding
to ${\cal T}$ and ${\cal P}_0$ in the fundamental representation of
$G'$ are given by constant unitary matrices $T$ and $P_0$,
respectively. We also denote the matrix representation of ${\cal P}_1$
by $P_1\equiv TP_0$. We refer to them as twist matrices, which must
satisfy the following constraints:
\begin{align}\label{s1z2bcmatcond1}
 (P_0)^2=(TP_0)^2=I, 
\end{align}
where $I$ is the unit matrix, in theories with a field belonging to the
fundamental representation.\footnote{More precisely, actions
of $(P_0)^2$ and $(TP_0)^2$ in representation spaces on any fields
contained in a theory must leave the fields unchanged for
consistency. 
For instance, if
only the adjoint representation exists in a theory, $(P_0)^2$ and
$(TP_0)^2$ can be chosen as nontrivial elements of the center
subgroup $\mathbb Z_n$ of $SU(n)$. 
In some theories with more than two compact extra
dimensions, a similar consideration to the above for the actions
of twist matrices shows that there exist nontrivial degrees of
freedom related to holonomy, so-called the 't Hooft
flux~\cite{tHooft,VG}, which we do not discuss in this paper.} 
In the following, we only examine the twist matrices on the fundamental representation of $G'$.
Our study can be
straightforwardly generalized to the other representation matrices,
which are obtained via tensor products of the fundamental (and
anti-fundamental) representations.

We can simplify the forms of the twist matrices by changing the basis
in a representation space of $G'$, which induces unitary
transformations on the twist matrices as $P_0\to WP_0W^\dag$ and
$T\to WTW^\dag$ with $WW^\dag=I$. Since $P_0$ is a unitary matrix,
eq.~\eqref{s1z2bcmatcond1} implies $P_0=P_0^{-1}=P_0^\dag$. Thus, by
taking a suitable choice of the basis, $P_0$ is diagonalized as
\begin{equation}\label{s1z2m01}
    P_0=
  \begin{pmatrix}
      -I_{n_1}&0\\
      0&I_{n_2}
  \end{pmatrix}=
  \begin{pmatrix}
      (P_0)_{(11)}&(P_0)_{(12)}\\
      (P_0)_{(21)}&(P_0)_{(22)}
  \end{pmatrix},  \qquad 
  T=
  \begin{pmatrix}
      (T)_{(11)}&(T)_{(12)}\\
      (T)_{(21)}&(T)_{(22)}
  \end{pmatrix}, 
\end{equation}
where $I_{n_k}$ $(k=1,2)$ are the $n_k \times n_k$ unit matrices, and we have introduced
$n_k\times n_l$ matrices $(P_0)_{(kl)}$ and $(T)_{(kl)}$. 
Note that $n_k$ are non-negative integers and satisfy
$n_1+n_2={\rm rank}(P_0)=n$.  From eq.~\eqref{s1z2m01}, we can write
$(P_0)_{(kl)}=(-1)^k\delta_{kl}I_{n_k}$.  In the following, we introduce
a notation as $(T)_{(kl)}=\submatrix{M}{k}{l}{[k-l]}$, where the
superscript $k-l=q$ is the ${\mathbb Z}_2$ charge defined by
$(P_0TP_0^{-1})_{(k\,k-q)}=(-1)^q\submatrix{M}{k}{k-q}{[q]}$.  We also
use the notation of
$\submatrix{M}{k}{l}{[k-l]}=\submatrix{M}{k}{l}{[k'-l']}
=\submatrix{M}{k'}{l'}{[k-l]}$ with $k'=k$
(mod 2) and $l'=l$ (mod 2).  This notation is more convenient for the
later discussions in ${\mathbb Z}_N$ $(N=3,4,6)$ orbifold models,
while it seems a little lengthy in ${\mathbb Z}_2$ orbifold models.

Keeping the diagonal form of $P_0$, we further simplify $T$ by unitary
transformations. Since $TP_0$ is unitary, eq.~\eqref{s1z2bcmatcond1}
implies $(TP_0)^\dag= TP_0$. Thus, by using
$(T^\dag)_{(kl)}= \submatrix{M}{l}{k}{[l-k]\dag}$, we find
\begin{align}\label{mmdagrels1}
 \submatrix{M}{k}{k-q}{[q]\dag}= (-1)^{-q}\submatrix{M}{k-q}{k}{[-q]}.
\end{align}
From the condition $TT^\dag =I$, it follows that
\begin{align}
  \delta_{q0}I_{n_k}= (TT^\dag)_{(kk-q)}
  =\sum_{q'}
  \submatrix{M}{k}{k+q'}{[-q']}
  \submatrix{M}{k-q}{k+q'}{[-q-q']\dag}
  =\sum_{q'}(-1)^{q'+q}
  \submatrix{M}{k}{k+q'}{[-q']}
  \submatrix{M}{k+q'}{k-q}{[q'+q]}, 
\end{align}
where the summation over $q'$ can be taken for any successive two
integers, e.g., $q'=0,1$ or $q'=1,2$.  The above equation implies
$\submatrix{M}{k}{k}{[0]}\submatrix{M}{k}{k-1}{[1]}
=\submatrix{M}{k}{k-1}{[1]}\submatrix{M}{k-1}{k-1}{[0]}$, which is
written in a more general form as
\begin{align}\label{mmz2rel1}
  \submatrix{M}{k}{k}{[0]}\submatrix{M}{k}{k-q}{[q]}
=\submatrix{M}{k}{k-q}{[q]}\submatrix{M}{k-q}{k-q}{[0]},
\end{align}
where $q=0$ gives the trivial relation.

From eq.~\eqref{mmdagrels1}, we find
$\submatrix{M}{k}{k}{[0]\dag}= \submatrix{M}{k}{k}{[0]} $. Thus,
$\submatrix{M}{k}{k}{[0]} $ is diagonalized by a unitary matrix, and
we can choose a basis where an $(i,j)$ element of
$\submatrix{M}{k}{k}{[0]}$ is given by
$(\submatrix{M}{k}{k}{[0]})_{ij}=a_k^i\delta_{ij}$
$(a_k^i\in {\mathbb R})$. In this basis, eq.~\eqref{mmz2rel1} gives
\begin{align}
  (a_k^i-a_{k-q}^j)(\submatrix{M}{k}{k-q}{[q]})_{ij}=0.
\end{align}
Thus, if $a_k^i\neq a_{k-q}^j$ is satisfied,
$(\submatrix{M}{k}{k-q}{[q]})_{ij}=0$ holds. Therefore, by using a
unitary transformation, we can take a basis where $P_0$ and $T$ are
written as the following block-diagonal forms:
\begin{align}\label{p0tbds1z21}
  &  P_0=
  \begin{pmatrix}
      P_0^{(0)}&&&\\
      &P_0^{(1)}&&\\
      &&\ddots &\\
      && &P_0^{(M)}
  \end{pmatrix}, 
\qquad  T=
  \begin{pmatrix}
      T^{(0)}&&&\\
      &T^{(1)}&&\\
      &&\ddots &\\
      && &T^{(M)}
  \end{pmatrix}, 
\end{align}
where $P_0^{(\lambda)}$ and $T^{(\lambda)}$ $(\lambda=0,1,\dots,M)$ are
\begin{align}\label{p0sub1}
&P_0^{(\lambda)}=
  \begin{pmatrix}
      (P_0^{(\lambda)})_{(11)}& (P_0^{(\lambda)})_{(12)}\\
      (P_0^{(\lambda)})_{(21)}& (P_0^{(\lambda)})_{(22)}
  \end{pmatrix}
=            \begin{pmatrix}
                -I_{n^{(\lambda)}_1}&0\\
                0&I_{n^{(\lambda)}_2}
            \end{pmatrix}, \\\label{t0sub1}
& T^{(\lambda)}=
    \begin{pmatrix}
    \submatrix{M}{1}{1}{(\lambda)[0]}&
    \submatrix{M}{1}{2}{(\lambda)[-1]}\\
    \submatrix{M}{2}{1}{(\lambda)[1]}&
    \submatrix{M}{2}{2}{(\lambda)[0]}                            
    \end{pmatrix}
=            \begin{pmatrix}
                a^{(\lambda)}I_{n^{(\lambda)}_1}&    \submatrix{M}{1}{2}{(\lambda)[-1]}\\
                  \submatrix{M}{2}{1}{(\lambda)[1]}&
a^{(\lambda)}I_{n^{(\lambda)}_2}
            \end{pmatrix}.
\end{align}
In the above equations, $n^{(\lambda)}_1$ and $n^{(\lambda)}_2$ are
non-negative integers, and $a^{(\lambda)}$ are real parameters that
satisfy $a^{(\lambda)}\neq a^{(\lambda')}$ for $\lambda\neq \lambda'$.
Note that a similar discussion as the derivation of
eq.~\eqref{mmdagrels1} gives the relation
$ \submatrix{M}{k}{k-q}{(\lambda)[q]\dag}=
(-1)^{-q}\submatrix{M}{k-q}{k}{(\lambda)[-q]} $.

Since $T$ is a unitary matrix,
$T^{(\lambda)}T^{(\lambda)\dag}=I_{n_1^{(\lambda)}+n_2^{(\lambda)}}$
also holds and gives 
$\sum_{q'} \submatrix{M}{k}{k+q'}{(\lambda)[-q']} 
\submatrix{M}{k-q}{k+q'}{(\lambda)[-q-q']\dag} 
= \delta_{q0}I_{n_k^{(\lambda)}}$. Thus, we obtain
\begin{align}
  (1-a^{(\lambda)2})I_{n_k^{(\lambda)}}
  =\submatrix{M}{k}{k+1}{(\lambda)[-1]}
  \submatrix{M}{k}{k+1}{(\lambda)[-1]\dag}, 
\end{align}
which implies that $a^{(\lambda)2}\leq 1$ is required as the diagonal elements
of the right-hand side of the above is non-negative.  Note that
$\submatrix{M}{k}{k+1}{(\lambda)[-1]}=0$ holds for $a^{(\lambda)2}=1$, which
means that $T^{(\lambda)}$ is already diagonal. We treat $a^{(\lambda)2}=1$ case
separately and take a basis such that $a^{(0)2}=1$ and $a^{(m)2}< 1$
for $m=1,\dots,M$ are satisfied. Then, $P_0^{(0)}$ and $T^{(0)}$ are
diagonal matrices whose eigenvalues are 1 or $-1$.

We focus on the diagonalization of $T^{(m)}$ ($m=1,\dots,M$).
Here, we recall theorems of the linear algebra that the rank of the
product of two matrices is not larger than the ranks of the two
matrices and that the rank of $m \times m'$ matrix is not larger than
$m$ nor $m'$.  
These tell us that
$n_k^{(m)}={\rm rank}(\submatrix{M}{k}{k+1}{(m)[-1]}
\submatrix{M}{k}{k+1}{(m)[-1]\dag})\leq {\rm
  rank}(\submatrix{M}{k}{k+1}{(m)[-1]})\leq n_{k+1}^{(m)}$. By
substituting $k+1$ for $k$, we also find
$n_{k+1}^{(m)}={\rm rank}(\submatrix{M}{k+1}{k}{(m)[1]}
\submatrix{M}{k+1}{k}{(m)[1]\dag})\leq {\rm
  rank}(\submatrix{M}{k+1}{k}{(m)[1]})\leq n_{k}^{(m)}$.  Therefore,
we conclude $n_1^{(m)}=n_2^{(m)}\equiv {r}^{(m)}$, which means
$\submatrix{M}{k}{k+1}{(m)[-1]}$ is an ${r}^{(m)}\times {r}^{(m)}$ square
matrix. Then, by using a unitary matrix that satisfies
$\submatrix{U}{k}{k+1}{(m)}\submatrix{U}{k}{k+1}{(m)\dag}=I_{{r}^{(m)}}$,
we can rewrite $\submatrix{M}{k}{k+1}{(m)[-1]}$ as
\begin{align}
  \submatrix{M}{k}{k+1}{(m)[-1]}=\sqrt{1-a^{(m)2}}
  \submatrix{U}{k}{k+1}{(m)}. 
\end{align}
Also, we find
$\submatrix{M}{k+1}{k}{(m)[1]}=- \submatrix{M}{k}{k+1}{(m)[-1]\dag}
=-\sqrt{1-a^{(m)2}}\submatrix{U}{k}{k+1}{(m)\dag}$.

From the above discussions, we can denote $P_0^{(m)}$ and $T^{(m)}$ by
\begin{align}
  P_0^{(m)}  =
  \begin{pmatrix}
      -I_{{r}^{(m)}}&    0\\
      0&I_{{r}^{(m)}}
  \end{pmatrix},
         \qquad 
  T^{(m)}  =
  \begin{pmatrix}
      \cos\theta^{(m)}I_{{r}^{(m)}}&
      \sin\theta^{(m)}\submatrix{U}{1}{2}{(m)}\\
      -\sin\theta^{(m)}\submatrix{U}{1}{2}{(m)\dag}&\cos\theta^{(m)}I_{{r}^{(m)}}
  \end{pmatrix},
\end{align}
where we have defined 
\begin{align}
    \cos\theta^{(m)}=a^{(m)}, \qquad 
    \sin\theta^{(m)}=\sqrt{1-a^{(m)2}}.
\end{align}
The angle $\theta^{(m)}$ can be chosen to satisfy
$0< \theta^{(m)}< \pi$. Note that
$\theta^{(m)}\neq \theta^{(m')}$ holds for $m \neq m'$ as
$a^{(m)}\neq a^{(m')}$ does. By using the unitary matrix:
\begin{align}
  V^{(m)}=
\begin{pmatrix}
    i\submatrix{U}{1}{2}{(m)\dag} &0\\
   0 &I_{{r}^{(m)}}
\end{pmatrix}, 
\end{align}
we can change the basis as 
\begin{align}\label{tmbch1}
  P_0^{(m)}
  \to V^{(m)}P_0^{(m)}V^{(m)\dag}=P_0^{(m)}=p_0\otimes I_{{r}^{(m)}},\qquad 
  T^{(m)}\to V^{(m)}T^{(m)}V^{(m)\dag}
  =t_1^{(m)}\otimes I_{{r}^{(m)}},
\end{align}
where $p_0$ and $t_1^{(m)}$ are defined by using the Pauli matrices
$\sigma_i$ as
\begin{align}
  p_0=-\sigma_3,\qquad 
  t_1^{(m)}=i \sin\theta^{(m)}
  \sigma_1+\cos\theta^{(m)}I_2
  =e^{i\theta^{(m)}\sigma_1}. 
\label{s1z2tmsig2}
\end{align}

As the final step to diagonalize the twist matrices, we consider a
$y$--dependent gauge transformation, under which $p_0$ and $t_1^{(m)}$
transform as
\begin{align}
&  p_0\to p_0'=\varOmega^{(m)}(-y)p_0\varOmega^{(m)\dag}(y)=p_0,\\
&  t_1^{(m)}\to t_1^{(m)}{}'=\varOmega^{(m)}(y+1)t_1^{(m)}\varOmega^{(m)\dag}(y)=(-1)^{l^{(m)}}I_2,
\end{align}
where
\begin{align}
  \varOmega^{(m)}(y)=\exp\left[i\left(-\theta^{(m)}+l^{(m)}\pi\right)y\sigma_1\right],
  \qquad l^{(m)}\in{\mathbb Z}.
\end{align}
Thus, we can choose a basis such that $P_0$ and $T$ are simultaneously
diagonalized, where the submatrices in eq.~\eqref{p0tbds1z21} are
given by
\begin{align}
  P_0^{(m)}=-\sigma_3\otimes I_{{r}^{(m)}}, \qquad 
  T^{(m)}=(-1)^{l^{(m)}}I_2\otimes I_{{r}^{(m)}},
  \quad {\rm for} \quad m=1,\dots,M,
\end{align}
in addition to the already diagonal ones $P_0^{(0)}$ and $T^{(0)}$.
It should be mentioned that there are several diagonal forms
of twist matrices depending on $l^{(m)}$ in an equivalence class. 

%
\section{Two-dimensional orbifolds}
\label{sec:2Dorb}
%

In the following sections, we discuss gauge theories on
$M^4\times T^2/{\mathbb Z}_N$ $(N=2,3,4,6)$. We here summarize the
basic properties of the 2D orbifolds.  
We particularly pay attention to relations satisfied by translation and
  rotation operations on $T^2/{\mathbb Z}_N$
  since these relations give rise to restriction of 
  the forms of twist matrices corresponding to the operations. 

A compactification on a 2D torus $T^2$ is obtained from the 2D
Euclidean space ${\mathbb R}^2$, which we call the universal covering
space, by modding out a 2D lattice $\bm \Lambda$ as
$T^2= {\mathbb R}^2/\bm \Lambda$. We denote the 2D lattice by
$\bm \Lambda=\{n_1\bm \lambda_1+n_2\bm \lambda_2~|~n_1,n_2\in{\mathbb Z}\}$, 
where $\bm \lambda_i$ $(i=1,2)$ are linearly
  independent basis vectors.  Coordinate vectors $\bm y$ and $\bm y'$
on $T^2$ are identified if $\bm y'-\bm y\in \bm \Lambda$ is satisfied.
In other words, any $\bm y$ on $T^2$ satisfies the following
identification:
\begin{align}\label{idenT2}
  \bm y\sim \bm y+n_1\bm \lambda_1+n_2\bm \lambda_2. 
\end{align}

It is convenient to use the complex coordinate system to deal with
$T^2$ and 2D orbifolds. Let $z$ be a complex coordinate $z=y^1+iy^2$,
where $y^1$ and $y^2$ are Cartesian coordinates on $T^2$. Then, the
identification in eq.~\eqref{idenT2} is expressed as
\begin{align}\label{ziden1}
  z\sim z+n_1+n_2\tau, 
\end{align}
where we have taken $|\bm \lambda_1|=1$ for simplicity of the notation
without loss of generality.  The geometry of $T^2$ is encoded in the
complex parameter $\tau$, which must satisfy ${\rm Im}\tau\neq 0$ to span $T^2$.
A fundamental region of $T^2$, which is an
independent region of the covering space under the identification in
eq.~\eqref{ziden1}, is now given by $\{p+q\tau~|~p,q\in[0,1)\}$.

It is natural to define the translations ${\cal T}_1$ and ${\cal T}_2$
as
\begin{align}\label{ztransdef1}
  {\cal T}_1:z\to z+1, \qquad 
  {\cal T}_2:z\to z+\tau. 
\end{align}
Then, the identification in eq.~\eqref{ziden1} is expressed as
$z\sim {\cal T}_1^{n_1}{\cal T}_2^{n_2}z$.  Note that
$[{\cal T}_1,{\cal T}_2]=0$ holds, as expected from the 2D
translational invariance.\footnote{The 2D translational invariance may
  be broken by, e.g., non-trivial field configurations that give
  magnetic flux~\cite{Bachas}.  While these possibilities are
  interesting, we do not consider them here.}

The orbifold $T^2/{\mathbb Z}_N$ is obtained from $T^2$ by
further modding out an Abelian discrete group ${\mathbb Z}_N$
$(N=2,3,4,6)$, whose elements are generated by the $N$--th root of
unity $e^{2\pi i/N}$ on the complex coordinate system.  We define the
${\mathbb Z}_N$ rotation ${\cal R}_0$ as
\begin{align}\label{idenZn}
  {\cal R}_0:z\to e^{2\pi i/N}z.
\end{align}
Then, the orbifold $T^2/{\mathbb Z}_N$ is given by imposing the
identification under the rotation~\eqref{idenZn} on the complex
coordinate $z$ on the $T^2$ torus: $z\sim {\cal R}_0z$. For $N=2$,
$\tau$ is arbitrarily chosen as long as ${\rm Im}\tau\neq 0$ is
satisfied.  On the other hand, $\tau$ is restricted by consistency in
the $N=3,4,6$ cases.  We hereafter take $\tau = e^{2\pi i/N}$ for
$N=3,4,6$.  From eq.~\eqref{idenZn}, we see $({\cal R}_0)^N={\cal I}$,
which is the identity operation.

For the $N=3,4,6$ cases, ${\cal T}_{2}$ can be rewritten as
${\cal T}_{2}={\cal R}_0{\cal T}_1{\cal R}_0^{-1}$.  More generally,
we can define the translations along $\tau^{m-1}=e^{2\pi i(m-1)/N}$
$(m\in \{1,\dots,N\})$ direction as
\begin{align}
{\cal T}_{m}={\cal R}_0^{m-1}{\cal T}_1{\cal R}_0^{1-m},
  \qquad {\cal T}_{m}:z\to z+\tau^{m-1}, \qquad {\rm for} \quad N=3,4,6.
\label{2D-Tn}
\end{align}
Since ${\cal T}_{m}$ are translations, they commute with each
other and obey $[{\cal T}_{m},{\cal T}_{m'}]=0$ for any pairs of
$m$ and $m'$. In addition, as $\tau$ is the $N$--th root of
unity for the $N=3,4,6$ cases, these translations satisfy the
relations shown in Table~\ref{T-translations}.
\begin{table}[t]
\bctr
\renewcommand{\arraystretch}{1.4}
\begin{tabular}{|l|l|c|}
    \hline
$T^2/{\mathbb Z}_N$ & relations among translations & $\tau$ \\ \hline
$T^2/{\mathbb Z}_3$ & $\prod_{m=1}^3{\cal T}_m={\cal I}$ & $e^{2\pi i/3}$\\
$T^2/{\mathbb Z}_4$ & $\prod_{m=1}^4{\cal T}_m={\cal I}$,~~~${\cal T}_1{\cal T}_3
={\cal T}_2{\cal T}_4={\cal I}$ & $e^{2\pi i/4}$\\
$T^2/{\mathbb Z}_6$ & $\prod_{m=1}^6{\cal T}_m={\cal I}$,~~
${\cal T}_1{\cal T}_4={\cal T}_2{\cal T}_5={\cal T}_3{\cal T}_6={\cal I}$,~~                                                 
${\cal T}_1{\cal T}_3{\cal T}_5={\cal T}_2{\cal T}_4{\cal T}_6={\cal I}$
& $e^{2\pi i/6}$
\\    \hline
\end{tabular}
\caption{Relations among translations}
\label{T-translations}
\ectr
\end{table}
These relations are summarized as
$({\cal T}_m {\cal R}_0^p)^{N/p} = {\cal I}$ with integers $p$ and
$N/p$~\cite{S&S}.  We can choose ${\cal T}_1$ and ${\cal T}_2$ as the
basis of the 2D translations. Using the above mentioned relations, any
${\cal T}_m$ can be expressed by ${\cal T}_1$ and ${\cal T}_2$ and,
thus, be expressed by ${\cal T}_1$ and ${\cal R}_0$ for the $N=3,4,6$
cases. On the other hand, for the $N=2$ case, ${\cal T}_2$ cannot be
written by ${\cal T}_1$ and ${\cal R}_0$ for arbitrary $\tau$
with ${\rm Im}\tau\neq 0$.

On the universal covering space of an orbifold $T^2/{\mathbb Z}_N$,
there exist invariant points under ${\cal R}_0$ up to the translations
in eq.~\eqref{ztransdef1}, called fixed points.  One can also define the
rotations around the fixed points on the orbifold as shown below. If
$z_{\rm F}$ is a fixed point, it follows that
\begin{align}\label{fxdptg1}
  z_{\rm F}=e^{2\pi i/N}z_{\rm F}+n_1+n_2\tau, \qquad n_1,n_2\in{\mathbb Z}. 
\end{align}
Let $z_{{\rm F},N}^{(n_1,n_2)}$ be the solutions of the above
equation. We obtain the following solutions depending on $N$:
\begin{align}\label{z23fp}
   & z_{{\rm F},2}^{(n_1,n_2)}={n_1+n_2\tau\over 2}\qquad ({\rm Im}\tau\neq 0),
   &&z_{{\rm F},3}^{(n_1,n_2)}={2n_1-n_2+(n_1+n_2)e^{2\pi i/3}\over 3},\\
   &z_{{\rm F},4}^{(n_1,n_2)}={n_1-n_2+(n_1+n_2)e^{2\pi i/4}\over 2},
   &&z_{{\rm F},6}^{(n_1,n_2)}=-n_2+(n_1+n_2)e^{2\pi i/6}.
\end{align}
     
Notice that any operation
${\cal T}_1^{n_1}{\cal T}_2^{n_2}{\cal R}_0$
  $(n_1,n_2\in{\mathbb Z})$ gives a ${\mathbb Z}_N$ rotation around a
fixed point $z_{{\rm F},N}^{(n_1,n_2)}$. This is understood
because the solutions of the equation
${\cal T}_1^{n_1}{\cal T}_2^{n_2}{\cal R}_0z-u=e^{2\pi i/N}(z-u)$ 
for $u$ are nothing but the fixed points defined in
eq.~\eqref{fxdptg1}.  Thus, the relation
$({\cal T}_1^{n_1}{\cal T}_2^{n_2}{\cal R}_0)^N={\cal I}$ is
satisfied.  Note that there is a ${\mathbb Z}_2$ subgroup in
${\mathbb Z}_4$. Also, there are ${\mathbb Z}_3$ and ${\mathbb Z}_2$
subgroups in ${\mathbb Z}_6$. Thus, ${\mathbb Z}_2$ operations in the
$T^2/{\mathbb Z}_4$ case and ${\mathbb Z}_3$ and ${\mathbb Z}_2$
operations in the $T^2/{\mathbb Z}_6$ case are naturally defined.  We
denote the fixed points under the subgroup ${\mathbb Z}_2$ in the
$N=4$ case by $z_{{\rm F},4,2}^{(n_1,n_2)}=(n_1+in_2)/2$.  The $\pi$ rotation
around $z_{{\rm F},4,2}^{(n_1,n_2)}$ is given by
${\cal T}_1^{n_1}{\cal T}_2^{n_2}{\cal R}_0^2$ in the $N=4$ case, which 
satisfies $({\cal T}_1^{n_1}{\cal T}_2^{n_2}{\cal R}_0^2)^2={\cal I}$.  Also,
the fixed points under the subgroup ${\mathbb Z}_2$ and
${\mathbb Z}_3$ in the $N=6$ case are denoted by
$z_{{\rm F},6,2}^{(n_1,n_2)}=(n_1+n_2e^{2\pi i/6})/2$ and
$z_{{\rm F},6,3}^{(n_1,n_2)}=[(n_1-n_2)+(n_1+2n_2)e^{2\pi i/6}]/3$, respectively.
The $\pi $ rotation around $z_{{\rm F},6,2}^{(n_1,n_2)}$ and the $2\pi /3$
rotation around $z_{{\rm F},6,3}^{(n_1,n_2)}$ are given by
${\cal T}_1^{n_1}{\cal T}_2^{n_2}{\cal R}_0^3$ and
${\cal T}_1^{n_1}{\cal T}_2^{n_2}{\cal R}_0^2$ in the $N=6$ case, where
$({\cal T}_1^{n_1}{\cal T}_2^{n_2}{\cal R}_0^3)^2={\cal I}$ and
$({\cal T}_1^{n_1}{\cal T}_2^{n_2}{\cal R}_0^2)^3={\cal I}$ are satisfied.

As the basis of the translations and rotations defined above, we can
choose $\{{\cal T}_1,{\cal T}_2,{\cal R}_0\}$ for the $N=2$ case and
$\{{\cal T}_1,{\cal R}_0\}$ for the $N=3,4,6$ cases. Any operations
are expressed by these basis operations. The translations ${\cal T}_m$
and the rotations around the fixed points
satisfy several relations shown above. These relations are redundant
but give sufficient ones that are satisfied by the basis
operations.

For the $N=3,4,6$ cases, except for ${\cal R}_0^N={\cal I}$, we can
derive the relations for the rotations discussed above using
the ones for the translations. For example, let us
consider ${\mathbb Z}_N$ $(N=3,4,6)$ rotations around arbitrary fixed
points, which are given by
${\cal T}_1^{n_1}{\cal T}_2^{n_2}{\cal R}_0$ and satisfy
$({\cal T}_1^{n_1}{\cal T}_2^{n_2}{\cal R}_0)^N={\cal I}$ as
discussed. This relation is derived by using the relations for the
translations as follows. First, notice that we get the relation
${\cal R}_0^k{\cal T}_1^i{\cal T}_2^j={\cal T}_{1+k}^i {\cal
  T}_{2+k}^j{\cal R}_0^{k}$, where ${\cal T}_{m+N}={\cal T}_m$, from
the definition of ${\cal T}_m$ given by eq.~\eqref{2D-Tn}.  Using this
relation, we can gather ${\cal R}_0$ in the following equation and
obtain
\begin{align}
  ({\cal T}_1^{n_1}{\cal T}_2^{n_2}{\cal R}_0)^N ={\cal T}_1^{n_1}{\cal
  T}_2^{n_1+n_2}\cdots {\cal T}_N^{n_1+n_2} {\cal T}_1^{n_2}{\cal R}_0^N
  =({\cal T}_1\cdots{\cal  T}_N)^{n_1+n_2}={\cal I}^{n_1+n_2}={\cal I}, 
\end{align}
where we have also used ${\cal R}_0^N={\cal I}$, $[{\cal T}_m,{\cal T}_{m'}]=0$ and
$\prod_{m=1}^N{\cal T}_m={\cal I}$. As another example, we consider
the $\pi$ rotation ${\cal T}_1^{n_1}{\cal T}_2^{n_2}{\cal R}_0^3$
in the $N=6$ case. We see
$({\cal T}_1^{n_1}{\cal T}_2^{n_2}{\cal R}_0^3)^2= {\cal
  T}_1^{n_1}{\cal T}_2^{n_2}{\cal T}_4^{n_1}{\cal T}_5^{n_2}
{\cal R}_0^6= ({\cal T}_1{\cal T}_4)^{n_1}({\cal T}_2{\cal
    T}_5)^{n_2}={\cal I} $, where we have used
${\cal R}_0^6={\cal I}$, $[{\cal T}_m,{\cal T}_{m'}]=0$ and
${\cal T}_1{\cal T}_4={\cal T}_2{\cal T}_5={\cal I}$.
Thus, the relation
$({\cal T}_1^{n_1}{\cal T}_2^{n_2}{\cal R}_0^3)^2={\cal I}$ is
derived.  As in these cases, the other relations for the rotations
discussed above are also derived using the relations for
the translations. In addition, we can also consider $N$
successive rotations around different fixed points, which are
always expressed by the translations as
$\prod_{a=1}^N ({\cal T}_1^{n_1^{(a)}}{\cal T}_2^{n_2^{(a)}}
{\cal R}_0) ={\cal T}_1^{n_1'}{\cal T}_2^{n_2'}$,
where $n_1^{(a)}$, $n_2^{(a)}$, $n_1'$ and $n_2'$ are integers. 
Such relations are also derived from the relations for the
translations.  Thus, it is sufficient to take care with
$[{\cal T}_m,{\cal T}_{m'}]=0$ $(m,m'\in\{1,\dots, N\})$, the
relations in Table~\ref{T-translations} and ${\cal R}_0^N={\cal I}$ as
independent ones for the $N=3,4,6$ cases.

In the following sections, we consider gauge theories on
$M^4\times T^2/{\mathbb Z}_N$. We denote coordinates on $M^4$ and
$T^2/{\mathbb Z}_N$ by $x^\mu$ ($\mu=0,1,2,3$) and $z=x^5+ix^6$, respectively.  
As in the $S^1/{\mathbb Z}_2$ case, the translations and the
${\mathbb Z}_N$ rotations accompany non-trivial twists in the
representation space of $G'=U(n)$,
  under which the Lagrangian is supposed to be invariant. We denote the twist
matrices corresponding to these operations by the Italic character
symbol, e.g., $T_m$ for ${\cal T}_m$, which are unitary matrices
belonging to the fundamental representation. These matrices are
constrained by the relations satisfied by the corresponding
translations and rotations. For example, the relation
$[{\cal T}_m,{\cal T}_{m'}]=0$ gives the constraint $[T_m,T_{m'}]=0$ for the
twist matrices.  As shown in the following sections, without loss of
generality, the twist matrices are simplified by taking a suitable
basis and gauge with the help of the constraints.

%
\section{$T^2/\mathbb{Z}_2$}
\label{sec:t2z2}
%

In this section, we discuss the twist matrices in the
$T^2/\mathbb{Z}_2$ case.  There are independent twist matrices $T_1$,
$T_2$ and $R_0$ satisfying the constraints:
\begin{align}\label{condt2zng1}
  [T_1,T_2]=0, \qquad  R_0^2= (T_1R_0)^2=(T_2R_0)^2=(T_1T_2R_0)^2=I,
\end{align}
where $I$ is the unit matrix.
With a similar discussion as in the
$S^1/{\mathbb Z}_2$ case given in section~\ref{sec:s1z2} to get
eqs.~\eqref{tmbch1} and~\eqref{s1z2tmsig2},
corresponding to eq.~\eqref{p0tbds1z21} for $P_0$ and $T_1$, 
we take block-diagonal
forms of $R_0$ and $T_1$. 
Then, the twist matrices are given by
\begin{gather}\label{r0tbdt2z21}
 R_0=
  \begin{pmatrix}
      R_0^{(0)}&&&\\
      &R_0^{(1)}&&\\
      &&\ddots &\\
      && &R_0^{(M)}
  \end{pmatrix}, 
\qquad  T_1=
  \begin{pmatrix}
      T_1^{(0)}&&&\\
      &T_1^{(1)}&&\\
      &&\ddots &\\
      && &T_1^{(M)}
  \end{pmatrix},\\\label{r0tbdt2z22}
    T_2=
  \begin{pmatrix}
      T_2^{(00)}&T_2^{(01)}&\dots&T_2^{(0M)}\\
      T_2^{(10)}&T_2^{(11)}&\dots&T_2^{(1M)}\\
      \vdots&\vdots&\ddots & \vdots\\
      T_2^{(M0)}&T_2^{(M1)} & \dots & T_2^{(MM)}
  \end{pmatrix},
\end{gather}
where we have denoted submatrices of $R_0$, $T_1$ and $T_2$
  by $R_0^{(\lambda)}$, $T_1^{(\lambda)}$ and $T_2^{(\lambda\lambda)}$
  $(\lambda=0,1,\dots,M)$, respectively.  As in
section~\ref{sec:s1z2}, we take $R_0^{(0)}$ and $T_1^{(0)}$ as
$r^{(0)}\times r^{(0)}$ diagonal matrices whose eigenvalues are 1 or $-1$.
Parameters such as $r^{(0)}$ which represent the size of matrices
are non-negative integers.
The same applies to the following ones.
The other submatrices $R_0^{(m)}$ and $T_1^{(m)}$
  $(m=1,\dots,M)$ are
defined to be $2r^{(m)}\times 2r^{(m)}$ ones and are 
given by
\begin{align}\label{r0t1mZ21}
  R_0^{(m)}&=            \begin{pmatrix}
                -I_{{r}^{(m)}}&0\\
                0&I_{{r}^{(m)}}
            \end{pmatrix}
                   =-\sigma_3\otimes I_{{r}^{(m)}}, \\\label{r0t1mZ22}
T_1^{(m)}&=
  \begin{pmatrix}
      \cos\theta^{(m)}I_{{r}^{(m)}}&
      i\sin\theta^{(m)}I_{{r}^{(m)}}\\
      i\sin\theta^{(m)}I_{{r}^{(m)}}&\cos\theta^{(m)}I_{{r}^{(m)}}
  \end{pmatrix}
                                      =e^{i\theta^{(m)}\sigma_1} \otimes I_{{r}^{(m)}},
\end{align}
where 
$0<\theta^{(m)}<\pi$ and
$\theta^{(m)}\neq \theta^{(m')}$ for $m\neq m'$.

Without disturbing the above structure of $R_0$ and $T_1$, we can
  simplify the form of $T_2$ by using the constraints in
  eq.~\eqref{condt2zng1} and unitary transformations.  As shown in
  appendix~\ref{ap:z201}, one finds that the submatrices $T_2^{(m0)}$
  and $T_2^{(0m)}$ vanish. Then, as discussed in
  appendix~\ref{ap:z202}, we can simplify $T_2^{(00)}$, 
  keeping diagonal forms of $R_0^{(0)}$ and
  $T_1^{(0)}$.  The submatrices $R_0^{(0)}$,
  $T^{(0)}$ and $T_2^{(00)}$ in eqs.~\eqref{r0tbdt2z21}
  and~\eqref{r0tbdt2z22} are written as
\begin{gather}
  R_0^{(0)}=\begin{pmatrix}
                R_0^{(0),1}&0\\
                0&R_0^{(0),2}
            \end{pmatrix},\qquad 
  T_1^{(0)}=\begin{pmatrix}
                T_1^{(0),1}&0\\
                0&T_1^{(0),2}
            \end{pmatrix}=  \begin{pmatrix}
                -I_{r^{(0)}_1}&0\\
                0&I_{r^{(0)}_2}
            \end{pmatrix},
\\
  T_2^{(00)}=  \begin{pmatrix}
                T_2^{(00),1}&0\\
                0&T_2^{(00),2}
            \end{pmatrix},
\end{gather}
where $r^{(0)}_1+r^{(0)}_2=r^{(0)}$, and 
$R_0^{(0),a}$, $T_1^{(0),a}$ and $T_2^{(00),a}$ $(a=1,2)$ are
$r^{(0)}_a\times r^{(0)}_a$ matrices.
The submatrices $R_0^{(0),a}$ $(a=1,2)$ are diagonal and
have eigenvalues $1$ or $-1$.  The submatrices
  $R_0^{(0),a}$, $T_1^{(0),a}$ and $T_2^{(00),a}$ are further
  rearranged as
\begin{align}
 R_0^{(0),a}&=
  \begin{pmatrix}
      R_{0}^{(0),a,0}&&&\\
      &R_{0}^{(0),a,1}&&\\
      &&\ddots &\\
      && &R_{0}^{(0),a,M_{0}^{(a)}}
  \end{pmatrix},
  \\
  T_1^{(0),a}&=
  \begin{pmatrix}
      T_1^{(0),a,0}&&&\\
      &T_1^{(0),a,1}&&\\
      &&\ddots &\\
      && &T_1^{(0),a,M_{0}^{(a)}}
  \end{pmatrix},
  \\
 T_2^{(00),a}&=
  \begin{pmatrix}
      T_2^{(00),a,0}&&&\\
      &T_2^{(00),a,1}&&\\
      &&\ddots &\\
      && &T_2^{(00),a,M_{0}^{(a)}}
  \end{pmatrix},
  \label{r0a0andt1t2}
\end{align}
where $R_{0}^{(0),a,0}$, $T_{1}^{(0),a,0}$ and $T_{2}^{(00),a,0}$ are
$r_a^{(0),0}\times r_a^{(0),0}$ diagonal matrices whose
eigenvalues are 1 or $-1$. In addition, for $m=1,\dots, M_{0}^{(a)}$,
the submatrices are given by 
\begin{align}
  R_{0}^{(0),a,m}= -\sigma_3 \otimes I_{r_a^{(0),m}}, \quad 
  T_{1}^{(0),a,m}=(-1)^aI_2\otimes I_{r_a^{(0),m}}, \quad 
  T_{2}^{(00),a,m}=e^{i\phi^{(0),a, m}\sigma_1}\otimes I_{r_a^{(0),m}}, 
\label{Z2R0(0)am2}
\end{align}
which are defined to be $2r_a^{(0),m}\times 2r_a^{(0),m}$ matrices.
The real parameters $\phi^{(0),a, m}$ 
  satisfy $0<\phi^{(0),a, m}<\pi$ and
  $\phi^{(0),a, m} \neq \phi^{(0),a, m'}$ for $m\neq m'$.

Let us start to discuss $T_2^{(mm')}$. As shown in appendix~\ref{ap:z203},
  there exist a basis where $T_2^{(mm')}=0$ for $m\neq m'$ is satisfied.
Then, the twist matrix $T_2$ becomes a diagonal form as
\begin{align}
  T_2=
  \begin{pmatrix}
      T_2^{(00)}&&&\\
      &T_2^{(1)}&&\\
      &&\ddots &\\
      && &T_2^{(M)}
  \end{pmatrix},
            \label{T2mZ22}
\end{align}
where $T_2^{(00)}$ and $T_2^{(m)}$ $(m=1,\dots,M)$ are
$r^{(0)}\times r^{(0)}$ and $2r^{(m)}\times 2r^{(m)}$ submatrices,
respectively.

The submatrices $R_0^{(m)}$ and $T^{(m)}_1$ in
  eq.~\eqref{r0tbdt2z21} and $T_2^{(m)}$ in eq.~\eqref{T2mZ22} are
  simplified by unitary transformations. As shown in
  appendix~\ref{ap:z204}, they are finally decomposed into $2\times 2$
  submatrices. The submatrices $R_0^{(m)}$, $T^{(m)}_1$ and $T_2^{(m)}$
  are written as
\begin{align}
  R_0^{(m)}&=
  \begin{pmatrix}
      R_0^{(m),1}&&&\\
      &R_0^{(m),2}&&\\
      &&\ddots &\\
      && &R_0^{(m),M^{(m)}}
  \end{pmatrix}, \\
  T_1^{(m)}&=
  \begin{pmatrix}
      T_1^{(m),1}&&&\\
      &T_1^{(m),2}&&\\
      &&\ddots &\\
      && &T_1^{(m),M^{(m)}}
  \end{pmatrix},
  \\
 T_2^{(m)}&=
  \begin{pmatrix}
      T_2^{(m),1}&&&\\
      &T_2^{(m),2}&&\\
      &&\ddots &\\
      && &T_2^{(m),M^{(m)}}
  \end{pmatrix},
            \label{bblockt2z2}
\end{align}
where $R_0^{(m),m'}$,  $T_1^{(m),m'}$ and
$T_2^{(m),m'}$ $(m'=1,\dots,
M^{(m)})$ are $2r^{(m),m'}\times 2r^{(m),m'}$ submatrices,
and $\sum_{m'=1}^{M^{(m)}}r^{(m),m'}=r^{(m)}$. 
These submatrices are given by
\begin{gather}
  R_0^{(m),m'}=-\sigma_3\otimes I_{r^{(m),m'}}, \qquad 
  T_1^{(m),m'}=e^{i\theta^{(m)}\sigma_1}\otimes I_{r^{(m),m'}}, 
  \label{bblockr0t1z22}\\
  T_2^{(m),m'}=
  \begin{pmatrix}
      \cos\phi^{(m),m'}I_{r^{(m),m'}}&
      i\sin\phi^{(m),m'}I_{r^{(m),m'}}\\
      i\sin\phi^{(m),m'}I_{r^{(m),m'}}&\cos\phi^{(m),m'}I_{r^{(m),m'}}
  \end{pmatrix}
  =e^{i\phi^{(m),m'}\sigma_1}\otimes I_{r^{(m),m'}}.
\label{Z2-T2mm'2}  
\end{gather}
The real parameters $\phi^{(m),m'}$ satisfy
  $\phi^{(m),m'}\neq \phi^{(m),m''}$ for $m'\neq m''$.

From the above discussions, the twist matrices are
  simultaneously 
rearranged to be block-diagonal with each diagonal blocks
being a $2 \times 2$ matrix. 
For $R_0^{(0),a,m}$, $T_1^{(0),a,m}$ and $T_2^{(00),a,m}$
in eq.~\eqref{Z2R0(0)am2}, the $2 \times 2$ matrices are given by
\begin{align}
 r_0 = \left(
\begin{array}{cc}
-1 & 0 \\
0 & 1
\end{array}
\right),
\qquad
 t_1 = \pm \left(
\begin{array}{cc}
1 & 0 \\
0 & 1 
\end{array}
\right),
\qquad
 t_2 = \left(
\begin{array}{cc}
b_2 & b_1 \\
b_1 & b_2 
\end{array}
\right),
\label{Z2-r0-1}
\end{align}
where $b_1$ is a pure imaginary number with ${\rm Im}\,b_1 > 0$
and $b_2$ is a real number
which satisfy $|b_1|^2 + b_2^2 = 1$.
For 
  ${R}_0^{(m),m'}$, ${T}_1^{(m),m'}$ and ${T}_2^{(m),m'}$
  in eqs.~\eqref{bblockr0t1z22} and \eqref{Z2-T2mm'2},
we can rearrange them
to be block-diagonal with each diagonal blocks
being a $2 \times 2$ matrix in the following form:
\begin{align}
 r_0 = \left(
\begin{array}{cc}
-1 & 0 \\
0 & 1
\end{array}
\right),
\qquad
 t_1 = \left(
\begin{array}{cc}
a_2 & a_1 \\
a_1 & a_2 
\end{array}
\right),
\qquad
 t_2 = \left(
\begin{array}{cc}
b_2 & b_1 \\
b_1 & b_2 
\end{array}
\right),
\label{Z2-r0-2}
\end{align}
where $a_1$ and $b_1$ are pure imaginary numbers with ${\rm Im}\,a_1 > 0$
and $a_2$ and $b_2$ are real numbers
which satisfy $|a_1|^2 + a_2^2 = 1$ and $|b_1|^2 + b_2^2 = 1$.
As discussed in section~\ref{Sec:GaugeTr}, these matrices are
simultaneously diagonalized by a gauge transformation.

The remaining $R_0^{(0),a,0}$, $T_1^{(0),a,0}$ and $T_2^{(00),a,0}$ 
are already diagonal and
can be treated as block-diagonal matrices with each
diagonal blocks being a $1\times1$ matrix.

%
\section{$T^2/\Z3$}
\label{sec:t2z3}
%

Next we examine the $T^2/\Z3$ orbifold.  The independent
twist matrices are $R_0$ and $T_1$, by which we can define the other
ones as 
\bequ
 {T}_2={R}_0{T}_1{R}_0^{-1}, \quad
 {T}_3={R}_0^{-1}{T}_1{R}_0, \quad
 {R}_1={T}_1{R}_0, \quad
 {R}_2= T_1T_2R_0={R}_0^{-1}{T}_1^{-1}{R}_0^{-1}. 
\label{Eq:Z3-T1T2R2}
\eequ
The twist matrices $R_1$ and $R_2$ correspond to $2\pi/3$ rotations around
  $z_{{\rm F},3}^{(1,0)}$ and $z_{{\rm F},3}^{(1,1)}$ 
defined in eq.~\eqref{z23fp}, respectively.
These satisfy the following relations to represent the corresponding 
operations, 
\beqn
 &&{R}_0^3={R}_1^3={R}_2^3={R}_0{R}_1{R}_2
  ={R}_1{R}_2{R}_0={R}_2{R}_0{R}_1={I},\\
 &&{T}_{m'}{T}_m={T}_m{T}_{m'}, \quad
 {T}_1{T}_2T_3=I, \quad
 {T}_{m+1}{R}_0={R}_0{T}_{m}, 
\label{Eq:Z3-relations}
\eeqn
where $I$ denotes the unit matrix 
and $T_{m+N} = T_m$ as before. 

Without loss of generality, the twist matrix $R_0$ can be diagonalized, 
with three possible eigenvalues: $\omega^k$ ($k=1,2,3$), where 
$\omega=e^{2\pi i/3}$. 
It is convenient to divide the twist matrices $R_0$ and $T_1$ into 
$3\times 3$ blocks as 
\bequ
 R_0=\begin{pmatrix}
       \omega I_{n_1}&&\\&\omega^2 I_{n_2}&\\&&I_{n_3}
     \end{pmatrix}
    =\begin{pmatrix}(R_0)_{(11)}&(R_0)_{(12)}&(R_0)_{(13)}\\
                    (R_0)_{(21)}&(R_0)_{(22)}&(R_0)_{(23)}\\
                    (R_0)_{(31)}&(R_0)_{(32)}&(R_0)_{(33)}\end{pmatrix},  \quad
 (R_0)_{(kl)}=\omega^k\delta_{kl}I_{n_k}, 
\eequ
where $n_k$ is a non-negative integer and $I_{n_k}$ 
denotes the $n_{k} \times n_{k}$ unit matrix, and 
\bequ
 T_1=\begin{pmatrix}(T_1)_{(11)}&(T_1)_{(12)}&(T_1)_{(13)}\\
                    (T_1)_{(21)}&(T_1)_{(22)}&(T_1)_{(23)}\\
                    (T_1)_{(31)}&(T_1)_{(32)}&(T_1)_{(33)}\end{pmatrix},  \quad
 (T_1)_{(kl)}=\submatrix{M}{k}{l}{[k-l]}. 
\eequ
As before, we have introduced $n_k\times n_l$ matrices 
$\submatrix{M}{k}{l}{[k-l]}$, and use a notation of
$\submatrix{M}{k}{l}{[k-l]}=\submatrix{M}{k}{l}{[k'-l']}=\submatrix{M}{k'}{l'}{[k-l]}$
with $k'=k$ (mod 3) and $l'=l$ (mod 3).
The upper index $k-l=q$ represents the charge of the $\Z3$ symmetry 
generated by $R_0$: 
$(R_0T_1R_0^{-1})_{(k\,k-q)}=\omega^{q}\submatrix{M}{k}{k-q}{[q]}$. 

As shown in appendix~\ref{app:t2z3-1}, the conditions $R_a^3=I$ ($a=0,1,2$), 
with the help of the relations in eq.~\eqref{Eq:Z3-T1T2R2} and 
$T_1^\dagger=T_1^{-1}$, lead to those among 
$\submatrix{M}{k}{l}{[k-l]}$ as
\bequ
 \submatrix{M}{k}{k-q}{[q]}\submatrix{M}{k-q}{k}{[-q]}
 = \submatrix{M}{k}{k+q}{[-q]}\submatrix{M}{k+q}{k}{[q]}, \quad
 \submatrix{M}{k}{k}{[0]}\submatrix{M}{k}{k-q}{[q]}
 = \submatrix{M}{k}{k-q}{[q]}\submatrix{M}{k-q}{k-q}{[0]}, 
\label{Eq:Z3-ConditionAmngM}
\eequ
which are further summarized in eq.~\eqref{ZN-qq'}. 
From the second relations in eq.~\eqref{Eq:Z3-ConditionAmngM}, we find 
that each $\submatrix{M}{k}{k}{[0]}$ commutes with its dagger to be a 
normal matrix. 
Though $\submatrix{M}{k}{k}{[0]}$ are not hermitian generally in this 
$T^2/\Z3$ case in contrast to the other $T_2/\Z{N}$ cases with even $N$, 
then they can be diagonalized by a unitary transformation. %

In appendix~\ref{app:t2z3-2}, we show that in the basis where 
$\submatrix{M}{k}{k}{[0]}$ are diagonalized as 
$(\submatrix{M}{k}{k}{[0]})_{ij}=a_k^i\delta_{ij}$,  
$(\submatrix{M}{k}{k-q}{[q]})_{ij}$ vanish if $a_k^i\neq a_{k-q}^j$, 
and thus the twist matrices $R_0$ and $T_1$ can be block-diagonalized as 
\beqn
& R_0=\begin{pmatrix}R_0^{(1)}&&&\\&R_0^{(2)}&&\\&&
                         \ddots&\\&&&&R_0^{(M)}
          \end{pmatrix},
 \quad
 R_0^{(m)}=\begin{pmatrix}
           \omega I_{n_1^{(m)}}&&\\
           &\omega^2 I_{n_2^{(m)}}&\\
           &&I_{n_3^{(m)}}
          \end{pmatrix}
\\
& T_1=\begin{pmatrix}T_1^{(1)}&&&\\&T_1^{(2)}&&\\&&
                         \ddots&\\&&&&T_1^{(M)}
          \end{pmatrix},
 \quad
 T_1^{(m)}=\begin{pmatrix}
           a^{(m)}I_{n_1^{(m)}} & \submatrix{M}{1}{2}{(m)[-1]} 
            & \submatrix{M}{1}{3}{(m)[1]} \\
           \submatrix{M}{2}{1}{(m)[1]}  & a^{(m)}I_{n_2^{(m)}} 
            & \submatrix{M}{2}{3}{(m)[-1]}  \\
           \submatrix{M}{3}{1}{(m)[-1]}  & \submatrix{M}{3}{2}{(m)[1]} 
            & a^{(m)}I_{n_3^{(m)}}
          \end{pmatrix},
\eeqn
where $n_k^{(m)}$ ($m = 1, 2, \cdots, M$) are non-negative integers and 
complex parameters $a^{(m)}$ satisfy $a^{(m)} \ne a^{(m')}$ for $m \ne m'$.

The unitarity conditions $T_1T_1^\dagger=T_1^\dagger T_1=I$ are 
examined in appendix~\ref{app:t2z3-3}. 
The condition $T_1T_1^\dagger=I$ derives 
\bequ
 \submatrix{M}{k}{k-1}{(m)[1]}\submatrix{M}{k}{k-1}{(m)[1]\dagger}
 + \submatrix{M}{k}{k+1}{(m)[-1]}\submatrix{M}{k}{k+1}{(m)[-1]\dagger}
 = (1-\abs{a^{(m)}}^2)I_{n_k^{(m)}},
\eequ
to shows that, when $\abs{a^{(m)}}=1$,
$\submatrix{M}{k}{k-q}{(m)[q]}=0$ for $q=\pm1$ 
and $T_1^{(m)}$ is already diagonal. 

The cases with $0<\abs{a^{(m)}}<1$ and $\abs{a^{(m)}}=0$ are studied 
respectively in appendix~\ref{app:t2z3-3-1} and \ref{app:t2z3-3-2}. 
In both cases, after some discussions, it is shown that 
$\submatrix{M}{k}{k-q}{(m)[q]}$ with $q=\pm1$ can be diagonalized. 
Then, we can rearrange $T_1^{(m)}$ further to be 
block-diagonal, where $R_0^{(m)}$ is still diagonal, with each diagonal 
blocks being a $3\times3$ matrix in the following form: 
\bequ
 r_0=\begin{pmatrix}
        \omega&0&0 \\ 0&\omega^2&0 \\ 0&0&1
   \end{pmatrix},\qquad 
 t_1=\begin{pmatrix}
    a_3&a_2&a_1 \\ a_1&a_3&a_2 \\ a_2&a_1&a_3
   \end{pmatrix},
\label{Z3-t1}
\eequ
where $a_1$, $a_2$ and $a_3$ are complex numbers which satisfy
$a_1^3 + a_2^3 + a_3^3 - 3 a_1 a_2 a_3 = 1$,
as seen from the combination of eqs.~\eqref{Eq:Z3-S1dagger-v2}, 
\eqref{Eq:Z3-S1dagger-v2-OffDiag} and \eqref{Eq:Z3-unitarity-diag}
derived from $T_1T_2T_3=T_1T_3T_2=I$ and $T_1 T_1^{\dagger} =I$,
and $|a_1|^2 + |a_2|^2 + |a_3|^2 = 1$ and 
$\overline{a_1}a_3 + \overline{a_3}a_2 + \overline{a_2}a_1 = 0$,
as seen from eqs.~\eqref{Eq:Z3-unitarity-diag} and \eqref{Eq:Z3-unitarity-offdiag},
respectively, derived from $T_1 T_1^{\dagger} =I$.
In section~\ref{Sec:GaugeTr}, it is shown that
$3\times3$ submatrices of this form are diagonalized 
by a suitable gauge transformation. 

In the case with $|a^{(m)}|=1$, $T_1^{(m)}$ is diagonal to be treated
as a block-diagonal matrix with each diagonal blocks
being a $1\times1$ matrix.

%
\section{$T^2/\Z4$}
\label{sec:t2z4}
%

We consider the $T^2/\Z4$ orbifold.
As before, we choose the twist matrices $R_0$ and $T_1$ as independent ones,
and show that they can become block-diagonal forms containing $4 \times 4$,
$2 \times 2$ and $1 \times 1$ matrices, using constraints and unitary transformations.
The details of a derivation are given in appendix~\ref{app:t2z4}.
Here, we present the outline.

The constraints restricting the forms of $R_0$ and $T_1$ are given by
\begin{align}
T_1^{\dagger} = T_1^{-1}, \qquad
T_{m'} T_m = T_m T_{m'}, \qquad
T_1 T_3 = I,
\label{Z4-constraints}
\end{align}
where $T_{m} = R_0^{m-1} T_1 R_0^{1-m}$ as explained in section~\ref{sec:2Dorb}.

Starting from the following $R_0$ and $T_1$,
\begin{align}
R_0=
  \begin{pmatrix}
      iI_{n_1}& & & \\
       &-I_{n_2}& & \\
       & &-iI_{n_3}& \\
       & & &I_{n_4}
  \end{pmatrix},\qquad
T_1=
  \begin{pmatrix}
      (T_1)_{(11)}&(T_1)_{(12)}&(T_1)_{(13)}&(T_1)_{(14)}\\
      (T_1)_{(21)}&(T_1)_{(22)}&(T_1)_{(23)}&(T_1)_{(24)}\\
      (T_1)_{(31)}&(T_1)_{(32)}&(T_1)_{(33)}&(T_1)_{(34)}\\
      (T_1)_{(41)}&(T_1)_{(42)}&(T_1)_{(43)}&(T_1)_{(44)}
  \end{pmatrix}, 
\label{Z4-R0T1}
\end{align}
with the $n_k \times n_k$ unit matrices $I_{n_k}$ and
$n_k \times n_l$ submatrices $(T_1)_{(kl)}$,
and using eq.~\eqref{C-Z4-(ai-aj)}
derived from $T_1^{\dagger} (= T_1^{-1} = T_3) = R_0^2 T_1 R_0^{-2}$
and $T_{m'} T_{m} = T_{m} T_{m'}$, 
we can rearrange $R_0$ and $T_1$ to be block-diagonal forms 
in a similar way to the $S^1/\Z2$ case in section~\ref{sec:s1z2},
such that
\begin{align}
R_0=
    \begin{pmatrix}
        R_0^{(1)} &          &        &         \\
                 & R_0^{(2)} &        &         \\
                 &          & \ddots &         \\
                 &          &        & R_0^{(M)}
    \end{pmatrix},\qquad
T_1=
    \begin{pmatrix}
        T_1^{(1)} &          &        &         \\
                 & T_1^{(2)} &        &         \\
                 &          & \ddots &         \\
                 &          &        & T_1^{(M)}
    \end{pmatrix},
\label{Z4-R0T1-block}
\end{align}
where $R_0^{(m)}$ and $T_1^{(m)}$ $(m = 1, 2, \cdots, M)$ are 
$n^{(m)} \times n^{(m)}$ matrices given by
\begin{align}
R_0^{(m)}=
    \begin{pmatrix}
        iI_{n_1^{(m)}} &          &        &         \\
                 & -I_{n_2^{(m)}} &        &         \\
                 &          & -iI_{n_3^{(m)}} &         \\
                 &          &        & I_{n_4^{(m)}}
    \end{pmatrix},
\label{Z4-R0m}
\end{align}
and
\begin{align}
&T_1^{(m)}=
    \begin{pmatrix}
        a^{(m)}I_{n_1^{(m)}} & \submatrix{M}{1}{2}{(m)[-1]} 
         & \submatrix{M}{1}{3}{(m)[-2]} & \submatrix{M}{1}{4}{(m)[1]} \\
        \submatrix{M}{2}{1}{(m)[1]} & a^{(m)}I_{n_2^{(m)}} 
         & \submatrix{M}{2}{3}{(m)[-1]} & \submatrix{M}{2}{4}{(m)[-2]}       \\
        \submatrix{M}{3}{1}{(m)[2]} & \submatrix{M}{3}{2}{(m)[1]} 
         & a^{(m)}I_{n_3^{(m)}} & \submatrix{M}{3}{4}{(m)[-1]} \\
        \submatrix{M}{4}{1}{(m)[-1]} & \submatrix{M}{4}{2}{(m)[2]}
         & \submatrix{M}{4}{3}{(m)[1]} & a^{(m)}I_{n_4^{(m)}}
    \end{pmatrix},
\label{Z4-T1m}
\end{align}
respectively.
Here, $n^{(m)} = \sum_{k=1}^4 n_k^{(m)}$ and
submatrices in $T_1^{(m)}$ are denoted
as $(T_1^{(m)})_{(kl)} = \submatrix{M}{k}{l}{(m)[k-l]}$
for $k \ne l$ and $(T_1^{(m)})_{(kk)} = \submatrix{M}{k}{k}{(m)[0]}=a^{(m)} I_{n_k^{(m)}}$
with real parameters $a^{(m)}$ satisfying $a^{(m)} \ne a^{(m')}$ for $m \ne m'$
and restricted to $-1 \leq a^{(m)} \leq 1$.
We use a notation of
$\submatrix{M}{k}{l}{(m)[k-l]}=\submatrix{M}{k}{l}{(m)[k'-l']}=\submatrix{M}{k'}{l'}{(m)[k-l]}$
with $k'=k$ (mod 4) and $l'=l$ (mod 4).
Parameters such as $n_{k}$ and $n_k^{(m)}$ which represent the size of matrices
are non-negative integers.
The same applies to the following ones.
Because $T_1^{(m)}$ is already diagonal for $a^{(m)}=\pm 1$, 
we focus on $-1 < a^{(m)} < 1$ hereafter.

Next, using eqs.~\eqref{C-Z4-Mdaggerm}--\eqref{C-Z4-1MM}
derived from the constraints in eq.~\eqref{Z4-constraints}, 
we can choose the basis such that $\submatrix{M}{k}{l}{(m)[k-l]}$
have the forms of eqs.~\eqref{C-Z4-Mmq=1}, \eqref{C-Z4-Mmq=-1} and \eqref{C-Z4-Mmq=2},
and rearrange $R_0^{(m)}$ and $T_1^{(m)}$ to be the forms of 
block-diagonal ones such as
$R_0^{(m)}=R_0^{(m)'}\oplus R_0^{(m)''}$ and $T_1^{(m)}=T_1^{(m)'}\oplus T_1^{(m)''}$
given by
\begin{align}
&R_0^{(m)}=
	\begin{pmatrix}
		R_0^{(m)'}& \\
		& R_0^{(m)''}\\
	\end{pmatrix}
=R_0^{(m)'}\oplus R_0^{(m)''},\\
& R_0^{(m)'}=
    \begin{pmatrix}
        iI_{r^{(m)}} &          &        &         \\
                 & -I_{r^{(m)}} &        &         \\
                 &          & -iI_{r^{(m)}} &         \\
                 &          &        & I_{r^{(m)}}
    \end{pmatrix},
\label{Z4-R0m'}\\
& R_0^{(m)''}=
    \begin{pmatrix}
        iI_{n_1^{(m)'}} &          &        &         \\
                 & -I_{n_2^{(m)'}} &        &         \\
                 &          & -iI_{n_1^{(m)'}} &         \\
                 &          &        & I_{n_2^{(m)'}}
    \end{pmatrix},
\label{Z4-R0m''}
\end{align}
and 
\begin{align}
&T_1^{(m)}=
	\begin{pmatrix}
		T_1^{(m)'}& \\
		& T_1^{(m)''}\\
	\end{pmatrix}
=T_1^{(m)'}\oplus T_1^{(m)''},\\
& T_1^{(m)'}=
    \begin{pmatrix}
        a^{(m)}I_{r^{(m)}} & {\hat M}^{(m)[1]}\submatrix{U}{1}{2}{(m)} 
& {\hat M}^{(m)[2]}\submatrix{U}{1}{3}{(m)} 
& {\hat M}^{(m)[1]}\submatrix{U}{1}{4}{(m)} \\
{\hat M}^{(m)[1]}\submatrix{U}{2}{1}{(m)} 
& a^{(m)}I_{r^{(m)}} 
& {\hat M}^{(m)[1]}\submatrix{U}{2}{3}{(m)} 
& {\hat M}^{(m)[2]}\submatrix{U}{2}{4}{(m)} \\
{\hat M}^{(m)[2]}\submatrix{U}{3}{1}{(m)} 
& {\hat M}^{(m)[1]}\submatrix{U}{3}{2}{(m)} 
& a^{(m)}I_{r^{(m)}} 
& {\hat M}^{(m)[1]}\submatrix{U}{3}{4}{(m)} \\
 {\hat M}^{(m)[1]}\submatrix{U}{4}{1}{(m)} 
& {\hat M}^{(m)[2]}\submatrix{U}{4}{2}{(m)} 
& {\hat M}^{(m)[1]}\submatrix{U}{4}{3}{(m)} & a^{(m)}I_{r^{(m)}}
    \end{pmatrix},
\label{Z4-T1m'}\\
&T_1^{(m)''}=
    \begin{pmatrix}
        0 & 0 & {\tilde U}_{1~\!\!3}^{(m)} & 0 \\
        0 & 0 & 0 & {\tilde U}_{2~\!\!4}^{(m)} \\
        {\tilde U}_{3~\!\!1}^{(m)} & 0 & 0 & 0\\
        0 & {\tilde U}_{4~\!\!2}^{(m)} & 0 & 0
    \end{pmatrix},
\label{Z4-T1m''0}
\end{align}
respectively.
Here, ${\hat M}^{(m)[1]}$ are $r^{(m)} \times r^{(m)}$ diagonal matrices
with positive elements, i.e., $({\hat M}^{(m)[1]})_{ii} > 0$, 
${\hat M}^{(m)[2]}$ are $r^{(m)} \times r^{(m)}$ diagonal matrices
with non-negative elements, i.e., $({\hat M}^{(m)[2]})_{ii} \ge 0$,
$\submatrix{U}{k}{l}{(m)}$ are $r^{(m)} \times r^{(m)}$ unitary matrices
and $\tilde{U}_{k~\!\!k-2}^{(m)}$ and $\tilde{U}_{k-2~\!\!k}^{(m)}$ are 
$n_k^{(m)'} \times n_k^{(m)'}$ unitary matrices.
$R_0^{(m)''}$ and $T_1^{(m)''}$ can appear only in the case of $a^{(m)}=0$.

We perform an appropriate unitary transformation
that make all submatrices in $T_1^{(m)'}$ and $T_1^{(m)''}$ diagonalized simultaneously
such as eqs.~\eqref{C-Z4-Tm'-tr} and \eqref{C-Z4-Tm''-tr},
keeping $R_0^{(m)}$ diagonal ones,
and rearrange the rows and columns in $R_0^{(m)}$ and $T_1^{(m)}$.
Then, we find that $R_0^{(m)}$ and $T_1^{(m)}$ can take simplified block-diagonal forms 
containing $4\times 4$, $2\times 2$ and $1\times 1$ matrices as the diagonal blocks.

From eqs.~\eqref{C-Z4-R0m'} and \eqref{C-Z4-Tm'-tr},
the $4\times 4$ submatrices of $R_0^{(m)}$ and $T_1^{(m)}$ 
have the forms of $r_0$ and $t_1$ presented as
\begin{align}
r_0= \left(
\begin{array}{cccc}
	i &  &  &  \\
	 & -1 &  & \\
	 &  & -i & \\
	 &  &  & 1 
\end{array}
\right),
\qquad
t_1 = \left(
\begin{array}{cccc}
a_4 & a_3 & a_2 & a_1 \\
a_1 & a_4 & a_3 & a_2 \\
a_2 & a_1 & a_4 & a_3 \\
a_3 & a_2 & a_1 & a_4 
\end{array}
\right),
\label{Z4-t1}
\end{align}
respectively, where $a_1$ and $a_3$ are complex numbers with $a_3 = - \overline{a_1}$, 
$a_2$ is a non-negative number and $a_4$ is a real number, and they satisfy
$2|a_1|^2 + a_2^2 + a_4^2 = 1$ and $2a_2 a_4 = a_1^2 + \overline{a_1}^2$.
In section~\ref{Sec:GaugeTr}, it is shown that
$4\times4$ submatrices of this form are diagonalized 
by a suitable gauge transformation. 

In case with $a^{(m)} = 0$, from eqs.~\eqref{C-Z4-R0m''} and \eqref{C-Z4-Tm''-tr},
$R_0^{(m)}$ and $T_1^{(m)}$ can contain 
the following type of $2 \times 2$ matrices as block-diagonal elements,
\begin{align}
r'_0 = i^{n'} \left(
\begin{array}{cc}
	-1 &  0  \\
	0 & 1  
\end{array}
\right),
\qquad
t'_1 = \left(
\begin{array}{cc}
0 & 1 \\
1 & 0 
\end{array}
\right),
\label{Z4-t'1}
\end{align}
where $n'$ is an integer.
Combining $r'_0$ with an odd $n'$ and that with an even $n'$,
we can make the $4 \times 4$ matrices $r_0$ and $t_1$ with $a_2 = 1$
and $a_1 = a_3 = a_4 = 0$ in eq.~\eqref{Z4-t1}, which are diagonalized simultaneously
by a suitable gauge transformation.
Hence, non-paired $r'_0$'s with $n'={\rm odd}$ alone or $n'={\rm even}$ alone remain
independent from $r_0$ and $t_1$,
and the number of diagonal-blocks composed of such
$r'_0$ and $t'_1$ is $|n_1^{(m)'}-n_2^{(m)'}|$.

In case with $a^{(m)}=\pm 1$, $R_0^{(m)}$ and $T_1^{(m)}$ can contain
$1\times1$ matrices as block-diagonal elements.

%
\section{$T^2/\Z6$}
\label{sec:t2z6}
%

We consider the $T^2/\Z6$ case.
As before, we choose the twist matrices $R_0$ and $T_1$ as independent ones,
and show that they can become block-diagonal forms containing $6 \times 6$,
$3 \times 3$, $2 \times 2$ and $1 \times 1$ matrices, 
using constraints and unitary transformations.
The details of derivations are given in appendix~\ref{app:t2z6}.
Here, we present the outline.

The constraints restricting the forms of $R_0$ and $T_1$ are given by
\begin{align}
T_1^{\dagger} = T_1^{-1}, \qquad
T_{m'} T_m = T_m T_{m'}, \qquad
T_1 T_4 = I, \qquad
T_1 T_3 T_5 = I,
\label{Z6-constraints}
\end{align}
where $T_{m} = R_0^{m-1} T_1 R_0^{1-m}$ as explained in section~\ref{sec:2Dorb}.

Starting from the following $R_0$ and $T_1$,
\begin{align}
R_0=
  \begin{pmatrix}
      \eta I_{n_1}& & & & & \\
       & \eta^2 I_{n_2}& & & & \\
       & &-I_{n_3}& & & \\
       & & &-\eta I_{n_4} & & \\
      & & & & -\eta^2 I_{n_5} & \\
      & & & & & I_{n_6}
  \end{pmatrix},
\end{align}
and
\begin{align}
T_1=
  \begin{pmatrix}
      (T_1)_{(11)}&(T_1)_{(12)}&(T_1)_{(13)}&(T_1)_{(14)}
&(T_1)_{(15)}&(T_1)_{(16)}\\
      (T_1)_{(21)}&(T_1)_{(22)}&(T_1)_{(23)}&(T_1)_{(24)}
&(T_1)_{(25)}&(T_1)_{(26)}\\
      (T_1)_{(31)}&(T_1)_{(32)}&(T_1)_{(33)}&(T_1)_{(34)}
&(T_1)_{(35)}&(T_1)_{(36)}\\
      (T_1)_{(41)}&(T_1)_{(42)}&(T_1)_{(43)}&(T_1)_{(44)}
&(T_1)_{(45)}&(T_1)_{(46)}\\
      (T_1)_{(51)}&(T_1)_{(52)}&(T_1)_{(53)}&(T_1)_{(54)}
&(T_1)_{(55)}&(T_1)_{(56)}\\
      (T_1)_{(61)}&(T_1)_{(62)}&(T_1)_{(63)}&(T_1)_{(64)}
&(T_1)_{(65)}&(T_1)_{(66)}
  \end{pmatrix}, 
\label{Z6-R0T1}
\end{align}
with $\eta = e^{2\pi i/6}$, the $n_k \times n_k$ unit matrices $I_{n_k}$
and $n_k \times n_l$ submatrices $(T_1)_{(kl)}$, and using eq.~\eqref{D-Z6-(ai-aj)}
derived from $T_1^{\dagger} (= T_1^{-1} = T_4) = R_0^3 T_1 R_0^{-3}$
and $T_{m'} T_m = T_m T_{m'}$,
we can rearrange $R_0$ and $T_1$ to be block-diagonal forms 
in a similar way to the $S^1/\Z2$ case in section~\ref{sec:s1z2},
such that
\begin{align}
R_0=
    \begin{pmatrix}
        R_0^{(1)} &          &        &         \\
                 & R_0^{(2)} &        &         \\
                 &          & \ddots &         \\
                 &          &        & R_0^{(M)}
    \end{pmatrix},\qquad
T_1=
    \begin{pmatrix}
        T_1^{(1)} &          &        &         \\
                 & T_1^{(2)} &        &         \\
                 &          & \ddots &         \\
                 &          &        & T_1^{(M)}
    \end{pmatrix},
\label{Z6-R0T1-block}
\end{align}
where $R_0^{(m)}$ and $T_1^{(m)}$ $(m = 1, 2, \cdots, M)$ are 
$n^{(m)} \times n^{(m)}$ matrices given by
\begin{align}
R_0^{(m)}=
    \begin{pmatrix}
    \eta I_{n_1^{(m)}}& & & & & \\
       &\eta^2 I_{n_2^{(m)}}& & & & \\
       & &-I_{n_3^{(m)}}& & & \\
       & & &-\eta I_{n_4^{(m)}} & & \\
      & & & & -\eta^2 I_{n_5^{(m)}} & \\
      & & & & & I_{n_6^{(m)}}
    \end{pmatrix},
\label{Z6-R0m}
\end{align}
and
\begin{align}
&T_1^{(m)}=
    \begin{pmatrix}
        a^{(m)}I_{n_1^{(m)}} & \submatrix{M}{1}{2}{(m)[-1]} 
& \submatrix{M}{1}{3}{(m)[-2]} & \submatrix{M}{1}{4}{(m)[-3]}
& \submatrix{M}{1}{5}{(m)[2]} & \submatrix{M}{1}{6}{(m)[1]} \\
       \submatrix{M}{2}{1}{(m)[1]} & a^{(m)}I_{n_2^{(m)}} 
& \submatrix{M}{2}{3}{(m)[-1]} & \submatrix{M}{2}{4}{(m)[-2]}  
& \submatrix{M}{2}{5}{(m)[-3]} & \submatrix{M}{2}{6}{(m)[2]}     \\
       \submatrix{M}{3}{1}{(m)[2]} & \submatrix{M}{3}{2}{(m)[1]} 
& a^{(m)}I_{n_3^{(m)}} & \submatrix{M}{3}{4}{(m)[-1]}
& \submatrix{M}{3}{5}{(m)[-2]} & \submatrix{M}{3}{6}{(m)[-3]} \\
       \submatrix{M}{4}{1}{(m)[3]} & \submatrix{M}{4}{2}{(m)[2]}
& \submatrix{M}{4}{3}{(m)[1]} & a^{(m)}I_{n_4^{(m)}}
& \submatrix{M}{4}{5}{(m)[-1]} & \submatrix{M}{4}{6}{(m)[-2]} \\
       \submatrix{M}{5}{1}{(m)[-2]} & \submatrix{M}{5}{2}{(m)[3]} 
& \submatrix{M}{5}{3}{(m)[2]} & \submatrix{M}{5}{4}{(m)[1]} &
a^{(m)}I_{n_5^{(m)}} & \submatrix{M}{5}{6}{(m)[-1]} \\
       \submatrix{M}{6}{1}{(m)[-1]} & \submatrix{M}{6}{2}{(m)[-2]} 
& \submatrix{M}{6}{3}{(m)[3]} & \submatrix{M}{6}{4}{(m)[2]} 
& \submatrix{M}{6}{5}{(m)[1]} & a^{(m)}I_{n_6^{(m)}} 
    \end{pmatrix},
\label{Z6-T1m}
\end{align}
respectively.
Here, $n^{(m)} = \sum_{k=1}^6 n_k^{(m)}$ and
submatrices in $T_1^{(m)}$ are denoted
as $(T_1^{(m)})_{(kl)} = \submatrix{M}{k}{l}{(m)[k-l]}$
for $k \ne l$ and $(T_1^{(m)})_{(kk)} = \submatrix{M}{k}{k}{(m)[0]}=a^{(m)} I_{n_k^{(m)}}$
with real parameters $a^{(m)}$ satisfying $a^{(m)} \ne a^{(m')}$ for $m \ne m'$
and restricted to $-1 \leq a^{(m)} \leq 1$.
We use a notation of
$\submatrix{M}{k}{l}{(m)[k-l]}=\submatrix{M}{k}{l}{(m)[k'-l']}=\submatrix{M}{k'}{l'}{(m)[k-l]}$
with $k'=k$ (mod 6) and $l'=l$ (mod 6).
Parameters such as $n_{k}$ and $n_k^{(m)}$ which represent the size of matrices
are non-negative integers.
The same applies to the following ones.
Because $T_1^{(m)}$ is already diagonal for $a^{(m)}=\pm 1$, 
we focus on $-1 < a^{(m)} < 1$ hereafter.

Next, using eqs.~\eqref{D-Z6-Mdaggerm}--\eqref{D-Z6-M=MM}
derived from the constraints in eq.~\eqref{Z6-constraints},
we can choose the basis such that $\submatrix{M}{k}{l}{(m)[k-l]}$
have the forms of eqs.~\eqref{D-Z6-Mmq=1}, \eqref{D-Z6-Mmq=-1} 
and \eqref{D-Z6-Mmq=2}--\eqref{D-Z6-Mmq=-2},
and rearrange $R_0^{(m)}$ and $T_1^{(m)}$ to be the forms of 
block-diagonal ones such as
$R_0^{(m)}=R_0^{(m)'}\oplus R_0^{(m)''}\oplus R_0^{(m)'''}$ 
and $T_1^{(m)}=T_1^{(m)'}\oplus T_1^{(m)''}\oplus T_1^{(m)'''}$,
where $R_0^{(m)'}$, $T_1^{(m)'}$,
$R_0^{(m)''}$, $T_1^{(m)''}$, $R_0^{(m)'''}$ and $T_1^{(m)'''}$ are 
also block-diagonal matrices whose submatrices are given by
\begin{align}
& (R_0^{(m)'})_{(k k-q)} = \eta^k \delta_{k k-q} I_{r^{(m)}}, \qquad
(T_1^{(m)'})_{(k k-q)}
={\hat M}^{(m)[q]}\submatrix{U}{k}{k-q}{(m)},
\label{Z6-submatrices'}\\
& (R_0^{(m)''})_{(k k-q)} = \eta^k \delta_{k k-q} I_{n_k^{(m)'}}, \qquad
(T_1^{(m)''})_{(k k-q)}
= -\frac{1}{3}I_{n_k^{(m)'}}\delta_{q~\!\! 0} 
+ \frac{2}{3}~\! \tilde{U}_{k~\!\!k-q}^{(m)} \delta_{q~\!\!\pm 2},
\label{Z6-submatrices''}\\
& (R_0^{(m)'''})_{(k k-q)} = \eta^k \delta_{k k-q} I_{n_k^{(m)''}}, \qquad
(T_1^{(m)'''})_{(k k-q)}
= -\frac{1}{2}I_{n_k^{(m)''}}\delta_{q~\!\! 0}
\pm \frac{\sqrt{3}}{2}~\! \tilde{U}_{k~\!\!k-q}^{(m)} \delta_{q~\!\!3},
\label{Z6-submatrices'''}
\end{align}
where ${\hat M}^{(m)[1]}$ are $r^{(m)} \times r^{(m)}$ diagonal matrices
with positive elements, i.e., $({\hat M}^{(m)[1]})_{ii} > 0$, 
${\hat M}^{(m)[2]}$ and ${\hat M}^{(m)[3]}$ are $r^{(m)} \times r^{(m)}$ diagonal matrices
with non-negative elements, i.e., $({\hat M}^{(m)[2]})_{ii} \ge 0$
and $({\hat M}^{(m)[3]})_{ii} \ge 0$,
${\hat M}^{(m)[-q]} = {\hat M}^{(m)[q]}$,
$\submatrix{U}{k}{k-q}{(m)}$ are $r^{(m)} \times r^{(m)}$ unitary matrices
and $\tilde{U}_{k~\!\!k-2}^{(m)}$ and $\tilde{U}_{k~\!\!k-3}^{(m)}$ are 
$n_k^{(m)'} \times n_k^{(m)'}$ and $n_k^{(m)''} \times n_k^{(m)''}$ unitary matrices,
respectively.
$R_0^{(m)''}$ and $T_1^{(m)''}$ can appear only in the case of $a^{(m)}=-1/3$,
and $R_0^{(m)'''}$ and $T_1^{(m)'''}$ can appear only in the case of $a^{(m)}=-1/2$.

We perform an appropriate unitary transformation
that make all submatrices in $T_1^{(m)'}$, $T_1^{(m)''}$ and $T_1^{(m)'''}$ 
diagonalized simultaneously
such as eqs.~\eqref{Z6-T1m'-tr}--\eqref{Z6-T1m'''-tr}, 
keeping $R_0^{(m)}$ diagonal ones,
and rearrange the rows and columns in $R_0^{(m)}$ and $T_1^{(m)}$.
Then, we find that $R_0^{(m)}$ and $T_1^{(m)}$ can take simplified block-diagonal forms 
containing $6\times 6$, $3\times 3$,  $2\times 2$ and $1\times 1$ matrices as the diagonal blocks.

From eqs.~\eqref{Z6-R0m'} and \eqref{Z6-T1m'-tr},
the $6 \times 6$ matrices of $R_0^{(m)}$ and $T_1^{(m)}$ 
have the form:
\begin{align}
r_0= \left(
\begin{array}{cccccc}
	\eta &  &  &  &  &\\
	 & \eta^2 &  &  &  &\\
	 &  & -1 &  &  &\\
	 &  &  & -\eta &  &\\
   &  &  &  & -\eta^2 &\\
   &  &  &  &  & 1
\end{array}
\right),
\qquad
t_1 = \left(
\begin{array}{cccccc}
a_6 & a_5 & a_4 & a_3 & a_2 & a_1 \\
a_1 & a_6 & a_5 & a_4 & a_3 & a_2 \\
a_2 & a_1 & a_6 & a_5 & a_4 & a_3 \\
a_3 & a_2 & a_1 & a_6 & a_5 & a_4 \\
a_4 & a_3 & a_2 & a_1 & a_6 & a_5 \\
a_5 & a_4 & a_3 & a_2 & a_1 & a_6
\end{array}
\right),
\label{Z6-t1}
\end{align}
where $a_1$, $a_2$, $a_4$ and $a_5$ are complex numbers
with $a_4 = \overline{a_2}$, $a_5 = - \overline{a_1}$, 
$a_3$ is a pure imaginary number, $a_6$ is a real number, and they satisfy
$2|a_1|^2 + 2|a_2|^2 + |a_3|^2 + a_6^2 = 1$,
$|a_1|^2 - |a_2|^2 - |a_3|^2 + a_6^2 = a_6$,
$2a_2 a_6 + \overline{a_2}^2 = a_1^2 - 2 \overline{a_1}a_3$,
$a_1 a_6 + a_3 \overline{a_2} + 2 a_2 \overline{a_1} = a_1$,
$-a_2 a_6 + a_1^2 + \overline{a_1} a_3 + \overline{a_2}^2 = a_2$
and $-2a_3 a_6 + a_1 a_2 - \overline{a_1} \overline{a_2} = a_3$.
We note that these relations are not independent of each other
and the number of independent parameters is two
as seen from the parametrization in eq.~\eqref{Gtr.t1-theta}.
In the next section, it is shown that
$6\times6$ submatrices of this form are diagonalized 
by a suitable gauge transformation.

In the case with $a^{(m)} = -1/3$, from eqs.~\eqref{Z6-R0m''} and \eqref{Z6-T1m''-tr},
$R_0^{(m)}$ and $T_1^{(m)}$ can contain 
following type of $3 \times 3$ matrices as block-diagonal elements,
\begin{align}
r'_0 = e^{2\pi n' i/6} \left(
\begin{array}{ccc}
	\omega &  0  & 0 \\
	0 & \omega^2 & 0 \\
  0 & 0 & 1
\end{array}
\right),
\qquad
t'_1 = \left(
\begin{array}{ccc}
-\frac{1}{3} & \frac{2}{3} & \frac{2}{3}\\
\frac{2}{3} & -\frac{1}{3} & \frac{2}{3}\\
\frac{2}{3} & \frac{2}{3} & -\frac{1}{3}
\end{array}
\right),
\label{Z6-t''1}
\end{align}
where $n'$ is an integer.

In the case with $a^{(m)} = -1/2$, from eqs.~\eqref{Z6-R0m'''} and \eqref{Z6-T1m'''-tr},
$R_0^{(m)}$ and $T_1^{(m)}$ can contain 
following type of $2 \times 2$ matrices as block-diagonal elements,
\begin{align}
r''_0 = e^{2\pi n'' i/6} \left(
\begin{array}{cc}
	-1 &  0  \\
	0 & 1  
\end{array}
\right),
\qquad
t''_1 = \left(
\begin{array}{cc}
-\frac{1}{2} & \pm\frac{\sqrt{3}}{2} i \\
\pm\frac{\sqrt{3}}{2} i & -\frac{1}{2}
\end{array}
\right),~~(\text{double sign in the same order}),
\label{Z6-t'''1}
\end{align}
where $n''$ is an integer.

In a similar way as the case of $T^2/\Z4$,
there can be block-diagonal elements which are diagonalized 
by a gauge transformation, with a suitable combination of
$3 \times 3$ matrices listed in eq.~\eqref{Z6-t''1} with different $n'$'s
or $2 \times 2$ matrices listed in eq.~\eqref{Z6-t'''1} with different $n''$'s.

In the case with $a^{(m)}=\pm 1$, $R_0^{(m)}$ and $T_1^{(m)}$ can contain
$1\times1$ matrices as block-diagonal elements.

%
\section{Gauge transformation}
\label{Sec:GaugeTr}
%

We show that $2 \times 2$ submatrices $t_1$ and $t_2$ on $T^2/\Z2$
and $N \times N$ submatrix $t_1$ on $T^2/\Z N$ ($N = 3, 4, 6$)
are diagonalized by suitable gauge transformations,
keeping $r_0$ the diagonal one.

In section~\ref{sec:t2z2}, we have found that $R_0$, $T_1$ and $T_2$ on $T^2/\Z2$ contain 
following $2 \times 2$ unitary matrices as members of block-diagonal ones,
\bequ
 r_0= X, 
 \qquad
 t_1= a_1 Y + a_2 I, 
 \qquad
 t_2= b_1 Y + b_2 I,
\label{Gtr.r0t1t2}
\eequ
where $X$ and $Y$ are defined by
\bequ
 X \equiv \left(
\begin{array}{cc}
-1 & 0 \\
0 & 1
\end{array}
\right),
\qquad
 Y \equiv \left(
\begin{array}{cc}
0 & 1 \\
1 & 0 
\end{array}
\right),
\label{Gtr.XY-2*2}
\eequ
respectively (see eqs.~\eqref{Z2-r0-1} and \eqref{Z2-r0-2}).
The coefficients $a_1$, $a_2$, $b_1$ and $b_2$ are numbers.
The matrices $t_1$ and $t_2$ are also written by
\bequ
 t_1 = e^{i \theta_{(1)} Y},
\qquad
 t_2 = e^{i \theta_{(2)} Y},
\label{Gtr.t1-2*2}
\eequ
where $\theta_{(1)}$ and $\theta_{(2)}$ are real numbers.
Using $\theta_{(1)}$ and $\theta_{(2)}$,
$a_1$, $a_2$, $b_1$ and $b_2$ are given by
\bequ
 a_1 = i\sin\theta_{(1)},~~ a_2 = \cos\theta_{(1)},~~
 b_1 = i\sin\theta_{(2)},~~ b_2 = \cos\theta_{(2)}.
\label{Gtr.abp}
\eequ
Then, $r_0$, $t_1$ and $t_2$ are transformed into the diagonal ones:
\begin{eqnarray}
&~& \tilde{r}_0 = \varOmega(-z, -\bar{z})
  r_0 \varOmega^{\dagger}(z, \bar{z}) = r_0,
\qquad
 \tilde{t}_1 = \varOmega(z + 1, \bar{z}+1) t_1 \varOmega^{\dagger}(z, \bar{z}) 
      = (-1)^{l_{(1)}} I,~~
\nonumber \\
&~& \tilde{t}_2 = \varOmega(z + \tau, \bar{z}+\bar{\tau}) t_2 \varOmega^{\dagger}(z, \bar{z}) 
      = (-1)^{l_{(2)}} I,
\label{Gtr.r0t1t2-prime}
\end{eqnarray}
under the gauge transformation whose transformation function is given by 
$\varOmega(z, \bar{z}) = e^{i\left(\beta z+ \bar{\beta}\bar{z}\right)Y}$
with 
$\beta 
= \frac{-\theta_{(1)} + l_{(1)}\pi}{2}\left(1 + \frac{{\rm Re}\tau}{{\rm Im}\tau}i\right)
- \frac{-\theta_{(2)} + l_{(2)}\pi}{2} \frac{i}{{\rm Im}\tau}$
($l_{(1)}$ and $l_{(2)}$:integers, ${\rm Im}\tau \ne 0$).

In the same way, $R_0$ and $T_1$ on $T^2/\Z N$ ($N = 3, 4, 6$) contain 
following $N \times N$ unitary matrices as members of block-diagonal ones,
\bequ
 r_0=X, 
 \qquad
 t_1=\sum_{p = 1}^{N} a_p Y^p, 
\label{Gtr.r0t1}
\eequ
where $X$ and $Y$ are defined by
\bequ
 X \equiv \left(
\begin{array}{ccccc}
\tau & & & &  \\
 & \tau^2 & & &   \\
 & & \ddots & &   \\
 & & & \tau^{N-1} & \\
 & & & & 1
\end{array}
\right),
\qquad
 Y \equiv \left(
\begin{array}{ccccc}
 & & & & 1 \\
1& & & &   \\
 &1& & &   \\
 & &\ddots & & \\
 & & &1& 
\end{array}
\right),
\label{Gtr.XY}
\eequ
respectively (see eqs.~\eqref{Z3-t1}, \eqref{Z4-t1} and \eqref{Z6-t1}).
Here, $\tau = e^{2\pi i/N}$, and 
$a_p$ ($p=1, 2, \cdots, N$) are numbers 
which satisfy specific relations, e.g.,
$a_1^3 + a_2^3 + a_3^3 - 3 a_1 a_2 a_3 = 1$,
$|a_1|^2 + |a_2|^2 + |a_3|^2 = 1$ and 
$\overline{a_1}a_3 + \overline{a_3}a_2 + \overline{a_2}a_1 = 0$
for $N=3$.
The specific relations among $a_p$ are rewritten compactly as
\begin{align}
& \abs{\alpha_j}^2 = 1, \quad (j=1, \cdots, N), & \quad \mbox{for } N=3, 4, 6,
\label{Gtr.|alpha|}\\
& \alpha_1 \alpha_2 \alpha_3 = 1, & \quad \mbox{for } N=3,
\label{Gtr.alpha-N=3}\\
& \alpha_1 \alpha_3 = \alpha_2 \alpha_4 = 1, & \quad \mbox{for } N=4,
\label{Gtr.alpha-N=4}\\
& \alpha_1 \alpha_4 = \alpha_2 \alpha_5 = \alpha_3 \alpha_6 = 1, \quad
\alpha_1 \alpha_3 \alpha_5 = \alpha_2 \alpha_4 \alpha_6 = 1, & \quad \mbox{for } N=6,
\label{Gtr.alpha-N=6}
\end{align}
where $\alpha_j \equiv \sum_{p=1}^{N} a_p \tau^{jp}$.
From them, $t_1$ is expressed as
\bequ
t_1 = \sum_{p=1}^N a_{p} Y^{p} = e^{i\left(\theta Y + \bar{\theta} Y^{N-1}\right)}.
\label{Gtr.t1Y}
\eequ
The derivation of eqs.~\eqref{Gtr.|alpha|} -- \eqref{Gtr.t1Y}
is given in appendix~\ref{app:ap}.

\begin{table}[t]
\bctr
\renewcommand{\arraystretch}{1.4}
\begin{tabular}{|c|l|l|c|l|}
\hline
$N$ & peculiar constraints & ~~~~~~$\beta$ & $\tilde{r}_0$ & ~~~$\tilde{t}_1$ \\ \hline
3 & $t_1t_2t_3 = I$ & $-\theta - \frac{2}{3} \pi \tilde{l}$ & $r_0$ 
& $\omega^{\tilde{l}} I$ \\
4 & $t_1t_2t_3t_4 = I$,~ $t_1 t_3 = t_2 t_4 = I$ 
& $-\theta + \frac{1+i}{2} \pi \tilde{l}$ & $r_0$ & $(-1)^{\tilde{l}} I$ \\
6 & $t_1t_2t_3t_4t_5t_6 = I$,~ $t_1t_4 = t_2 t_5 = t_3 t_6 = I$,~ 
$t_1 t_3 t_5 = t_2 t_4 t_6 = I$
& $-\theta$ & $r_0$ & $I$ \\ \hline
\end{tabular}
\caption{Peculiar constraints and gauge transformed matrices}
\label{T-gauge}
\ectr
\end{table}

Based on eq.~\eqref{Gtr.t1Y}, we find that $r_0$ and $t_1$ are transformed as
\bequ
 \tilde{r}_0 = \varOmega(\tau z, \bar{\tau}~\!\bar{z})
  r_0 \varOmega^{\dagger}(z, \bar{z}),
\qquad
 \tilde{t}_1 = \varOmega(z + 1, \bar{z}+1) t_1 \varOmega^{\dagger}(z, \bar{z}),
\label{Gtr.r0t1-prime}
\eequ
under a gauge transformation,
and $\tilde{r}_0$ and $\tilde{t}_1$ become diagonal ones,
using the gauge transformation function:
\bequ
 \varOmega(z, \bar{z}) 
= e^{i\left(\beta z Y + \bar{\beta}\bar{z} Y^{N-1}\right)}
\label{Gtr.Omega}
\eequ
with a suitable value $\beta$ including an integer $\tilde{l}$, 
as shown in Table~\ref{T-gauge}.

In contrast, $R_0$ and $T_1$ on $T^2/\Z4$ can contain $2 \times 2$ matrices:
\begin{align}
r'_0 = i^{n'} \left(
\begin{array}{cc}
	-1 &  0  \\
	0 & 1  
\end{array}
\right),
\qquad
t'_1 = \left(
\begin{array}{cc}
0 & 1 \\
1 & 0 
\end{array}
\right), \quad (n':{\rm integer}),
\label{Gtr-Z4-t'1}
\end{align}
as block-diagonal elements, and $t'_1$ cannot become diagonal ones,
keeping $r'_0$ diagonal ones,
by the use of a gauge transformation.
In the same way, $R_0$ and $T_1$ on $T^2/\Z6$
can contain $3 \times 3$ matrices:
\begin{align}
r'_0 = e^{2\pi n' i/6} \left(
\begin{array}{ccc}
	\omega &  0  & 0 \\
	0 & \omega^2 & 0 \\
  0 & 0 & 1
\end{array}
\right),
\qquad
t'_1 = \left(
\begin{array}{ccc}
-\frac{1}{3} & \frac{2}{3} & \frac{2}{3}\\
\frac{2}{3} & -\frac{1}{3} & \frac{2}{3}\\
\frac{2}{3} & \frac{2}{3} & -\frac{1}{3}
\end{array}
\right), \quad (n':{\rm integer}),
\label{Gtr-Z6-t''1}
\end{align} 
and $2 \times 2$ matrices: 
\begin{align}
r''_0 = e^{2\pi n'' i/6}\left(
\begin{array}{cc}
	-1 &  0  \\
	0 &  1 
\end{array}
\right),
\qquad
t''_1 = \left(
\begin{array}{cc}
-\frac{1}{2} & \pm\frac{\sqrt{3}}{2} i \\
\pm\frac{\sqrt{3}}{2} i & -\frac{1}{2}
\end{array}
\right), 
\label{Gtr-Z6-t'''1}\\
 (n'':{\rm integer},~~\text{double sign in the same order}),
\nonumber
\end{align}
as block-diagonal elements,
and $t'_1$ and $t''_{1}$ cannot become diagonal ones,
keeping $r'_0$ and $r''_0$ diagonal ones,
by the use of a gauge transformation.
Note that the matrices given in eq.~\eqref{Gtr-Z4-t'1}
and those in eqs.~\eqref{Gtr-Z6-t''1} and \eqref{Gtr-Z6-t'''1} 
satisfy every relation
required from transformation properties on $T^2/\Z4$ and $T^2/\Z6$, respectively.

The absence of gauge transformations that make $t'_1$ and/or $t''_1$ diagonal ones
is understood from the fact that
there are no continuous Wilson line phases relating to such block-diagonal elements.
The continuous Wilson line phases are non-integrable phases of, for instance, 
$W_1=e^{ig(\langle A_z \rangle + \langle A_{\bar{z}} \rangle)}T_1$
  and $W_2=e^{ig(\tau \langle A_z \rangle + \bar \tau\langle A_{\bar{z}} \rangle)}T_2$
with the vacuum expectation values (VEVs)
of zero modes in
$A_z = (A_5 - iA_6)/2$ and $A_{\bar{z}} (= \overline{A_{z}}) = (A_5 + iA_6)/2$.
Here $A_5$ and $A_6$ are the gauge fields along the coordinates $y^1$ and $y^2$,
respectively. We remind that $|\bm \lambda_1|=1$ is taken. 
The VEV $\langle A_z \rangle$ should change as
\begin{align}
R_0 \langle A_z \rangle R_0^{-1} = \tau \langle A_z \rangle,
\label{Gtr-<Az>}
\end{align}
under $\Z N$ rotation.
Actually, there are no non-vanishing VEVs
that satisfy the above relation
for $r'_0$ in eqs.~\eqref{Gtr-Z4-t'1} and \eqref{Gtr-Z6-t''1}
and for $r''_0$ in eq.~\eqref{Gtr-Z6-t'''1}.

%
\section{Physical implications}
\label{Sec:PhysImp}

%
\subsection{Rank reduction}
\label{Sec:GaugeTr-rank}

Here, we discuss a rank reduction of gauge group.
The physical symmetries are determined
in cooperation of boundary conditions of fields and the dynamics of the Wilson line phases,
by the Hosotani mechanism~\cite{H1,H2,HHH&K}.
In concrete terms, starting from boundary conditions specified by
the twist matrices $(R_0, T_1, T_2)$, we calculate the effective potential $V_{\rm eff}$ for
the Wilson line phases, 
which is flat at the tree level%
\footnote{We note that there can appear zero modes of $A_z$ which have 
nonvanishing quartic terms at the tree level, coming from the 
commutation relations in the field strength $F_{z\bar z}$, 
or $F_{56}$~\cite{treepot6D}. 
Their VEVs make $F_{z\bar z}$, namely the magnetic field, nonvanishing 
and thus are not gauge equivalent with the vanishing VEV, 
$\VEV{A_z}=0$, which results in $\VEV{F_{z\bar z}}=0$. 
This means that these VEVs can not be gauged away, neither have the 
phase nature.
}
and generated by the quantum corrections. 
And then, minimization of $V_{\rm eff}$ tells us the VEV, $\langle A_z \rangle$. %
After performing the gauge transformation with $\varOmega(z, \bar{z})
= e^{ig(\langle A_z \rangle z + \langle A_{\bar{z}} \rangle \bar{z})}$,
$\langle A_z \rangle$ is shifted to $\langle A'_z \rangle = 0$
and the twist matrices are transformed as
\begin{align}
&(R_0, T_1, T_2) \to (R_0^{\rm sym}, T_1^{\rm sym}, T_2^{\rm sym})
\nonumber \\
& = (\varOmega(-z, -\bar{z})R_0\varOmega^{\dagger}(z, \bar{z}), 
\varOmega(z+1, \bar{z}+1)T_1\varOmega^{\dagger}(z, \bar{z}), 
\varOmega(z+\tau, \bar{z}+\bar{\tau})T_2\varOmega^{\dagger}(z, \bar{z})).
\label{Gtr-R0-sym}
\end{align}
We note that, based on the definition of continuous Wilson line phases
as eigenvalues of $-i\ln W_1$ and $-i\ln W_2$,
they are gauge invariant because a change of $\langle A_z \rangle$
is canceled out by that of $T_1$ and $T_2$.
The physical gauge symmetry $\mathcal{H}^{\rm sym}$ 
is spanned by the generators $T^a$
that commute with $(R_0^{\rm sym}, T_1^{\rm sym}, T_2^{\rm sym})$:
\begin{align}
\mathcal{H}^{\rm sym} = \{T^a; [T^a, R_0^{\rm sym}]
=[T^a, T_1^{\rm sym}]=[T^a, T_2^{\rm sym}]=0\}.
\label{Gtr-Ta}
\end{align}
Then the rank of unbroken gauge group 
agrees with the number of Cartan subalgebras that commute with 
$R_0^{\rm sym}$ and $T_{m}^{\rm sym}$,
where $m=1, 2$ for $T^2/\Z 2$
and $m=1$ for $T^2/\Z N$ ($N=3, 4, 6$).
Thus the reduction of rank does not occur
if $R_0^{\rm sym}$ and $T_{m}^{\rm sym}$ are diagonal.
In other words,
when diagonal $R_0$ and $T_{m}$ belong to the equivalence class,
there are symmetry-enhanced points in the parameter space of 
the VEVs of the Wilson line phases, 
while the rank is reduced on a generic point.
In contrast, the reduction always occurs in the presence of $t'_1$ on $T^2/\Z 4$
and $t'_1$ and/or $t''_{1}$ on $T^2/\Z 6$,
because they take discrete values and cannot be diagonalized.
Note that the $2\times2$ matrices ($3\times3$ matrix), 
$t'_1$ in eq.~\eqref{Gtr-Z4-t'1} and 
$t''_1$ in eq.~\eqref{Gtr-Z6-t'''1} 
($t'_1$ in eq.~\eqref{Gtr-Z6-t''1}), 
are contained in the expression of the $t_1$ in eq.~\eqref{Gtr.t1-2*2} 
(in eq.~\eqref{Gtr.r0t1} with $N=3$). 
While the matrix $t_1$ in the $\Z2$ ($\Z3$) orbifold can be diagonalized,
keeping $r_0$ a diagonal form,
thanks to a singular gauge transformation 
that shifts the zero modes of $A_z$,
these degrees are projected out by the additional elements of the 
enlarged group in the $\Z4$ and $\Z6$ orbifolds, as examined 
in eq.~\eqref{Gtr-<Az>}. 
This observation indicates that the rank reduction by the 
$t'_1$ and/or $t''_{1}$, or the discrete Wilson line phase, can be 
realized only for $\Z N$ orbifolds with $N$ being a nonprime number.

In refs.~\cite{S&S,FN&W}, the rank reduction of gauge group has been
studied in orbifold construction.  It is pointed that the rank is not
reduced by the discrete Wilson lines.  In these references, the rank
reduction is discussed mainly focusing on the $\Z2$ orbifolds with
$N=2$ being a prime number, and
the possibility of the discrete Wilson lines that do not commute with
the rotation $R_0$ has not been studied.
Our results show counter examples to the claim. 

%
\subsection{Examples}
\label{Sec:GaugeTr-examples}

Finally, we show some examples of the boundary conditions for illustration 
purpose, in the $T^2/\Z4$ case. 

The first one is an $SU(6)$ model with 
\bequ
 R_0=\begin{pmatrix}
       -1&&&&&\\&1&&&&\\&&1&&&\\&&&1&&\\&&&&1&\\ &&&&&1
     \end{pmatrix},  \quad
 T_1=\begin{pmatrix}
       1&&&&&\\&1&&&&\\&&1&&&\\&&&-1&&\\&&&&-1&\\ &&&&&-1
     \end{pmatrix}, 
\eequ
by which the $SU(6)$ symmetry is broken down to 
$SU(3)\times SU(2)\times U(1)^2$. 
We note that, when these boundary conditions are chosen in an $S^1/\Z2$ case, 
the continuous Wilson line generically induces further symmetry breaking 
$SU(2)\times U(1)\to U(1)$, which may be applied to the electroweak symmetry 
breaking~\cite{GHU-HHK&Y, Lim:2007jv}. 
In contrast, in the $T^2/\Z4$ case, there are no degrees of the continuous 
Wilson line phases and the rank is not reduced.\footnote{
In the case where the $(1,1)$ component of $R_0$, $-1$, is replaced
by $i$, there appears a zero mode which can not be gauged away, 
discussed in the previous footnote. 
When it acquires a nonvanishing VEV, the rank is reduced~\cite{treepot6D}.
}

To realize the degrees of the continuous Wilson line in the $T^2/\Z4$ case, 
we modify the first example to $SU(8)$ model with 
\bequ
 R_0=\begin{pmatrix}
       i&&&&&&&\\&-1&&&&&&\\&&-i&&&&&\\&&&1&&&&\\
       &&&&1&&&\\&&&&&1&&\\&&&&&&1&\\ &&&&&&&1
     \end{pmatrix},  \quad
 T_1=\begin{pmatrix}
       1&&&&&&&\\&1&&&&&&\\&&1&&&&&\\&&&1&&&&\\
       &&&&1&&&\\&&&&&-1&&\\&&&&&&-1&\\ &&&&&&&-1
     \end{pmatrix}, 
\eequ
by which the $SU(8)$ symmetry is broken down to 
$SU(3)\times SU(2)\times U(1)^4$. 
Due to the degrees of the Wilson line phases, in the basis where 
$\VEV{A_z}$ is gauged away, the upper left $4\times4$ 
submatrix in $T_1$ is written in general as the $t_1$ given in 
eq.~\eqref{Z4-t1}. 
Generically, then, the further symmetry breaking 
$SU(2)\times U(1)^3\to U(1)$ occurs, which might be identified with the 
electroweak symmetry breaking. 
In this second example, the rank can be reduced with the 
continuous Wilson line phases. 

The third one is for the rank reduction without the continuous Wilson line 
phases: 
$SU(7)$ model with 
\begin{align}
  R_0=\left(
  \begin{array}{cc|ccccc}
    -1&0&&&&&\\
    0&1&&&&&\\ \hline
    &&1&&&&\\
    &&&1&&&\\
    &&&&1&&\\
    &&&&&1&\\
    &&&&&&1
  \end{array}\right), \qquad
  T_1=\left(
  \begin{array}{cc|ccccc}
    0&1&&&&&\\
    1&0&&&&&\\ \hline
    &&1&&&&\\
    &&&1&&&\\
    &&&&-1&&\\
    &&&&&-1&\\
    &&&&&&-1
  \end{array}\right),
\end{align}
by which the $SU(7)$ symmetry is broken down to 
$SU(3)\times SU(2)\times U(1)^2$. 
In this example, the rank is reduced without the degrees of the continuous 
Wilson line phases. 

%
\section{Conclusions and discussions}
\label{sec:concl}
%

We have studied the existence of diagonal representatives 
in each equivalence class of the twist matrices
in $SU(n)$ or $U(n)$ gauge theories compactified
on the orbifolds $T^2/\Z N$ ($N = 2, 3, 4, 6$), 
supposing that the theory has a global $G'=U(n)$ symmetry.
Using constraints, unitary transformations and gauge transformations,
we have examined whether the twist matrices can simultaneously
become diagonal or not.
We have shown that at least one diagonal representative necessarily exists
in each equivalence class on $T^2/\Z 2$ and $T^2/\Z 3$, 
but the twist matrices on $T^2/\Z 4$ and $T^2/\Z 6$
can contain not only diagonal matrices but also non-diagonal $2 \times 2$ ones 
($t'_1$ in eq.~\eqref{Gtr-Z4-t'1}) and
non-diagonal $3 \times 3$ and $2 \times 2$ ones
($t'_1$ in eq.~\eqref{Gtr-Z6-t''1} and $t''_1$ in eq.~\eqref{Gtr-Z6-t'''1}), respectively,
as members of block-diagonal submatrices.

The absence of gauge transformations that make $t'_1$ 
and/or $t''_1$ diagonal ones
is related to the absence of
Wilson line phases with respect to such block-diagonal ones.
When $R_0$ and $T_{m}$ contain diagonal elements alone, 
there are symmetry-enhanced points in the parameter space of 
the continuous Wilson line phases, 
while the rank is reduced on a generic point. 
In contrast, the reduction occurs in the presence of $t'_1$ on $T^2/\Z 4$
and $t'_1$ and/or $t''_1$ on $T^2/\Z 6$.
We note that these matrices have discrete parameters, and thus 
  our results show examples
that the rank-reducing symmetry breaking is caused 
by the discrete Wilson line phases
in $\Z N$ orbifolds with $N$ being a nonprime number.

In this article, we restrict our gauge group into $SU(n)$ or $U(n)$ to ensure the 
unitary transformations that diagonalize submatrices in our calculation
are a part of the symmetry transformations. 
The extension to more general gauge group is an important issue to be addressed 
in a future work. 
Furthermore, there remains the arbitrariness problem of which type of boundary conditions
should be chosen without relying on phenomenological information,
and it would be challenging to find a mechanism or principle
that determine boundary conditions of fields.

\appendix

\section{Derivation of block-diagonal forms on $T^2/\Z2$}
\label{app:t2z2}

%
%

In this section, we derive the results shown in
  section~\ref{sec:t2z2}.  We take the twist matrices $R_0$ and $T_1$
  as shown in eqs.~\eqref{r0tbdt2z21}, \eqref{r0t1mZ21}
    and~\eqref{r0t1mZ22}.  The matrix $T_2$ is now expressed
by the submatrices $T_2^{(\lambda\lambda')}$
  $(\lambda,\lambda'=0,\dots,M)$ as
\begin{align}
    T_2=
  \begin{pmatrix}
      T_2^{(00)}&T_2^{(01)}&\dots&\\
      T_2^{(10)}&T_2^{(11)}&\dots&\\
      \vdots&\vdots&\ddots &
  \end{pmatrix}. 
\end{align}

\subsection{Derivation of $T_2^{(0m)}=T_2^{(m0)}=0$}
\label{ap:z201}
Let us start to discuss how $T_2$ is simplified through unitary
transformations 
that keep the forms of $R_0$
and $T_1$.  First, we examine the condition $[T_1,T_2]=0$ in
eq.~\eqref{condt2zng1} to restrict the forms of $T_2$.  From the
above definitions, we obtain the following expressions:
\begin{align}
   T_1 T_2=
  \begin{pmatrix}
     T_1^{(0)} T_2^{(00)}&T_1^{(0)}T_2^{(01)}&\dots&\\
      T_1^{(1)}T_2^{(10)}&T_1^{(1)}T_2^{(11)}&\dots&\\
      \vdots&\vdots&\ddots &
  \end{pmatrix}, \qquad 
   T_2 T_1=
  \begin{pmatrix}
      T_2^{(00)}T_1^{(0)}&T_2^{(01)}T_1^{(1)}&\dots&\\
      T_2^{(10)}T_1^{(0)}&T_2^{(11)}T_1^{(1)}&\dots&\\
      \vdots&\vdots&\ddots &
  \end{pmatrix}, 
\end{align}
which imply the relations $T_1^{(0)}T_2^{(00)}=T_2^{(00)}T_1^{(0)}$,
$T_1^{(0)}T_2^{(0m)}=T_2^{(0m)}T_1^{(m)}$,
$T_1^{(m)}T_2^{(m0)}=T_2^{(m0)}T_1^{(0)}$ and
$T_1^{(m)}T_2^{(mm')}=T_2^{(mm')}T_1^{(m')}$ for $m, m'=1,\dots,M$. From
the second and third relations, we can derive that $T_2^{(0m)}$ and
$T_2^{(m0)}$ vanish. To see this, let us examine
$T_1^{(0)}T_2^{(0m)}=T_2^{(0m)}T_1^{(m)}$, which implies
\begin{align}
  \sum_{k=1}^{r^{(0)}}(T_1^{(0)})_{ik}(T_2^{(0m)})_{kj}=
  \sum_{k=1}^{2{r}^{(m)}}(T_2^{(0m)})_{ik}(T_1^{(m)})_{kj},
\end{align}
where $(i,j)$ elements of a matrix $A$ are denoted by $(A)_{ij}$.
Since the elements of $T_1^{(0)}$ take $(T_1^{(0)})_{ik}=(-1)^{l_i}\delta_{ik}$ with
$l_i\in{\mathbb Z}$, the above equation is rearranged as 
\begin{align}\label{t20mto0}
  \sum_{k=1}^{2{r}^{(m)}}(T_2^{(0m)})_{ik}\left[(T_1^{(m)})_{kj}-(-1)^{l_i}\delta_{kj}\right]=0.
\end{align}
Let us define the $2r^{(m)}\times 2r^{(m)}$ matrix $\tilde T_1^{(m)}$
as $(\tilde T_1^{(m)})_{kj}=(T_1^{(m)})_{kj}-(-1)^{l_i}\delta_{kj}$,
which is also expressed as follows:
\begin{align}
\tilde T_1^{(m)}&=
  \begin{pmatrix}
      \cos\theta^{(m)}-(-1)^{l_i}&
      i\sin\theta^{(m)}\\
      i\sin\theta^{(m)}&\cos\theta^{(m)}-(-1)^{l_i}
  \end{pmatrix}\otimes I_{{r}^{(m)}}.  
\end{align}
Parameters such as $2r^{(m)}$ which represent the size of matrices
are non-negative integers.
The same applies to the following ones.
Since $\cos\theta^{(m)}\neq \pm 1$, $\det \tilde T_1^{(m)}\neq 0$
holds. Thus, there exists the inverse matrix of $\tilde
T_1^{(m)}$. Therefore, from eq.~\eqref{t20mto0}, we conclude
$T_2^{(0m)}=0$. One also finds $T_2^{(m0)}=0$ through a similar
examination with the relation
$T_1^{(m)}T_2^{(m0)}=T_2^{(m0)}T_1^{(0)}$.

\subsection{Decomposition of $T_2^{(00)}$ into $2\times 2$ submatrices}
\label{ap:z202}
Next, let us focus on the relation
$T_1^{(0)}T_2^{(00)}=T_2^{(00)}T_1^{(0)}$. We can choose a basis such that 
$R_0^{(0)}$ and $T_1^{(0)}$ are given by 
\begin{align}
  R_0^{(0)}=\begin{pmatrix}
                R_0^{(0),1}&0\\
                0&R_0^{(0),2}
            \end{pmatrix},\qquad 
  T_1^{(0)}=\begin{pmatrix}
                T_1^{(0),1}&0\\
                0&T_1^{(0),2}
            \end{pmatrix}=  \begin{pmatrix}
                -I_{r^{(0)}_1}&0\\
                0&I_{r^{(0)}_2}
            \end{pmatrix},
\end{align}
where we have introduced $r^{(0)}_1$ and $r^{(0)}_2$
$(r^{(0)}_1+r^{(0)}_2=r^{(0)})$. The submatrix $R_0^{(0),a}$ $(a=1,2)$
is diagonal and has eigenvalues $1$ or $-1$.  Then, from
$T_1^{(0)}T_2^{(00)}=T_2^{(00)}T_1^{(0)}$, one sees that $T_2^{(00)}$
takes the block-diagonal form as
\begin{align}
  T_2^{(00)}=  \begin{pmatrix}
                T_2^{(00),1}&0\\
                0&T_2^{(00),2}
            \end{pmatrix}. 
\end{align}
From the constraints $R_0^2=(T_2R_0)^2=I$ in
eq.~\eqref{condt2zng1}, we see that
\begin{align}\label{r0t20rel1}
  (R_0^{(0),a})^2=
  (T_2^{(00),a}R_0^{(0),a})^2=I_{r^{(0)}_a}, \qquad a=1,2.
\end{align}
If we make $R_0^{(0),a}$ and $T_2^{(00),a}$ correspond to $P_0$ and
$T$ appeared in section~\ref{sec:s1z2}, eq.~\eqref{r0t20rel1}
corresponds to eq.~\eqref{s1z2bcmatcond1}. In addition, since
$T_1^{(0),a}$ are proportional to the unit matrices, if we take
unitary transformations to simplify the form of $T_2^{(00),a}$,
$T_1^{(0),a}$ is unchanged. Thus, as we have done in
section~\ref{sec:s1z2}, we can simplify the form of $T_2^{(00),a}$
keeping the diagonal form of $R_0^{(0),a}$. As a result, we obtain
the simplified forms of the twist matrices $R_0^{(0),a}$,
  $T_1^{(0),a}$ and $T_2^{(00),a}$ as follows:
\begin{align}
    R_0^{(0),a}&=
  \begin{pmatrix}
      R_{0}^{(0),a,0}&&&\\
      &R_{0}^{(0),a,1}&&\\
      &&\ddots &\\
      && &R_{0}^{(0),a,M_{0}^{(a)}}
  \end{pmatrix},
  &&  R_{0}^{(0),a,m}= -\sigma_3 \otimes I_{r_a^{(0),m}}, \\
  T_1^{(0),a}&=
  \begin{pmatrix}
      T_1^{(0),a,0}&&&\\
      &T_1^{(0),a,1}&&\\
      &&\ddots &\\
      && &T_1^{(0),a,M_{0}^{(a)}}
  \end{pmatrix},&&
  T_{1}^{(0),a,m}=(-1)^aI_2\otimes I_{r_a^{(0),m}},  \\
  T_2^{(00),a}&=
  \begin{pmatrix}
      T_2^{(00),a,0}&&&\\
      &T_2^{(00),a,1}&&\\
      &&\ddots &\\
      && &T_2^{(00),a,M_{0}^{(a)}}
  \end{pmatrix},
&&  T_{2}^{(00),a,m}=e^{i\phi^{(0),a, m}\sigma_1}\otimes I_{r_a^{(0),m}}, 
\end{align}
where $R_{0}^{(0),a,0}$, $T_{1}^{(0),a,0}$ and $T_{2}^{(00),a,0}$ are
$r_a^{(0),0}\times r_a^{(0),0}$ diagonal matrices whose
eigenvalues are 1 or $-1$.
The real parameters $\phi^{(0),a, m}$ 
  satisfy $0<\phi^{(0),a, m}<\pi$ and
  $\phi^{(0),a, m} \neq \phi^{(0),a, m'}$ for $m\neq m'$.

\subsection{Block-diagonal form of $T_2$}
\label{ap:z203}
Let us start to discuss $T_2^{(mm')}$ $(m,m'=1,\dots ,M)$.  First, we
define
\begin{align}
\qquad T_2^{(mm')}=
            \begin{pmatrix}
                (T_2^{(mm')})_{(11)}&(T_2^{(mm')})_{(12)}\\
                (T_2^{(mm')})_{(21)}&(T_2^{(mm')})_{(22)}   
            \end{pmatrix}, 
\end{align}
where $(T_2^{(mm')})_{(kl)}$ is an $r^{(m)}\times r^{(m')}$ matrix.
Also, we can rewrite $T_2^{(mm')}$ by introducing
$r^{(m)}\times r^{(m')}$ matrices $T_{2,\mu}^{(mm')}$ $(\mu=0,1,2,3)$
as follows:
\begin{align}
  T_2^{(mm')}=\sum_{\mu=0}^3 \sigma_\mu\otimes T_{2,\mu}^{(mm')}, 
\end{align}
where $\sigma_\mu=(\sigma_0,\sigma_i)$ and $\sigma_0=I_2$.
Since $T_1$ and $T_2$ commute with each other, it follows that
\begin{align}
  \begin{split}
&  \left(e^{i\theta^{(m)}\sigma_1}-e^{i\theta^{(m')}\sigma_1}\right)
  (\sigma_0\otimes T_{2,0}^{(mm')}+\sigma_1\otimes T_{2,1}^{(mm')})\\
&\qquad   +
  \left(e^{i\theta^{(m)}\sigma_1}-e^{-i\theta^{(m')}\sigma_1}\right)
  (\sigma_2\otimes T_{2,2}^{(mm')}+\sigma_3\otimes T_{2,3}^{(mm')})
  =0.
\end{split}
\end{align}
This equation implies $T_{2,\mu}^{(mm')}=0$ for $m\neq m'$ and
$T_{2,2}^{(mm)}=T_{2,3}^{(mm)}=0$. Thus, $T_2$ takes the
block-diagonal form.  For convenience, we introduce the notation as
$T_{2,0}^{(mm)}=A_{2}^{(m)}$ and $T_{2,1}^{(mm)}=B_{2}^{(m)}$. Then,
$T_2$ is given by
\begin{align}
  T_2=
  \begin{pmatrix}
      T_2^{(00)}&&&\\
      &T_2^{(1)}&&\\
      &&\ddots &\\
      && &T_2^{(M)}
  \end{pmatrix}, \qquad {\rm where} \qquad
            T_2^{(m)}=
            \sigma_0\otimes A_{2}^{(m)}+\sigma_1\otimes B_{2}^{(m)}.
            \label{T2mZ2}
\end{align}

\subsection{Decomposition of $T_2^{(m)}$ into $2\times 2$ submatrices}
\label{ap:z204}

From $(T_2R_0)^2=I$ in eq.~\eqref{condt2zng1}, it follows that $(T_2R_0)^\dag= T_2R_0$,
which implies $(T_2^{(m)}R_0^{(m)})^\dag= T_2^{(m)}R_0^{(m)}$. 
Using $R_0^{(m)}=-\sigma_3\otimes I_{{r}^{(m)}}$
in eq.~\eqref{r0t1mZ21}, we find
$A_2^{(m)\dag}=A_2^{(m)}$ and $B_2^{(m)\dag}=-B_2^{(m)}$. Thus, by
using a unitary transformation that keeps the forms of $R_0^{(m)}$ and
$T_1^{(m)}$, we can diagonalize $A_2^{(m)}$ as
$A_2^{(m)}\to \hat A_{2}^{(m)}$, where the element of
$\hat A_{2}^{(m)}$ is given by
$(\hat A_{2}^{(m)})_{ij}=\tilde a^{(m)i}_2\delta_{ij}$ 
$(\tilde a^{(m)i}_2 \in {\mathbb R})$.  
In this new
basis, let us redefine $B_2^{(m)}$ as
$T_2^{(m)}= \sigma_0\otimes \hat A_{2}^{(m)}+\sigma_1\otimes
B_{2}^{(m)}$. Then, from the condition $T_2T_2^\dag=I$, we get
\begin{align}
  \sigma_0\otimes (I_{r^{(m)}}-\hat A_{2}^{(m)}\hat A_{2}^{(m)}
+B_{2}^{(m)}B_{2}^{(m)}
  )+\sigma_1\otimes (\hat A_{2}^{(m)}B_{2}^{(m)}-B_{2}^{(m)}\hat A_{2}^{(m)})=0.
\end{align}
The term proportional to $\sigma_1$ in the above implies
$(\tilde a^{(m)i}_2-\tilde a^{(m)j}_2)(B_{2}^{(m)})_{ij}=0$.  Thus,
using a unitary transformation that trivially acts on $\sigma_0$ and
$\sigma_1$ in eq.~\eqref{T2mZ2}, we can move to a basis such that
$R_0^{(m)}$, $T_1^{(m)}$ and $T_2^{(m)}$ take the
  block-diagonal forms as
\begin{gather}
            \label{bblockt2z2_ap1}
  R_0^{(m)}=
  \begin{pmatrix}
      R_0^{(m),1}&&&\\
      &R_0^{(m),2}&&\\
      &&\ddots &\\
      && &R_0^{(m),M^{(m)}}
  \end{pmatrix}, \qquad R_0^{(m),m'}=-\sigma_3\otimes I_{r^{(m),m'}}, \\
            \label{bblockt2z2_ap2}
  T_1^{(m)}=
  \begin{pmatrix}
      T_1^{(m),1}&&&\\
      &T_1^{(m),2}&&\\
      &&\ddots &\\
      && &T_1^{(m),M^{(m)}}
  \end{pmatrix},\qquad   T_1^{(m),m'}=e^{i\theta^{(m)}\sigma_1}\otimes I_{r^{(m),m'}}, 
  \\
    T_2^{(m)}=
  \begin{pmatrix}
      T_2^{(m),1}&&&\\
      &T_2^{(m),2}&&\\
      &&\ddots &\\
      && &T_2^{(m),M^{(m)}}
  \end{pmatrix}.
            \label{bblockt2z2_ap3}
\end{gather}
  
The submatrices of $T_2^{(m)}$ in eq.~\eqref{bblockt2z2_ap3} can be
written more explicitly as
\begin{align}
            T_2^{(m),m'}=
a_{2}^{(m),m'}  \sigma_0\otimes I_{r^{(m),m'}}+\sigma_1\otimes
  B_{2}^{(m),m'},
\end{align}
where $a_{2}^{(m),m'}\in{\mathbb R}$ and $B_{2}^{(m),m'}$ is an
$r^{(m),m'}\times r^{(m),m'}$ matrix, which satisfies
$B_{2}^{{(m),m'}^\dag}=-B_{2}^{(m),m'}$.  
Now, we again use the condition $T_2T_2^\dag=I$ to get
$(1-a_{2}^{(m),m'2})I_{r^{(m),m'}}=B_{2}^{(m),m'}B_{2}^{{(m),m'}^\dag}$,
which implies $a_{2}^{(m),m'2}\leq 1$. From the above,
$B_{2}^{(m),m'}$ is rewritten by using a unitary hermitian matrix
$U^{(m),m'}$, where 
$U^{(m),m'}U^{{(m),m'}^\dag}
=U^{{(m),m'}^\dag}U^{(m),m'}=I_{r^{(m),m'}}$, as
$B_{2}^{(m),m'}=i\sqrt{1-a_{2}^{(m),m'2}}U^{(m),m'}$.
Keeping the forms of $R_0$ and $T_1$ in eqs.~\eqref{bblockt2z2_ap1}
and~\eqref{bblockt2z2_ap2}, we
can diagonalize $U^{(m),m'}$ by the basis change as
$T_2^{(m),m'} \to (\sigma_0\otimes W^{(m),m'}) T_2^{(m),m'} (\sigma_0\otimes W^{(m),m'})^\dag$, 
where $W^{(m),m'}U^{(m),m'}W^{{(m),m'}^\dag}$ is chosen to
be a diagonal matrix.  Note that the possible eigenvalues of
$U^{(m),m'}$ are $\pm 1$.  Thus, by introducing a parameter
$0\leq \phi^{(m),m'}< 2\pi$, $T_2^{(m),m'}$ is written as
\begin{align}
    T_2^{(m),m'}=
  \begin{pmatrix}
      \cos\phi^{(m),m'}I_{r^{(m),m'}}&
      i\sin\phi^{(m),m'}I_{r^{(m),m'}}\\
      i\sin\phi^{(m),m'}I_{r^{(m),m'}}&\cos\phi^{(m),m'}I_{r^{(m),m'}}
  \end{pmatrix}
  =e^{i\phi^{(m),m'}\sigma_1}\otimes I_{r^{(m),m'}}.
\end{align}
\section{Derivation of block-diagonal forms on $T^2/\Z3$}
\label{app:t2z3}

%
%

Here, we show the details of the calculation for the $T^2/\Z3$ 
orbifold. 
As discussed at the beginning of section~\ref{sec:t2z3}, 
we choose $R_0$ and $T_1$ as the independent twist matrices, and 
work on a basis where these are written as 
\bequ
 R_0=\begin{pmatrix}
       \omega I_{n_1}&&\\&\omega^2 I_{n_2}&\\&&I_{n_3}
     \end{pmatrix},  \quad
 T_1=\begin{pmatrix}(T_1)_{(11)}&(T_1)_{(12)}&(T_1)_{(13)}\\
                    (T_1)_{(21)}&(T_1)_{(22)}&(T_1)_{(23)}\\
                    (T_1)_{(31)}&(T_1)_{(32)}&(T_1)_{(33)}\end{pmatrix},  \quad
 (T_1)_{(kl)}=\submatrix{M}{k}{l}{[k-l]}. 
\eequ
Here, $n_k$ ($k=1,2,3$) is a non-negative integer,
$\submatrix{M}{k}{l}{[k-l]}$ are $n_k\times n_l$ matrices, and 
we adopt a notation of
$\submatrix{M}{k}{l}{[k-l]}=\submatrix{M}{k}{l}{[k'-l']}=\submatrix{M}{k'}{l'}{[k-l]}$
with $k'=k$ (mod 3) and $l'=l$ (mod 3).

%
\subsection{$R_a^3=I$}
\label{app:t2z3-1}

Now, we investigate the conditions $R_a^3=I$ ($a=0,1,2$). 
It is trivial for $a=0$. 
For $a=1,2$, it is convenient to examine $R_1^\dagger=R_1^2$ and 
$R_2=R_2^{\dagger2}$ as $R_a^3$ is rather complicated. 
With the help of the relations  ${R}_1={T}_1{R}_0$ and 
${R}_2={R}_0^{-1}{T}_1^{-1}{R}_0^{-1}$ shown in eq.~\eqref{Eq:Z3-T1T2R2},
these conditions lead to expressions of $T_1^\dagger$: 
$T_1^\dagger=R_0T_1R_0T_1R_0=R_0^\dagger T_1R_0^\dagger T_1R_0^\dagger$, 
which is derived also from $T_1T_2T_3=T_1T_3T_2=I$. 
Since $(T_1^\dagger)_{(k\,k-q)}= {(T_1)_{(k-q\,k)}}^\dagger$, we find
\begin{align}
& {\submatrix{M}{k-q}{k}{[-q]\dagger}} 
 = \sum_{q'} \omega^k\submatrix{M}{k}{k+q'}{[-q']}\omega^{k+q'}
            \submatrix{M}{k+q'}{k-q}{[q+q']}\omega^{k-q}
 = \sum_{q'} \omega^{q'-q}\submatrix{M}{k}{k+q'}{[-q']}
            \submatrix{M}{k+q'}{k-q}{[q+q']}
\nonumber \\
& ~~~~~~~~~\! 
 = \sum_{q'} \omega^{-k}\submatrix{M}{k}{k+q'}{[-q']}\omega^{-k-q'}
            \submatrix{M}{k+q'}{k-q}{[q+q']}\omega^{-k+q}
 = \sum_{q'} \omega^{-q'+q}\submatrix{M}{k}{k+q'}{[-q']}
            \submatrix{M}{k+q'}{k-q}{[q+q']},
\label{Eq:Z3-S1dagger}
\end{align}
where the summation over $q'$ can be taken for any successive three integers.
The equality $\sum_{q'} \omega^{q'-q}\submatrix{M}{k}{k+q'}{[-q']}\submatrix{M}{k+q'}{k-q}{[q+q']}
= \sum_{q'} \omega^{-q'+q}\submatrix{M}{k}{k+q'}{[-q']}\submatrix{M}{k+q'}{k-q}{[q+q']}$ 
requests that, with $\bar\omega=\omega^2$
\beqn
 (\omega-\bar\omega)\submatrix{M}{k}{k+1}{[-1]}\submatrix{M}{k+1}{k}{[1]}
 + (\bar\omega-\omega)\submatrix{M}{k}{k-1}{[1]}\submatrix{M}{k-1}{k}{[-1]}
 =0, 
 &&\quad \mbox{for } q=0, \\ 
 (\omega-\bar\omega)\submatrix{M}{k}{k-1}{[1]}\submatrix{M}{k-1}{k-1}{[0]}
 + (\bar\omega-\omega)\submatrix{M}{k}{k}{[0]}\submatrix{M}{k}{k-1}{[1]}
 =0, 
 &&\quad \mbox{for } q=1, \\ 
 (\omega-\bar\omega)\submatrix{M}{k}{k}{[0]}\submatrix{M}{k}{k+1}{[-1]}
 + (\bar\omega-\omega)\submatrix{M}{k}{k+1}{[-1]}\submatrix{M}{k+1}{k+1}{[0]}
 =0, 
 &&\quad \mbox{for } q=-1. 
\eeqn
These are summarized as 
\beqn
 &&\submatrix{M}{k}{k-q}{[q]}\submatrix{M}{k-q}{k}{[-q]}
 = \submatrix{M}{k}{k+q}{[-q]}\submatrix{M}{k+q}{k}{[q]}, 
\label{Eq:Z3-CommuationOfCandD}\\
 &&\submatrix{M}{k}{k}{[0]}\submatrix{M}{k}{k-q}{[q]}
 = \submatrix{M}{k}{k-q}{[q]}\submatrix{M}{k-q}{k-q}{[0]}, 
\label{Eq:Z3-CommuationOfAandCD}
\eeqn
which happen to hold also for $q=0$ (trivially), and are further 
summarized in eq.~\eqref{ZN-qq'}. 

Note that the second relations indicate that $\submatrix{M}{k}{k}{[0]}$ 
commutes with a product 
$\submatrix{M}{k}{k+q'}{[-q']}\submatrix{M}{k+q'}{k}{[q']}$, 
which appears in the expression of ${\submatrix{M}{k}{k}{[0]\dagger}}$
in eq.~\eqref{Eq:Z3-S1dagger} for $q=0$. 
This means that each $\submatrix{M}{k}{k}{[0]}$ commutes with its 
dagger, 
$[\submatrix{M}{k}{k}{[0]},{\submatrix{M}{k}{k}{[0]\dagger}}]=0$, 
to be a normal matrix. 
Then $\submatrix{M}{k}{k}{[0]}$ can be 
diagonalized by a unitary transformation. %

%
\subsection{Block-diagonal form}
\label{app:t2z3-2}

Since the unitary transformation that diagonalizes $\submatrix{M}{k}{k}{[0]}$ 
does not modify $R_0$, we may move to a basis 
with $(\submatrix{M}{k}{k}{[0]})_{ij}=a_k^i\delta_{ij}$. 
In this basis, the requirement in eq.~\eqref{Eq:Z3-CommuationOfAandCD} is 
written as 
\bequ
 (a_k^i-a_{k-q}^j)(\submatrix{M}{k}{k-q}{[q]})_{ij}=0,
\eequ
leading to $(\submatrix{M}{k}{k-q}{[q]})_{ij}=0$ if $a_k^i\neq a_{k-q}^j$. 
Thus, we see that the unitary matrices can be block-diagonalized as, 
\beqn
& R_0=\begin{pmatrix}R_0^{(1)}&&&\\&R_0^{(2)}&&\\&&
                         \ddots&\\&&&&R_0^{(M)}
          \end{pmatrix},
 \quad
 R_0^{(m)}
 =\begin{pmatrix}(R_0^{(m)})_{(11)}&(R_0^{(m)})_{(12)}&(R_0^{(m)})_{(13)}\\
                 (R_0^{(m)})_{(21)}&(R_0^{(m)})_{(22)}&(R_0^{(m)})_{(23)}\\
                 (R_0^{(m)})_{(31)}&(R_0^{(m)})_{(32)}&(R_0^{(m)})_{(33)}
  \end{pmatrix},  
\\
& T_1=\begin{pmatrix}T_1^{(1)}&&&\\&T_1^{(2)}&&\\&&
                         \ddots&\\&&&&T_1^{(M)}
          \end{pmatrix},
 \quad
 T_1^{(m)}
 =\begin{pmatrix}(T_1^{(m)})_{(11)}&(T_1^{(m)})_{(12)}&(T_1^{(m)})_{(13)}\\
                 (T_1^{(m)})_{(21)}&(T_1^{(m)})_{(22)}&(T_1^{(m)})_{(23)}\\
                 (T_1^{(m)})_{(31)}&(T_1^{(m)})_{(32)}&(T_1^{(m)})_{(33)}
  \end{pmatrix},
\eeqn
with 
\bequ
 (R_0^{(m)})_{(kl)}=\omega^k\delta_{kl}I_{n_k^{(m)}}, \quad
 (T_1^{(m)})_{(k\,k-q)}=\submatrix{M}{k}{k-q}{(m)[q]}, \quad
 \submatrix{M}{k}{k}{(m)[0]}=a^{(m)}I_{n_k^{(m)}}, 
\eequ
where $n_k^{(m)}$ ($m = 1, 2, \cdots, M$) are non-negative integers, 
as before. 
To be more concrete, 
\bequ
 R_0^{(m)}=\begin{pmatrix}
           \omega I_{n_1^{(m)}}&&\\
           &\omega^2 I_{n_2^{(m)}}&\\
           &&I_{n_3^{(m)}}
          \end{pmatrix}, \quad 
 T_1^{(m)}=\begin{pmatrix}
           a^{(m)}I_{n_1^{(m)}} & \submatrix{M}{1}{2}{(m)[-1]} 
            & \submatrix{M}{1}{3}{(m)[1]} \\
           \submatrix{M}{2}{1}{(m)[1]}  & a^{(m)}I_{n_2^{(m)}} 
            & \submatrix{M}{2}{3}{(m)[-1]}  \\
           \submatrix{M}{3}{1}{(m)[-1]}  & \submatrix{M}{3}{2}{(m)[1]} 
            & a^{(m)}I_{n_3^{(m)}}
          \end{pmatrix},
\label{Eq:Z3-R0mT1m}
\eequ
where parameters $a^{(m)}$ satisfy $a^{(m)} \ne a^{(m')}$ for $m \ne m'$.

For later convenience, we rewrite the relation in eq.~\eqref{Eq:Z3-S1dagger}
as 
\beqn
 \overline{a^{(m)}}I_{n_k^{(m)}}
   = a^{(m)2}I_{n_k^{(m)}} 
    - \submatrix{M}{k}{k-1}{(m)[1]}\submatrix{M}{k-1}{k}{(m)[-1]},  \qquad\,
 &&\quad \mbox{for } q=0, 
\label{Eq:Z3-S1dagger-v2}\\
 \submatrix{M}{k-q}{k}{(m)[-q]\dagger}  
 = - a^{(m)}\submatrix{M}{k}{k-q}{(m)[q]}  
   + \submatrix{M}{k}{k+q}{(m)[-q]}\submatrix{M}{k+q}{k-q}{(m)[-q]},    
 &&\quad \mbox{for } q=\pm1.  
\label{Eq:Z3-S1dagger-v2-OffDiag}
\eeqn

%
\subsection{$T_1T_1^\dagger=T_1^{\dagger}T_1=I$}
\label{app:t2z3-3}

Next, we examine a unitarity condition $T_1T_1^\dagger=I$, 
which implies for each block as 
\bequ
 (T_1^{(m)}T_1^{(m)\dagger})_{(k \,k-q)}
 = \sum_{q'}\submatrix{M}{k}{k+q'}{(m)[-q']}
           \submatrix{M}{k-q}{k+q'}{(m)[-q'-q]\dagger}
 = \delta_{q0}I_{n_k^{(m)}}. 
\eequ
To be more concrete, 
\beqn
 \submatrix{M}{k}{k-1}{(m)[1]}\submatrix{M}{k}{k-1}{(m)[1]\dagger}
 + \submatrix{M}{k}{k+1}{(m)[-1]}\submatrix{M}{k}{k+1}{(m)[-1]\dagger}
 = (1-\abs{a^{(m)}}^2)I_{n_k^{(m)}},
 &&\quad \mbox{for } q=0, 
\label{Eq:Z3-unitarity-diag}\\ 
 a^{(m)} \submatrix{M}{k-q}{k}{(m)[-q]\dagger} 
 + \overline{a^{(m)}} \submatrix{M}{k}{k-q}{(m)[q]}
 + \submatrix{M}{k}{k+q}{(m)[-q]}\submatrix{M}{k-q}{k+q}{(m)[q]\dagger}  
 =0,
 &&\quad \mbox{for } q=\pm1. 
\label{Eq:Z3-unitarity-offdiag} 
\eeqn
The condition for $q=0$ in eq.~\eqref{Eq:Z3-unitarity-diag} shows that 
$\submatrix{M}{k}{k\mp1}{(m)[\pm1]}$ (double sign in the same order)
vanishes when $\abs{a^{(m)}}=1$. 
In this case, $T_1^{(m)}$ is already diagonal, 
i.e., $(T_1^{(m)})_{(kl)}= a^{(m)}\delta_{kl}I_{n_k^{(m)}}$,
($\abs{a^{(m)}}=1$),
 and thus we work on the case $\abs{a^{(m)}}<1$ below. 

%
\subsubsection{$0<\abs{a^{(m)}}<1$}
\label{app:t2z3-3-1}

In the case with $0<\abs{a^{(m)}}<1$, the relation in eq.~\eqref{Eq:Z3-S1dagger-v2} tells us 
that $\submatrix{M}{k}{k-1}{(m)[1]}\submatrix{M}{k-1}{k}{(m)[-1]}$ 
is proportional to $I_{n_k^{(m)}}$ with a non-vanishing proportional constant.  
This indicates that 
 $n_1^{(m)}={\rm rank}(\submatrix{M}{1}{3}{(m)[1]}
                       \submatrix{M}{3}{1}{(m)[-1]})
 \leq{\rm rank}(\submatrix{M}{3}{1}{(m)[-1]})
 \leq n_3^{(m)}={\rm rank}(\submatrix{M}{3}{2}{(m)[1]}
                           \submatrix{M}{2}{3}{(m)[-1]})
 \leq n_2^{(m)}={\rm rank}(\submatrix{M}{2}{1}{(m)[1]}
                           \submatrix{M}{1}{2}{(m)[-1]})
 \leq n_1^{(m)}$, and thus that $n_1^{(m)}=n_2^{(m)}=n_3^{(m)} 
 (\equiv {r}^{(m)})$.
Then, $\submatrix{M}{k}{k-q}{(m)[q]}$ is an ${r}^{(m)}\times {r}^{(m)}$ matrix 
of rank ${r}^{(m)}$ and thus is invertible, and the above proportional 
relation shows 
$\submatrix{M}{k-1}{k}{(m)[-1]}\propto(\submatrix{M}{k}{k-1}{(m)[1]})^{-1}$.

Here, we note that a unitary transformation that diagonalizes a hermitian 
matrix $\submatrix{M}{k}{k-1}{(m)[1]} \submatrix{M}{k}{k-1}{(m)[1]\dagger}$ 
does not modify $R_0^{(m)}$ nor $\submatrix{M}{k}{k}{(m)[0]}$ and thus we can 
operate it freely. 
After the transformation, the relation in eq.~\eqref{Eq:Z3-unitarity-diag} 
tells us that 
$\submatrix{M}{k}{k+1}{(m)[-1]}\submatrix{M}{k}{k+1}{(m)[-1]\dagger}$ is also 
diagonalized simultaneously. 
Since 
$\submatrix{M}{k-1}{k}{(m)[-1]}\propto(\submatrix{M}{k}{k-1}{(m)[1]})^{-1}$, 
we can express each $\submatrix{M}{k}{k-1}{(m)[1]}$ by a product of a diagonal 
matrix ${\hat M}_k^{(m)[1]}$ with positive 
diagonal elements and a possible unitary matrix $\submatrix{U}{k}{}{(m)}$ that 
commutes with ${\hat M}_k^{(m)[1]}$, in this basis. 
Note that $\submatrix{U}{k}{}{(m)}$ is a diagonal phase matrix 
if ${\hat M}_k^{(m)[1]}$ has no degeneracy. 
We express $\submatrix{M}{k-1}{k}{(m)[-1]}= 
{\hat M}_k^{(m)[-1]}\submatrix{U}{k}{}{(m)\dagger}$ 
with ${\hat M}_k^{(m)[-1]}\propto 
({\hat M}_k^{(m)[1]})^{-1}$. 

In order to find constraints on ${\hat M}_k^{(m)[1]}$ and 
$\submatrix{U}{k}{}{(m)}$, let us examine 
$\submatrix{M}{k}{k-1}{(m)[1]}\submatrix{M}{k}{k-1}{(m)[1]\dagger}
(=({\hat M}_k^{(m)[1]})^2)$ and 
$\submatrix{M}{k+1}{k}{(m)[1]\dagger}\submatrix{M}{k+1}{k}{(m)[1]}
(=({\hat M}_{k+1}^{(m)[1]})^2)$. 
The relations in eq.~\eqref{Eq:Z3-S1dagger-v2-OffDiag} show that they 
are written as 
\beqn
 \submatrix{M}{k}{k-1}{(m)[1]}\submatrix{M}{k}{k-1}{(m)[1]\dagger}
 =\submatrix{M}{k}{k-1}{(m)[1]}
  (-a^{(m)}\submatrix{M}{k-1}{k}{(m)[-1]}
   +\submatrix{M}{k-1}{k-2}{(m)[1]}\submatrix{M}{k-2}{k}{(m)[1]}), 
\label{Eq:Z3-MMdaggerFromR^3-1}\\
 \submatrix{M}{k+1}{k}{(m)[1]\dagger}\submatrix{M}{k+1}{k}{(m)[1]}
 =(-a^{(m)}\submatrix{M}{k}{k+1}{(m)[-1]}
   +\submatrix{M}{k}{k-1}{(m)[1]}\submatrix{M}{k-1}{k+1}{(m)[1]})
  \submatrix{M}{k+1}{k}{(m)[1]}, 
\label{Eq:Z3-MMdaggerFromR^3-2}
\eeqn
whose last terms in the two right-hand sides are common apparently. 
In addition, the first terms in the right-hand sides are also common 
due to the relation in eq.~\eqref{Eq:Z3-CommuationOfCandD}, and thus 
we find 
${\hat M}_k^{(m)[1]}={\hat M}_{k+1}^{(m)[1]}$. 
Since this holds for all $k$, ${\hat M}_k^{(m)[1]}$ are 
common for all $k$, and thus 
the possible unitary matrix $\submatrix{U}{k}{}{(m)}$ commutes with 
${\hat M}_{k'}^{(m)[1]}$. 

Note that $\submatrix{M}{k}{k+1}{(m)[-1]}\submatrix{U}{k+1}{}{(m)}
(={\hat M}_{k+1}^{(m)[-1]})$, 
in addition to 
$\submatrix{M}{k+1}{k}{(m)[1]\dagger}\submatrix{U}{k+1}{}{(m)}
(={\hat M}_{k+1}^{(m)[1]})$, is diagonal. 
Then, the relation in eq.~\eqref{Eq:Z3-S1dagger-v2-OffDiag} requires 
that 
$\submatrix{M}{k}{k-1}{(m)[1]} \submatrix{M}{k-1}{k+1}{(m)[1]} 
 \submatrix{U}{k+1}{}{(m)}
={\hat M}_k^{(m)[1]}{\hat M}_{k-1}^{(m)[1]}
 \submatrix{U}{k}{}{(m)} \submatrix{U}{k-1}{}{(m)} \submatrix{U}{k+1}{}{(m)}$ 
is also diagonal, and so is the combination 
$\submatrix{U}{k}{}{(m)}\submatrix{U}{k-1}{}{(m)}\submatrix{U}{k+1}{}{(m)}$. 
In particular, we set 
$\submatrix{U}{2}{}{(m)}\submatrix{U}{1}{}{(m)}\submatrix{U}{3}{}{(m)}
=\hat{\varTheta}^{(m)3}$ where $\hat{\varTheta}^{(m)}$ is a diagonal phase matrix. 
Using this observation, we see that the following unitary transformation 
of $T^{(m)}_1$, which does not change $R^{(m)}_0$, diagonalizes 
$\submatrix{M}{k}{k-q}{(m)[q]}$:
\beqn
 &&T_1^{(m)} =
 \begin{pmatrix}
   a^{(m)}I_{{r}^{(m)}} 
   & {\hat M}_2^{(m)[-1]}\submatrix{U}{2}{}{(m)\dagger}  
   & {\hat M}_1^{(m)[1]}\submatrix{U}{1}{}{(m)} \\
   {\hat M}_2^{(m)[1]}\submatrix{U}{2}{}{(m)} 
   & a^{(m)}I_{{r}^{(m)}} 
   & {\hat M}_3^{(m)[-1]}\submatrix{U}{3}{}{(m)\dagger}   \\
   {\hat M}_1^{(m)[-1]}\submatrix{U}{1}{}{(m)\dagger}  
   & {\hat M}_3^{(m)[1]}\submatrix{U}{3}{}{(m)} 
   &a^{(m)}I_{{r}^{(m)}}
 \end{pmatrix}
\nonumber \\
&& \quad\to\quad V^{(m)}T_1^{(m)}V^{(m)\dagger}
= \begin{pmatrix}
   a^{(m)}I_{{r}^{(m)}} 
   & {\hat M}^{(m)[-1]}\hat{\varTheta}^{(m)\dagger}  
   & {\hat M}^{(m)[1]}\hat{\varTheta}^{(m)} \\
   {\hat M}^{(m)[1]}\hat{\varTheta}^{(m)} 
   & a^{(m)}I_{{r}^{(m)}} 
   & {\hat M}^{(m)[-1]}\hat{\varTheta}^{(m)\dagger} \\
   {\hat M}^{(m)[-1]}\hat{\varTheta}^{(m)\dagger} 
   & {\hat M}^{(m)[1]}\hat{\varTheta}^{(m)} 
   &a^{(m)}I_{{r}^{(m)}}
 \end{pmatrix},
\label{Eq:Z3-T1final}\\&&
 \qquad{\rm with}\quad 
 V^{(m)}=
 \begin{pmatrix}
   \hat{\varTheta}^{(m)\dagger} \submatrix{U}{2}{}{(m)} &&\\ 
   &I_{{r}^{(m)}}& \\
   && \hat{\varTheta}^{(m)} \submatrix{U}{3}{}{(m)\dagger}
 \end{pmatrix},
\eeqn
where $k$ in ${\hat M}_k^{(m)[\mp1]}$ is omitted 
in $V^{(m)}T_1^{(m)}V^{(m)\dagger}$ because ${\hat M}_k^{(m)[\mp1]}$
are independent of $k$.

%
\subsubsection{$a^{(m)}=0$}
\label{app:t2z3-3-2}

In the case with $a^{(m)}=0$, the conditions in eqs.~\eqref{Eq:Z3-unitarity-diag} 
and \eqref{Eq:Z3-unitarity-offdiag} are 
simplified as 
\bequ
 \submatrix{M}{k}{k-1}{(m)[1]}\submatrix{M}{k}{k-1}{(m)[1]\dagger}
 + \submatrix{M}{k}{k+1}{(m)[-1]}\submatrix{M}{k}{k+1}{(m)[-1]\dagger}
 = I_{n_k^{(m)}}, \quad
 \submatrix{M}{k}{k+q}{(m)[-q]}\submatrix{M}{k-q}{k+q}{(m)[q]\dagger}  
 =0, 
\label{Eq:Z3-unitarity-a=0}
\eequ
where $q=\pm1$. 
Similarly, the conditions from $T_1^\dagger T_1=I$ are given as 
\bequ
 \submatrix{M}{k+1}{k}{(m)[1]\dagger}\submatrix{M}{k+1}{k}{(m)[1]}
 + \submatrix{M}{k-1}{k}{(m)[-1]\dagger}\submatrix{M}{k-1}{k}{(m)[-1]}
 = I_{n_k^{(m)}}, \quad
 \submatrix{M}{k}{k+q}{(m)[-q]\dagger}\submatrix{M}{k}{k-q}{(m)[q]}  
 =0. 
\label{Eq:Z3-TdaggerT-a=0}
\eequ
In addition, those in eqs.~\eqref{Eq:Z3-S1dagger-v2} and 
\eqref{Eq:Z3-S1dagger-v2-OffDiag} are
\bequ
 \submatrix{M}{k}{k-1}{(m)[1]}\submatrix{M}{k-1}{k}{(m)[-1]}=0, \quad
 \submatrix{M}{k-q}{k}{(m)[-q]\dagger}=
 \submatrix{M}{k}{k+q}{(m)[-q]}\submatrix{M}{k+q}{k-q}{(m)[-q]}. 
\label{Eq:Z3-R^3-a=0}
\eequ

Multiplying $\submatrix{M}{k}{k-1}{(m)[1]}$ from the right to the first 
condition in eq.~\eqref{Eq:Z3-unitarity-a=0}, we get 
\bequ
 \submatrix{M}{k}{k-1}{(m)[1]}\submatrix{M}{k}{k-1}{(m)[1]\dagger}
 \submatrix{M}{k}{k-1}{(m)[1]}
 = \submatrix{M}{k}{k-1}{(m)[1]}, 
\eequ
as $\submatrix{M}{k}{k+1}{(m)[-1]\dagger}\submatrix{M}{k}{k-1}{(m)[1]}=0$. 
This indicates that, in the basis where 
$\submatrix{M}{k}{k-1}{(m)[1]}\submatrix{M}{k}{k-1}{(m)[1]\dagger}$ 
are diagonal, which can be chosen without loss of generality, 
they can be written as 
\bequ
\submatrix{M}{k}{k-1}{(m)[1]}\submatrix{M}{k}{k-1}{(m)[1]\dagger}=
 \begin{pmatrix}
   I_{{r_k}^{(m)}} &\\ &0 \\
 \end{pmatrix},
\label{Eq.Z3-MMdagger(a=0)}
\eequ
and then 
\bequ
\submatrix{M}{k}{k+1}{(m)[-1]}\submatrix{M}{k}{k+1}{(m)[-1]\dagger}=
 \begin{pmatrix}
   0 &\\ & I_{{n_k}^{(m)}-{r_k}^{(m)}} \\
 \end{pmatrix},
\eequ
where ${r_k}^{(m)}$ are the rank of $\submatrix{M}{k}{k-1}{(m)[1]}$. 
We note that ${\rm rank}(\submatrix{M}{k}{k+1}{(m)[-1]})=
{n_k}^{(m)}-{r_k}^{(m)}$.

The linear algebra tells us that the first conditions in 
eq.~\eqref{Eq:Z3-R^3-a=0} imply that 
${\rm rank}(\submatrix{M}{k}{k-1}{(m)[1]})
+{\rm rank}(\submatrix{M}{k-1}{k}{(m)[-1]})
-{n_{k-1}}^{(m)}<{\rm rank}(0)=0$, and thus 
$0>{r_k}^{(m)}+{n_{k-1}}^{(m)}-{r_{k-1}}^{(m)}-{n_{k-1}}^{(m)}
 ={r_k}^{(m)}-{r_{k-1}}^{(m)}$. 
This means ${r_k}^{(m)}<{r_{k-1}}^{(m)}<{r_{k-2}}^{(m)}= {r_{k+1}}^{(m)}<{r_k}^{(m)}$ 
and thus ${r_k}^{(m)}$ are independent of $k$. 

Since the above discussion also holds after flip the sign of $q$ in 
$\submatrix{M}{k}{k-q}{(m)[q]}$, we see that 
${\rm rank}(\submatrix{M}{k}{k+1}{(m)[-1]})={n_k}^{(m)}-{r_k}^{(m)}$, 
and thus ${n_k}^{(m)}$ are also independent of $k$. 

Now, we discuss 
$\submatrix{M}{k}{k+q}{(m)[-q]\dagger}\submatrix{M}{k}{k+q}{(m)[-q]}$. 
As in the previous $0<\abs{a^{(m)}}<1$ case, the second 
conditions in eq.~\eqref{Eq:Z3-R^3-a=0} derive the conditions 
in eqs.~\eqref{Eq:Z3-MMdaggerFromR^3-1} and 
\eqref{Eq:Z3-MMdaggerFromR^3-2} with $a^{(m)}=0$. 
In this case, the right hand sides of these equations are common and 
we see that 
$\submatrix{M}{k+1}{k}{(m)[1]\dagger}\submatrix{M}{k+1}{k}{(m)[1]}$
are diagonal from eq.~\eqref{Eq.Z3-MMdagger(a=0)}, 
and thus from the first conditions in eq.\eqref{Eq:Z3-TdaggerT-a=0},
$\submatrix{M}{k-1}{k}{(m)[-1]\dagger}\submatrix{M}{k-1}{k}{(m)[-1]}$
are also diagonal in the present basis. 
This means that we can write as 
\bequ
\submatrix{M}{k}{k-1}{(m)[1]}=
 \begin{pmatrix}
   U_k^{\prime(m)} &\\ & 0 \\
 \end{pmatrix},\quad
\submatrix{M}{k-1}{k}{(m)[-1]}=
 \begin{pmatrix}
   0 &\\ & U_k^{\prime\prime(m)\dagger} \\
 \end{pmatrix},
\eequ
where $U_k^{\prime(m)}$ and $U_k^{\prime\prime(m)}$ are 
$r_k^{(m)}\times r_k^{(m)}$ and 
$(n_k^{(m)}-r_k^{(m)})\times (n_k^{(m)}-r_k^{(m)})$ unitary matrices, 
respectively. 

Here, we can rearrange $T_1^{(m)}$ (and of course also $R_0^{(m)}$) into 
two block-diagonal parts where both the diagonal blocks are written 
in a similar form as in eq.~\eqref{Eq:Z3-R0mT1m}, 
but one with $a^{(m)}=0$ and $\submatrix{M}{k-1}{k}{(m)[-1]}=0$ 
and the other one with  $a^{(m)}=0$ and 
$\submatrix{M}{k}{k-1}{(m)[1]}=0$. 

The degrees of freedom of the unitary matrices $U_k^{\prime(m)}$ and 
$U_k^{\prime\prime(m)}$ are treated in a similar way as in the previous 
case discussed above eq.~\eqref{Eq:Z3-T1final}. 
At the end, we get the same form as in the last matrix 
in eq.~\eqref{Eq:Z3-T1final} but with constraints of $a^{(m)}=0$, 
${\hat M}^{(m)[q]}=I_{r^{(m)}}$ and ${\hat M}^{(m)[-q]}=0$ where 
$q=\pm1$.

\section{Derivation of block-diagonal forms on $T^2/\Z4$}
\label{app:t2z4}

In this appendix, we explain the details of block-diagonalization
of twist matrices on $T^2/\Z4$ given in section~\ref{sec:t2z4}.

We can start with the following $R_0$ and $T_1$, without loss of generality,
\begin{align}
R_0=
  \begin{pmatrix}
      iI_{n_1}& & & \\
       &-I_{n_2}& & \\
       & &-iI_{n_3}& \\
       & & &I_{n_4}
  \end{pmatrix},\qquad
T_1=
  \begin{pmatrix}
      (T_1)_{(11)}&(T_1)_{(12)}&(T_1)_{(13)}&(T_1)_{(14)}\\
      (T_1)_{(21)}&(T_1)_{(22)}&(T_1)_{(23)}&(T_1)_{(24)}\\
      (T_1)_{(31)}&(T_1)_{(32)}&(T_1)_{(33)}&(T_1)_{(34)}\\
      (T_1)_{(41)}&(T_1)_{(42)}&(T_1)_{(43)}&(T_1)_{(44)}
  \end{pmatrix}, 
\label{C-Z4-R0T1}
\end{align}
where $I_{n_k}$ are $n_k \times n_k$ unit matrices and
$(T_1)_{(kl)}$ are $n_k \times n_l$ submatrices.
We denote the submatrices as $(T_1)_{(kl)} = \submatrix{M}{k}{l}{[k-l]}$
and use a notation of
$\submatrix{M}{k}{l}{[k-l]}=\submatrix{M}{k}{l}{[k'-l']}=\submatrix{M}{k'}{l'}{[k-l]}$
with $k'=k$ (mod 4) and $l'=l$ (mod 4).
The upper index $k-l=q$ represents the charge of the $\Z4$ symmetry 
generated by $R_0$: 
$(R_0T_1R_0^{-1})_{(k\,k-q)}=i^{q}\submatrix{M}{k}{k-q}{[q]}$. 
Parameters such as $n_k$ which represent the size of matrices
are non-negative integers.
The same applies to the following ones.

\subsection{Block-diagonalization of $T_1$ and relations on submatrices}
\label{C1}

First, let us restrict the form of $T_1$ and derive relations on submatrices of $T_1$,
using constraints $T_1^{\dagger} = T_1^{-1}$, $T_{m'} T_m = T_m T_{m'}$ and $T_1 T_3 = I$
where $T_{m} = R_0^{m-1} T_1 R_0^{1-m}$
as explained in section~\ref{sec:2Dorb}.

From $T_1^{\dagger} (= T_1^{-1} = T_3) = R_0^2 T_1 R_0^{-2}$, 
we derive the relation:
\begin{align}
\submatrix{M}{k-q}{k}{[-q]\dagger}=(-1)^{q}\submatrix{M}{k}{k-q}{[q]}.
\label{C-Z4-Mdagger}
\end{align}
Taking $q=0$ in eq.~\eqref{C-Z4-Mdagger}, 
we find that $\submatrix{M}{k}{k}{[0]\dagger}=\submatrix{M}{k}{k}{[0]}$,
and $\submatrix{M}{k}{k}{[0]}$ is diagonalized as $(\submatrix{M}{k}{k}{[0]})_{ij}
=a_k^i\delta_{ij}$ $(a_k^i\in {\mathbb R})$
by a unitary transformation,
without modifying $R_0$,
where $(\submatrix{M}{k}{k}{[0]})_{ij}$
are ($i,j$) elements of $\submatrix{M}{k}{k}{[0]}$. 

As shown in appendix~\ref{app:MM}, 
from $T_{m'} T_{m} = T_{m} T_{m'}$, i.e., 
$T_1 R_0^{m-m'} T_1 = R_0^{m-m'} T_1 R_0^{m'-m} T_1 R_0^{m-m'}$, 
we derive the relation:
\begin{align}
\submatrix{M}{k}{k-q'}{[q']}\submatrix{M}{k-q'}{k-q}{[q-q']}
=\submatrix{M}{k}{k-q+q'}{[q-q']}\submatrix{M}{k-q+q'}{k-q}{[q']}.
\label{C-Z4-qq'}
\end{align}
Setting $q'=0$ in eq.~\eqref{C-Z4-qq'} 
and using $(\submatrix{M}{k}{k}{[0]})_{ij}=a_k^i\delta_{ij}$,
we obtain the relation:
\begin{align}
(a_k^i-a_{k-q}^j)(\submatrix{M}{k}{k-q}{[q]})_{ij}=0,
\label{C-Z4-(ai-aj)}
\end{align}
and rearrange $T_1$ to be a block-diagonal matrix, 
while keeping $R_0$ a diagonal one:
\begin{align}
R_0=
    \begin{pmatrix}
        R_0^{(1)} &          &        &         \\
                 & R_0^{(2)} &        &         \\
                 &          & \ddots &         \\
                 &          &        & R_0^{(M)}
    \end{pmatrix},\qquad
T_1=
    \begin{pmatrix}
        T_1^{(1)} &          &        &         \\
                 & T_1^{(2)} &        &         \\
                 &          & \ddots &         \\
                 &          &        & T_1^{(M)}
    \end{pmatrix},
\label{C-Z4-R0T1-block}
\end{align}
where $R_0^{(m)}$ and $T_1^{(m)}$ $(m = 1, 2, \cdots, M)$ are 
$n^{(m)} \times n^{(m)}$ matrices given by
\begin{align}
R_0^{(m)}=
    \begin{pmatrix}
        iI_{n_1^{(m)}} &          &        &         \\
                 & -I_{n_2^{(m)}} &        &         \\
                 &          & -iI_{n_3^{(m)}} &         \\
                 &          &        & I_{n_4^{(m)}}
    \end{pmatrix},
\label{C-Z4-R0m}
\end{align}
and
\begin{align}
&T_1^{(m)}=
    \begin{pmatrix}
        a^{(m)}I_{n_1^{(m)}} & \submatrix{M}{1}{2}{(m)[-1]} 
         & \submatrix{M}{1}{3}{(m)[-2]} & \submatrix{M}{1}{4}{(m)[1]} \\
        \submatrix{M}{2}{1}{(m)[1]} & a^{(m)}I_{n_2^{(m)}} 
         & \submatrix{M}{2}{3}{(m)[-1]} & \submatrix{M}{2}{4}{(m)[-2]}       \\
        \submatrix{M}{3}{1}{(m)[2]} & \submatrix{M}{3}{2}{(m)[1]} 
         & a^{(m)}I_{n_3^{(m)}} & \submatrix{M}{3}{4}{(m)[-1]} \\
        \submatrix{M}{4}{1}{(m)[-1]} & \submatrix{M}{4}{2}{(m)[2]}
         & \submatrix{M}{4}{3}{(m)[1]} & a^{(m)}I_{n_4^{(m)}}
    \end{pmatrix},
\label{C-Z4-T1m}
\end{align}
respectively.
Here, $n^{(m)} = \sum_{k=1}^4 n_k^{(m)}$ and
submatrices in $T_1^{(m)}$ are denoted
as $(T_1^{(m)})_{(kl)} = \submatrix{M}{k}{l}{(m)[k-l]}$
for $k \ne l$ and $(T_1^{(m)})_{(kk)} = \submatrix{M}{k}{k}{(m)[0]}=a^{(m)} I_{n_k^{(m)}}$.
Real parameters $a^{(m)}$ satisfy $a^{(m)} \ne a^{(m')}$ for $m \ne m'$.
Submatrices obey the following relations 
that come from $T_1^{\dagger} = T_3$ and $T_{m'} T_{m} = T_{m} T_{m'}$,
\begin{align}
& \submatrix{M}{k-q}{k}{(m)[-q]\dagger}=(-1)^{q}\submatrix{M}{k}{k-q}{(m)[q]},
\label{C-Z4-Mdaggerm}\\
& \submatrix{M}{k}{k-q'}{(m)[q']}\submatrix{M}{k-q'}{k-q}{(m)[q-q']}
=\submatrix{M}{k}{k-q+q'}{(m)[q-q']}\submatrix{M}{k-q+q'}{k-q}{(m)[q']}.
\label{C-Z4-qq'm}
\end{align}

From $T_1 T_3 = T_1 R_0^2 T_1 R_0^{-2} = I$, 
we derive the relation:
\begin{align}
\sum_{q'} (-1)^{q-q'}
\submatrix{M}{k}{k-q'}{(m)[q']}\submatrix{M}{k-q'}{k-q}{(m)[q-q']} 
= \delta_{k~\!\!k-q} I_{n_k^{(m)}},
\label{C-Z4-1MM}
\end{align}
where the summation over $q'$ can be taken for any successive four integers.
Setting $q=0$ and $q=2$ in eq.~\eqref{C-Z4-1MM} and using eqs.~\eqref{C-Z4-Mdaggerm}
and \eqref{C-Z4-qq'm},
we obtain the relations:
\begin{align}
& 2\submatrix{M}{k}{k-1}{(m)[1]}\submatrix{M}{k}{k-1}{(m)[1]\dagger}
+\submatrix{M}{k}{k-2}{(m)[2]}\submatrix{M}{k}{k-2}{(m)[2]\dagger}
=\left(1-a^{(m)2}\right)I_{n_k^{(m)}},
& \quad \mbox{for } q=0,
\label{C-Z4-MM0m}\\
& 2a^{(m)}\submatrix{M}{k}{k-2}{(m)[2]}
=\submatrix{M}{k}{k-1}{(m)[1]}\submatrix{M}{k-1}{k-2}{(m)[1]}
+\submatrix{M}{k}{k+1}{(m)[-1]}\submatrix{M}{k+1}{k-2}{(m)[-1]},
& \quad \mbox{for } q=2.
\label{C-Z4-MM2m}
\end{align}
The relations obtained by setting $q=1$ and $q=3$ in eq.~\eqref{C-Z4-1MM} 
are not independent of eq.~\eqref{C-Z4-qq'm}, because they are also derived from
linear combinations of eq.~\eqref{C-Z4-qq'm}.
From eq.~\eqref{C-Z4-MM0m}, we find that $0 \leq a^{(m)2} \leq 1$.
Note that $T_1^{(m)}$ is already diagonal for $a^{(m)}=\pm 1$,
and then $R_0$ and $T_1$ contain
$n^{(m)} \times n^{(m)}$ diagonal matrices whose submatrices
are given by $(R_0^{(m)})_{(kl)} = i^k \delta_{kl} I_{n_k^{(m)}}$
and $(T_1^{(m)})_{(kl)} = \pm \delta_{kl} I_{n_k^{(m)}}$
for $a^{(m)}=\pm 1$ (double sign in the same order)
as members of block-diagonal matrices.
Hereafter, we focus on $-1 < a^{(m)} < 1$.

\subsection{Restriction of $\submatrix{M}{k}{l}{(m)[k-l]}$}
\label{C2}

Next, let us restrict the form of $\submatrix{M}{k}{l}{(m)[k-l]}$,
using relations obtained in the previous subsection.

For an arbitrary complex matrix $A$ with a rank $r$,
$A A^{\dagger}$ is a hermitian matrix,
and is rearranged to be the block-diagonal
form of $(AA^\dagger)'_{\rm D} \oplus {\bm 0}$
where $(AA^\dagger)'_{\rm D}$ is an $r \times r$ diagonal matrix
with positive elements
and ${\bm 0}$ is a submatrix whose elements are zero,
after a diagonalization by a unitary transformation.
Then $(AA^\dagger)'_{\rm D}=(A^\dagger A)'_{\rm D}$ holds,
and $A$ is written as $A = \hat{A}_r {U} \oplus {\bm 0}$,
where $\hat{A}_r$ is a diagonal $r \times r$ matrix
with non-vanishing real positive elements
and ${U}$ is an $r \times r$ unitary matrix
which commutes with $\hat{A}_r$, 
i.e., $[\hat{A}_r, U]=0$,
as seen from appendix~\ref{ap:mmdag}.

Using eq.~\eqref{C-Z4-Mdaggerm}, eq.~\eqref{C-Z4-qq'm} with $q=0$ 
is rewritten as
\begin{align}
\submatrix{M}{k}{k-q'}{(m)[q']}\submatrix{M}{k}{k-q'}{(m)[q']\dagger}
=\submatrix{M}{k+q'}{k}{(m)[q']\dagger}\submatrix{M}{k+q'}{k}{(m)[q']},
\label{C-Z4-qq'm0}
\end{align}
and the following relation on the rank of submatrices is obtained
\begin{align}
& {\rm rank}(\submatrix{M}{k}{k-q'}{(m)[q']})
= {\rm rank}(\submatrix{M}{k}{k-q'}{(m)[q']\dagger})
= {\rm rank}(\submatrix{M}{k}{k-q'}{(m)[q']}\submatrix{M}{k}{k-q'}{(m)[q']\dagger})
\nonumber \\
& = {\rm rank}(\submatrix{M}{k+q'}{k}{(m)[q']\dagger}\submatrix{M}{k+q'}{k}{(m)[q']})
= {\rm rank}(\submatrix{M}{k+q'}{k}{(m)[q']}).
\label{C-Z4-qq'm0-rank}
\end{align}
As seen from eq.~\eqref{C-Z4-MM0m}, 
$\submatrix{M}{k}{k-1}{(m)[1]}\submatrix{M}{k}{k-1}{(m)[1]\dagger}$
and $\submatrix{M}{k}{k-2}{(m)[2]}\submatrix{M}{k}{k-2}{(m)[2]\dagger}$
are diagonalized simultaneously by some unitary matrices $W_k^{(m)}$ such that
\begin{align}
\left(W_k^{(m)} \submatrix{M}{k}{k-q'}{(m)[q']}
\submatrix{M}{k}{k-q'}{(m)[q']\dagger} W_k^{(m)\dagger}\right)_{\rm R}
= \left(\submatrix{M}{k}{k-q'}{(m)[q']}\submatrix{M}{k}{k-q'}{(m)[q']\dagger}\right)'_{\rm D} 
\oplus {\bm 0},
\quad (q' =1, 2),
\label{C-Z4-WkM}
\end{align}
where $(\cdots)_{\rm R}$ means 
that a suitable rearrangement by interchanges of rows and/or those of columns has been done.
From eq.~\eqref{C-Z4-qq'm0}, we find that
$\submatrix{M}{k+1}{k}{(m)[1]\dagger}\submatrix{M}{k+1}{k}{(m)[1]}$
and $\submatrix{M}{k+2}{k}{(m)[2]\dagger}\submatrix{M}{k+2}{k}{(m)[2]}$
are also diagonalized simultaneously by $W_k^{(m)}$ such that
\begin{align}
\left(W_k^{(m)} \submatrix{M}{k+q'}{k}{(m)[q']\dagger}
\submatrix{M}{k+q'}{k}{(m)[q']} W_k^{(m)\dagger}\right)_{\rm R}
= \left(\submatrix{M}{k+q'}{k}{(m)[q']\dagger}\submatrix{M}{k+q'}{k}{(m)[q']}\right)'_{\rm D} 
\oplus {\bm 0},
\quad (q' =1, 2),
\label{C-Z4-WkMdagger}
\end{align}
and obtain the relation:
\begin{align}
\left(\submatrix{M}{k}{k-q'}{(m)[q']}\submatrix{M}{k}{k-q'}{(m)[q']\dagger}\right)'_{\rm D}
= \left(\submatrix{M}{k+q'}{k}{(m)[q']\dagger}\submatrix{M}{k+q'}{k}{(m)[q']}\right)'_{\rm D}
\quad (q' =1, 2).
\label{C-Z4-qq'm0D}
\end{align}
Replacing $k$ with $k+q'$ in eq.~\eqref{C-Z4-WkM}
and $k$ with $k-q'$ in eq.~\eqref{C-Z4-WkMdagger}, we derive the relations:
\begin{align}
\left(W_{k+q'}^{(m)} \submatrix{M}{k+q'}{k}{(m)[q']}
\submatrix{M}{k+q'}{k}{(m)[q']\dagger} W_{k+q'}^{(m)\dagger}\right)_{\rm R}
= \left(\submatrix{M}{k+q'}{k}{(m)[q']}\submatrix{M}{k+q'}{k}{(m)[q']\dagger}\right)'_{\rm D} 
\oplus {\bm 0},
\quad (q' =1, 2),
\label{C-Z4-Wk+q'Mdagger}
\end{align}
and
\begin{align}
\left(W_{k-q'}^{(m)} \submatrix{M}{k}{k-q'}{(m)[q']\dagger}
\submatrix{M}{k}{k-q'}{(m)[q']} W_{k-q'}^{(m)\dagger}\right)_{\rm R}
= \left(\submatrix{M}{k}{k-q'}{(m)[q']\dagger}\submatrix{M}{k}{k-q'}{(m)[q']}\right)'_{\rm D} 
\oplus {\bm 0},
\quad (q' =1, 2),
\label{C-Z4-Wk-q'Mdagger}
\end{align}
respectively.
Here $\left(\submatrix{M}{k}{k-q'}{(m)[q']}\submatrix{M}{k}{k-q'}{(m)[q']\dagger}\right)'_{\rm D}
= \left(\submatrix{M}{k}{k-q'}{(m)[q']\dagger}\submatrix{M}{k}{k-q'}{(m)[q']}\right)'_{\rm D}$
and $\left(\submatrix{M}{k+q'}{k}{(m)[q']\dagger}\submatrix{M}{k+q'}{k}{(m)[q']}\right)'_{\rm D}
= \left(\submatrix{M}{k+q'}{k}{(m)[q']}\submatrix{M}{k+q'}{k}{(m)[q']\dagger}\right)'_{\rm D}$
hold on and, using eq.~\eqref{C-Z4-qq'm0D}, we obtain the relation:
\begin{align}
\left(\submatrix{M}{k}{k-q'}{(m)[q']}\submatrix{M}{k}{k-q'}{(m)[q']\dagger}\right)'_{\rm D}
= \left(\submatrix{M}{k+q'}{k}{(m)[q']}\submatrix{M}{k+q'}{k}{(m)[q']\dagger}\right)'_{\rm D}
\quad (q' =1, 2).
\label{C-Z4-qq'm0Dk}
\end{align}
Note that the rearrangement for $q'=1$ is generally different from that for $q'=2$
in eq.~\eqref{C-Z4-WkM}, but the same rearrangement has been done 
in eqs.~\eqref{C-Z4-WkMdagger}, \eqref{C-Z4-Wk+q'Mdagger} and \eqref{C-Z4-Wk-q'Mdagger} 
as that in eq.~\eqref{C-Z4-WkM} for each $q'$.
From eqs.~\eqref{C-Z4-WkM} and \eqref{C-Z4-Wk-q'Mdagger}, we see
that both $\submatrix{M}{k}{k-q'}{(m)[q']}\submatrix{M}{k}{k-q'}{(m)[q']\dagger}$
and $\submatrix{M}{k}{k-q'}{(m)[q']\dagger}\submatrix{M}{k}{k-q'}{(m)[q']}$ ($q' = 1, 2$)
become diagonal by the biunitary transformation
$W_k^{(m)} \submatrix{M}{k}{k-q'}{(m)[q']} W_{k-q'}^{(m)\dagger}$
without modifying $R_0^{(m)}$ and $\submatrix{M}{k}{k}{(m)[0]}$,
and hereafter we take a basis where they are already diagonal.

Using eq.~\eqref{C-Z4-qq'm0Dk} with $q' = 1$ iteratively, we derive the relation:
\begin{align}
& \left(\submatrix{M}{k}{k-1}{(m)[1]}\submatrix{M}{k}{k-1}{(m)[1]\dagger}\right)'_{\rm D}
= \left(\submatrix{M}{k+1}{k}{(m)[1]}\submatrix{M}{k+1}{k}{(m)[1]\dagger}\right)'_{\rm D}
= \left(\submatrix{M}{k+2}{k+1}{(m)[1]}
\submatrix{M}{k+2}{k+1}{(m)[1]\dagger}\right)'_{\rm D}
\nonumber \\
& = \left(\submatrix{M}{k+3}{k+2}{(m)[1]}
\submatrix{M}{k+3}{k+2}{(m)[1]\dagger}\right)'_{\rm D}.
\label{C-Z4-qq'm0Dq=1}
\end{align}
From eq.~\eqref{C-Z4-qq'm0Dq=1},
we find that 
$\left(\submatrix{M}{k}{k-1}{(m)[1]}\submatrix{M}{k}{k-1}{(m)[1]\dagger}\right)'_{\rm D}$
are independent of $k$ and $\submatrix{M}{k}{k-1}{(m)[1]}$ have a common rank 
$r^{(m)} (= {\rm rank}(\submatrix{M}{k}{k-1}{(m)[1]}))$.  
Then, based on the argument in appendix~\ref{ap:mmdag},
we choose the basis such that $\submatrix{M}{k}{k-1}{(m)[1]}$ have the form:
\begin{align}
\submatrix{M}{k}{k-1}{(m)[1]}&=\left(
\begin{array}{ccc|cc}
 &           &                           &  \\
                 & {\hat M}^{(m)[1]}\submatrix{U}{k}{k-1}{(m)} &  & \text{\large{0}} \\
                 &           &  &  \\
\hline
 & & & \\
 & \text{\large{0}} & & \text{\large{0}} \\
\end{array}
\right),
\label{C-Z4-Mmq=1}
\end{align}
where ${\hat M}^{(m)[1]}$ are $r^{(m)} \times r^{(m)}$ diagonal matrices
with positive elements, i.e., $({\hat M}^{(m)[1]})_{ii} > 0$, and
$\submatrix{U}{k}{k-1}{(m)}$ are $r^{(m)} \times r^{(m)}$ unitary matrices
that take a block-diagonal form such as $U$ in eq.~\eqref{A-ArU}.
From eqs.~\eqref{C-Z4-Mdaggerm} and \eqref{C-Z4-Mmq=1}, 
$\submatrix{M}{k-1}{k}{(m)[-1]}$ are given by
\begin{align}
\submatrix{M}{k-1}{k}{(m)[-1]}&=-\submatrix{M}{k}{k-1}{(m)[1]\dagger}=\left(
\begin{array}{ccc|cc}
 &           &                           &  \\
                 & {\hat M}^{(m)[1]}\submatrix{U}{k-1}{k}{(m)} &  & \text{\large{0}} \\
                 &           &  &  \\
\hline
 & & & \\
 & \text{\large{0}} & & \text{\large{0}} \\
\end{array}
\right),
\label{C-Z4-Mmq=-1}
\end{align}
where $-\submatrix{U}{k}{k-1}{(m)\dagger}$ are denoted as $\submatrix{U}{k-1}{k}{(m)}$.

From eqs.~\eqref{C-Z4-qq'm0-rank} and \eqref{C-Z4-qq'm0Dk} with $q' = 2$, 
we derive the relations:
\begin{align}
& {\rm rank}(\submatrix{M}{k}{k-2}{(m)[2]})
= {\rm rank}(\submatrix{M}{k+2}{k}{(m)[2]}), 
\label{C-Z4-qq'm0-rankq=2}\\
& \left(\submatrix{M}{k}{k-2}{(m)[2]}\submatrix{M}{k}{k-2}{(m)[2]\dagger}\right)'_{\rm D}
= \left(\submatrix{M}{k+2}{k}{(m)[2]}\submatrix{M}{k+2}{k}{(m)[2]\dagger}\right)'_{\rm D}.
\label{C-Z4-qq'm0Dq=2}
\end{align}
Using eq.~\eqref{C-Z4-MM0m} and the fact that 
$\left(\submatrix{M}{k}{k-1}{(m)[1]}\submatrix{M}{k}{k-1}{(m)[1]\dagger}\right)'_{\rm D}$
are independent of $k$ and 
$\submatrix{M}{k}{k-1}{(m)[1]}\submatrix{M}{k}{k-1}{(m)[1]\dagger}$
are diagonal in the present basis,i.e., 
$\submatrix{M}{k}{k-1}{(m)[1]}\submatrix{M}{k}{k-1}{(m)[1]\dagger} 
= ({\hat M}^{(m)[1]})^2 \oplus {\bm 0}$,
we find that
the $r^{(m)} \times r^{(m)}$ submatrices in
$\left(\submatrix{M}{k}{k-2}{(m)[2]}\submatrix{M}{k}{k-2}{(m)[2]\dagger}\right)'_{\rm D}$
corresponding to $({\hat M}^{(m)[1]})^2$
are also independent of $k$ and 
$\submatrix{M}{k}{k-2}{(m)[2]}\submatrix{M}{k}{k-2}{(m)[2]\dagger}$
are also diagonal, and then
the form of $\submatrix{M}{k}{k-2}{(m)[2]}$ can be written as
\begin{align}
\submatrix{M}{k}{k-2}{(m)[2]}&=\left(
\begin{array}{ccc|cc}
 &           &                           &  \\
                 & {\hat M}^{(m)[2]}\submatrix{U}{k}{k-2}{(m)} &  & \text{\large{0}} \\
                 &           &  &  \\
\hline
 & & & \\
 & \text{\large{0}} &  & \sqrt{1-a^{(m)2}}~\! 
\tilde{U}_{k~\!\!k-2}^{(m)}\\
\end{array}
\right),
\label{C-Z4-Mmq=2}
\end{align}
where ${\hat M}^{(m)[2]}$ are $r^{(m)} \times r^{(m)}$ diagonal matrices
with non-negative elements, i.e., $({\hat M}^{(m)[2]})_{ii} \ge 0$, and
$\submatrix{U}{k}{k-2}{(m)}$ are $r^{(m)} \times r^{(m)}$ unitary matrices, i.e., 
$\submatrix{U}{k}{k-2}{(m)}\submatrix{U}{k-2}{k}{(m)}
=\submatrix{U}{k-2}{k}{(m)}\submatrix{U}{k}{k-2}{(m)}=I_{r^{(m)}}$.
Here $\submatrix{U}{k}{k-2}{(m)\dagger}$ are denoted as $\submatrix{U}{k-2}{k}{(m)}$.
We also use the notation $\tilde{U}_{k-2~\!\!k}^{(m)}
= \tilde{U}_{k~\!\!k-2}^{(m)\dagger}$ in the following.
Using eqs.~\eqref{C-Z4-MM0m} and \eqref{C-Z4-Mmq=2}, we derive the relations
$\tilde{U}_{k~\!\!k-2}^{(m)}\tilde{U}_{k~\!\!k-2}^{(m)\dagger}=I_{n_k^{(m)'}}$
where $n_{k}^{(m)'} = n_{k}^{(m)}-r^{(m)}$.
Replacing $k$ with $k+2$ for 
$\tilde{U}_{k~\!\!k-2}^{(m)}\tilde{U}_{k~\!\!k-2}^{(m)\dagger}=I_{n_k^{(m)'}}$,
we obtain $\tilde{U}_{k+2~\!\!k}^{(m)}\tilde{U}_{k+2~\!\!k}^{(m)\dagger}=I_{n_{k+2}^{(m)'}}$
where $n_{k+2}^{(m)'} = n_{k+2}^{(m)}-r^{(m)}$.
Then, we obtain the relations:
\begin{align}
{\rm rank}(\submatrix{M}{k}{k-2}{(m)[2]})={\rm rank}({\hat M}^{(m)[2]})+ n_k^{(m)'}, \quad
{\rm rank}(\submatrix{M}{k+2}{k}{(m)[2]})={\rm rank}({\hat M}^{(m)[2]})+ n_{k+2}^{(m)'}.
\label{C-Z4-rankMk+2}
\end{align}
From eqs.~\eqref{C-Z4-qq'm0-rankq=2} and \eqref{C-Z4-rankMk+2},
we find that $n_k^{(m)'} = n_{k+2}^{(m)'}$, i.e.,
$n_1^{(m)'} = n_3^{(m)'}$ and $n_2^{(m)'} = n_4^{(m)'}$, and then
$\tilde{U}_{k~\!\!k-2}^{(m)}$ are 
$n_k^{(m)'} \times n_k^{(m)'}$ unitary matrices.
Thus, $\submatrix{M}{k}{k-2}{(m)[2]}$ 
are $n_k^{(m)} \times n_k^{(m)}$ matrices 
where $n_1^{(m)} = n_3^{(m)}$ and $n_2^{(m)} = n_4^{(m)}$.
The $(2, 2)$ block in the right-hand side of eq.~\eqref{C-Z4-Mmq=2}
can appear for $a^{(m)} = 0$ as will be seen from eq.~\eqref{C-Z4-qq'mD-hat-tilde}.

Inserting eqs.~\eqref{C-Z4-Mmq=1}, \eqref{C-Z4-Mmq=-1} and \eqref{C-Z4-Mmq=2} into eq.~\eqref{C-Z4-MM0m},
we obtain the relation:
\begin{align}
2({\hat M}^{(m)[1]})^2 + ({\hat M}^{(m)[2]})^2
=\left(1-a^{(m)2}\right)I_{{r}^{(m)}}.
\label{C-Z4-MM0m-M}
\end{align}
From eq.~\eqref{C-Z4-MM0m-M},
the submatrices of ${\hat M}^{(m)[q]}\submatrix{U}{k}{k-q}{(m)}$ ($q=1, 2$)
are described as $({\hat M}^{(m)[q]}\submatrix{U}{k}{k-q}{(m)})_{(k'l')}
= m^{(m)[q]}_{k'}\submatrix{u}{k}{k-q}{(m, k')} \delta_{k'l'}$
with non-negative numbers $m^{(m)[q]}_{k'}$ and unitary matrices $\submatrix{u}{k}{k-q}{(m, k')}$,
i.e., $\submatrix{U}{k}{k-q}{(m)}$ can have off-diagonal
elements in a block-diagonal matrix
whose corresponding block-diagonal one in ${\hat M}^{(m)[q]}$
is proportional to the unit matrix.
Then we find that the commutation relation $[{\hat M}^{(m)[q]}, \submatrix{U}{k}{k-q'}{(m)}]=0$
holds for any integers $q$ and $q'$, 
with ${\hat M}^{(m)[0]} = a^{(m)}I_{r^{(m)}}$,
$\submatrix{U}{k}{k}{(m)}=I_{r^{(m)}}$, ${\hat M}^{(m)[-q]}={\hat M}^{(m)[q]}$
and $\submatrix{U}{k-q}{k}{(m)}=(-1)^q \submatrix{U}{k}{k-q}{(m)\dagger}$.

Inserting eqs.~\eqref{C-Z4-Mmq=1}, \eqref{C-Z4-Mmq=-1} and \eqref{C-Z4-Mmq=2} 
into eq.~\eqref{C-Z4-qq'm}, we obtain the relation
$\submatrix{U}{k}{k-q'}{(m)}\submatrix{U}{k-q'}{k-q}{(m)}
= \submatrix{U}{k}{k-q+q'}{(m)}\submatrix{U}{k-q+q'}{k-q}{(m)}$
and write down the relations:
\begin{align}
\submatrix{U}{k-q}{k-q-q'}{(m)}\submatrix{U}{k-q-q'}{k}{(m)}
= \submatrix{U}{k-q}{k+q'}{(m)}\submatrix{U}{k+q'}{k}{(m)}
~~~{\rm and}~~~
\submatrix{U}{k}{k-q'}{(m)}\submatrix{U}{k-q'}{k+q}{(m)}
= \submatrix{U}{k}{k+q+q'}{(m)}\submatrix{U}{k+q+q'}{k+q}{(m)},
\label{C-Z4-U-rel}
\end{align}
where $k$ and $q$ are replaced with $k-q$ and $-q$ in the first relation,
and $q$ is replaced with $-q$ in the second one.
Exactly speaking, the relations~\eqref{C-Z4-U-rel} are obtained
except for the parts involving $m^{(m)[2]}_{k'}=0$. 
Because unitary submatrices for $m^{(m)[2]}_{k'}=0$, e.g., $\submatrix{u}{k}{k-2}{(m, k')}$,
are arbitrary and they can be taken to make the relations also hold for
the parts relating to $m^{(m)[2]}_{k'}=0$, we choose them in this way
and write down following relations in the same fashion that
unitary submatrices for $m^{(m)[2]}_{k'}=0$ can be contained as in eq.~~\eqref{C-Z4-U-rel}. 
Multiplying $\submatrix{U}{k}{k-q}{(m)}$ and $\submatrix{U}{k+q}{k}{(m)}$
by the above relations, respectively, we obtain the relations:
\begin{align} 
& \submatrix{U}{k}{k-q}{(m)}\submatrix{U}{k-q}{k-q-q'}{(m)}\submatrix{U}{k-q-q'}{k}{(m)}
= \submatrix{U}{k}{k-q}{(m)}\submatrix{U}{k-q}{k+q'}{(m)}\submatrix{U}{k+q'}{k}{(m)},
\label{C-Z4-U-rel1-1} \\
& \submatrix{U}{k}{k-q'}{(m)}\submatrix{U}{k-q'}{k+q}{(m)}\submatrix{U}{k+q}{k}{(m)}
= \submatrix{U}{k}{k+q+q'}{(m)}\submatrix{U}{k+q+q'}{k+q}{(m)}\submatrix{U}{k+q}{k}{(m)}.
\label{C-Z4-U-rel1-2}
\end{align}
Using eqs.~\eqref{C-Z4-U-rel1-1} and \eqref{C-Z4-U-rel1-2}, we derive the relation:
\begin{align}
J_{kk}^{(m)} \equiv 
\submatrix{U}{k}{k-1}{(m)}\submatrix{U}{k-1}{k-2}{(m)}\submatrix{U}{k-2}{k}{(m)}
= \submatrix{U}{k}{k-1}{(m)}\submatrix{U}{k-1}{k+1}{(m)}\submatrix{U}{k+1}{k}{(m)}
= \submatrix{U}{k}{k+2}{(m)}\submatrix{U}{k+2}{k+1}{(m)}\submatrix{U}{k+1}{k}{(m)}.
\label{C-Z4-U-rel2}
\end{align}
Inserting eqs.~\eqref{C-Z4-Mmq=1}, \eqref{C-Z4-Mmq=-1} and \eqref{C-Z4-Mmq=2}
into eq.~\eqref{C-Z4-MM2m}, we obtain the relations:
\begin{align}
& 2 a^{(m)}{\hat M}^{(m)[2]} = ({\hat M}^{(m)[1]})^2 (J_{kk}^{(m)} + J_{kk}^{(m)\dagger}),
\label{C-Z4-qq'mD-hat}\\
& 2 a^{(m)} \sqrt{1-a^{(m)2}}~\! \tilde{U}_{k~\!\!k-2}^{(m)} = 0,
\label{C-Z4-qq'mD-hat-tilde}
\end{align}
and find $n_k^{(m)'}=0$ for $-1 < a^{(m)} < 1$ 
from eq.~\eqref{C-Z4-qq'mD-hat-tilde}
because $\tilde{U}_{k~\!\!k-2}^{(m)}$ are unitary matrices.

In this way, the form of $\submatrix{M}{k}{l}{(m)[k-l]}$
is restricted as eqs.~\eqref{C-Z4-Mmq=1}, \eqref{C-Z4-Mmq=-1} and \eqref{C-Z4-Mmq=2}
with ${\hat M}^{(m)[1]}$ and ${\hat M}^{(m)[2]}$ satisfying eqs.~\eqref{C-Z4-MM0m-M} and
\eqref{C-Z4-qq'mD-hat}, $\submatrix{U}{k}{l}{(m)}$ satisfying eqs.~\eqref{C-Z4-U-rel} and
\eqref{C-Z4-qq'mD-hat}, and $\tilde{U}_{k~\!\!k-2}^{(m)}$ satisfying 
eq.~\eqref{C-Z4-qq'mD-hat-tilde}.

\subsection{Diagonalization of submatrices in $T_1^{(m)}$ and rearrangement}
\label{C3}

Based on eqs.~\eqref{C-Z4-Mmq=1}, \eqref{C-Z4-Mmq=-1} and \eqref{C-Z4-Mmq=2},
we can rearrange $R_0^{(m)}$ and $T_1^{(m)}$ to be the form of 
block-diagonal ones such as
$R_0^{(m)}=R_0^{(m)'}\oplus R_0^{(m)''}$ and $T_1^{(m)}=T_1^{(m)'}\oplus T_1^{(m)''}$,
where $R_0^{(m)'}$, $T_1^{(m)'}$, $R_0^{(m)''}$ and $T_1^{(m)''}$ are given by
\begin{align}
& R_0^{(m)'}=
    \begin{pmatrix}
        iI_{r^{(m)}} &          &        &         \\
                 & -I_{r^{(m)}} &        &         \\
                 &          & -iI_{r^{(m)}} &         \\
                 &          &        & I_{r^{(m)}}
    \end{pmatrix},
\label{C-Z4-R0m'}\\
& T_1^{(m)'}=
    \begin{pmatrix}
        a^{(m)}I_{r^{(m)}} & {\hat M}^{(m)[1]}\submatrix{U}{1}{2}{(m)} 
& {\hat M}^{(m)[2]}\submatrix{U}{1}{3}{(m)} 
& {\hat M}^{(m)[1]}\submatrix{U}{1}{4}{(m)} \\
{\hat M}^{(m)[1]}\submatrix{U}{2}{1}{(m)} 
& a^{(m)}I_{r^{(m)}} 
& {\hat M}^{(m)[1]}\submatrix{U}{2}{3}{(m)} 
& {\hat M}^{(m)[2]}\submatrix{U}{2}{4}{(m)} \\
{\hat M}^{(m)[2]}\submatrix{U}{3}{1}{(m)} 
& {\hat M}^{(m)[1]}\submatrix{U}{3}{2}{(m)} 
& a^{(m)}I_{r^{(m)}} 
& {\hat M}^{(m)[1]}\submatrix{U}{3}{4}{(m)} \\
 {\hat M}^{(m)[1]}\submatrix{U}{4}{1}{(m)} 
& {\hat M}^{(m)[2]}\submatrix{U}{4}{2}{(m)} 
& {\hat M}^{(m)[1]}\submatrix{U}{4}{3}{(m)} & a^{(m)}I_{r^{(m)}}
    \end{pmatrix},
\label{C-Z4-T1m'}\\
& R_0^{(m)''}=
    \begin{pmatrix}
        iI_{n_1^{(m)'}} &          &        &         \\
                 & -I_{n_2^{(m)'}} &        &         \\
                 &          & -iI_{n_1^{(m)'}} &         \\
                 &          &        & I_{n_2^{(m)'}}
    \end{pmatrix},
\label{C-Z4-R0m''}
\end{align}
and 
\begin{align}
T_1^{(m)''}=
    \begin{pmatrix}
        0 & 0 & {\tilde U}_{1~\!\!3}^{(m)} & 0 \\
        0 & 0 & 0 & {\tilde U}_{2~\!\!4}^{(m)} \\
        {\tilde U}_{3~\!\!1}^{(m)} & 0 & 0 & 0\\
        0 & {\tilde U}_{4~\!\!2}^{(m)} & 0 & 0
    \end{pmatrix},
\label{C-Z4-T1m''0}
\end{align}
respectively.

For the convenience of a later calculation, 
we write down the following relation
obtained by setting $k=2$ in eq.~\eqref{C-Z4-U-rel2},
\begin{align}
J_{22}^{(m)}
=\submatrix{U}{2}{1}{(m)}\submatrix{U}{1}{4}{(m)}\submatrix{U}{4}{2}{(m)}
=\submatrix{U}{2}{1}{(m)}\submatrix{U}{1}{3}{(m)}\submatrix{U}{3}{2}{(m)}
=\submatrix{U}{2}{4}{(m)}\submatrix{U}{4}{3}{(m)}\submatrix{U}{3}{2}{(m)}.
\label{C-Z4-U-rel-ex}
\end{align}

Let us perform a unitary transformation $V^{(m)}T_1^{(m)}V^{(m)\dagger}$
with $V^{(m)}=V^{(m)'} \oplus V^{(m)''}$,
where $V^{(m)'}$ and $V^{(m)''}$ are given by
\begin{align}
&V^{(m)'}=
    \begin{pmatrix}
         {\hat{\varTheta}}^{(m)[1]\dagger}U^{(m)}\submatrix{U}{2}{1}{(m)} &  &  &  \\
         & U^{(m)} &  & \\
         &  & -{\hat{\varTheta}}^{(m)[1]}U^{(m)}\submatrix{U}{2}{3}{(m)} &  \\
         &  &  & U^{(m)}\submatrix{U}{2}{4}{(m)}
    \end{pmatrix},
\label{C-Z4-Vm'}
\end{align}
and
\begin{align}
&V^{(m)''}=
    \begin{pmatrix}
        {\tilde U}_{3~\!\!1}^{(m)} &  &  &  \\
         & {\tilde U}_{4~\!\!2}^{(m)} &  & \\
         &  & I_{n_1^{(m)'}} &  \\
         &  &  & I_{n_2^{(m)'}}
    \end{pmatrix},
\label{C-Z4-Vm''}
\end{align}
respectively.
Here $U^{(m)}$ are unitary matrices that make $J_{22}^{(m)}$ diagonal ones
$\hat{J}_{22}^{(m)}(=U^{(m)} J_{22}^{(m)} U^{(m)\dagger})$ and
$\hat{\varTheta}^{(m)[1]}$ are diagonal matrices
whose squares agree with $\hat{J}_{22}^{(m)}$.
It is understood that $U^{(m)}$ as well as $\submatrix{U}{k}{k-q}{(m)}$
commute with ${\hat M}^{(m)[1]}$ and ${\hat M}^{(m)[2]}$
from the fact that $J_{22}^{(m)}$ can have off-diagonal elements only in 
a block-diagonal matrix whose corresponding block-diagonal one in 
${\hat M}^{(m)[1]}$ and ${\hat M}^{(m)[2]}$
is proportional to the unit matrix.
Then, we find that $R_0^{(m)'}$ and $R_0^{(m)''}$ remain unchanged
and, using $\submatrix{U}{k-q}{k}{(m)}=(-1)^q\submatrix{U}{k}{k-q}{(m)\dagger}$,
${\tilde U}_{k-2~\!\!k}^{(m)}={\tilde U}_{k~\!\!k-2}^{(m)\dagger}$ and
eq.~\eqref{C-Z4-U-rel-ex}, 
$T_1^{(m)'}$ and $T_1^{(m)''}$ are transformed as
\begin{align}
&V^{(m)'}T_1^{(m)'}V^{(m)'\dagger}\nonumber\\
&= \begin{pmatrix}
        a^{(m)}I_{r^{(m)}} & -{\hat M}^{(m)[1]}{\hat{\varTheta}}^{(m)[1]\dagger}
&  {\hat M}^{(m)[2]} & {\hat M}^{(m)[1]}{\hat{\varTheta}}^{(m)[1]} \\
 {\hat M}^{(m)[1]}{\hat{\varTheta}}^{(m)[1]} & a^{(m)}I_{r^{(m)}} 
& -{\hat M}^{(m)[1]}{\hat{\varTheta}}^{(m)[1]\dagger} 
& {\hat M}^{(m)[2]} \\
 {\hat M}^{(m)[2]} & {\hat M}^{(m)[1]}{\hat{\varTheta}}^{(m)[1]} 
& a^{(m)}I_{r^{(m)}} & -{\hat M}^{(m)[1]}{\hat{\varTheta}}^{(m)[1]\dagger} \\
 -{\hat M}^{(m)[1]}{\hat{\varTheta}}^{(m)[1]\dagger}
& {\hat M}^{(m)[2]} & {\hat M}^{(m)[1]}{\hat{\varTheta}}^{(m)[1]} & a^{(m)}I_{r^{(m)}}
    \end{pmatrix},
\label{C-Z4-Tm'-tr}
\end{align}
and 
\begin{align}
V^{(m)''}T_1^{(m)''}V^{(m)''\dagger}
=   \begin{pmatrix}
        0 & 0 &  I_{n_1^{(m)'}} & 0 \\
        0 & 0 & 0 & I_{n_2^{(m)'}} \\
        I_{n_1^{(m)'}} & 0 & 0 & 0\\
        0 & I_{n_2^{(m)'}} & 0 & 0
    \end{pmatrix},
\label{C-Z4-Tm''-tr}
\end{align}
respectively.
Using $\hat{\varTheta}^{(m)[1]}$ and $\hat{\varTheta}^{(m)[1]\dagger}$,
eq.~\eqref{C-Z4-qq'mD-hat} is rewritten as
\begin{align}
2 a^{(m)}{\hat M}^{(m)[2]} = ({\hat M}^{(m)[1]}{\hat{\varTheta}}^{(m)[1]})^2 
+({\hat M}^{(m)[1]}{\hat{\varTheta}}^{(m)[1]\dagger})^2.
\label{C-Z4-qq'mD-hat-M}
\end{align}

\section{Derivation of block-diagonal forms on $T^2/\Z6$}
\label{app:t2z6}

In this appendix, we explain the details of block-diagonalization
of twist matrices on $T^2/\Z6$ given in section~\ref{sec:t2z6}.

We can start with the following $R_0$ and $T_1$, without loss of generality,
\begin{align}
R_0=
  \begin{pmatrix}
      \eta I_{n_1}& & & & & \\
       & \eta^2 I_{n_2}& & & & \\
       & &-I_{n_3}& & & \\
       & & &-\eta I_{n_4} & & \\
      & & & & -\eta^2 I_{n_5} & \\
      & & & & & I_{n_6}
  \end{pmatrix},
\end{align}
and
\begin{align}
T_1=
  \begin{pmatrix}
      (T_1)_{(11)}&(T_1)_{(12)}&(T_1)_{(13)}&(T_1)_{(14)}
&(T_1)_{(15)}&(T_1)_{(16)}\\
      (T_1)_{(21)}&(T_1)_{(22)}&(T_1)_{(23)}&(T_1)_{(24)}
&(T_1)_{(25)}&(T_1)_{(26)}\\
      (T_1)_{(31)}&(T_1)_{(32)}&(T_1)_{(33)}&(T_1)_{(34)}
&(T_1)_{(35)}&(T_1)_{(36)}\\
      (T_1)_{(41)}&(T_1)_{(42)}&(T_1)_{(43)}&(T_1)_{(44)}
&(T_1)_{(45)}&(T_1)_{(46)}\\
      (T_1)_{(51)}&(T_1)_{(52)}&(T_1)_{(53)}&(T_1)_{(54)}
&(T_1)_{(55)}&(T_1)_{(56)}\\
      (T_1)_{(61)}&(T_1)_{(62)}&(T_1)_{(63)}&(T_1)_{(64)}
&(T_1)_{(65)}&(T_1)_{(66)}
  \end{pmatrix}, 
\label{D-Z6-R0T1}
\end{align}
where $\eta = e^{2\pi i/6}$, $I_{n_k}$ are $n_k \times n_k$ unit matrices and
$(T_1)_{(kl)}$ are $n_k \times n_l$ submatrices.
We denote the submatrices as $(T_1)_{(kl)} = \submatrix{M}{k}{l}{[k-l]}$
and use a notation of 
$\submatrix{M}{k}{l}{[k-l]}=\submatrix{M}{k}{l}{[k'-l']}=\submatrix{M}{k'}{l'}{[k-l]}$
with $k'=k$ (mod 6) and $l'=l$ (mod 6).
The upper index $k-l=q$ represents the charge of the $\Z6$ symmetry 
generated by $R_0$: 
$(R_0T_1R_0^{-1})_{(k\,k-q)}=\eta^{q}\submatrix{M}{k}{k-q}{[q]}$.
Parameters such as $n_k$ which represent the size of matrices
are non-negative integers.
The same applies to the following ones.

\subsection{Block-diagonalization of $T_1$ and relations on submatrices}
\label{D1}

First, let us restrict the form of $T_1$ and derive relations on submatrices of $T_1$,
using constraints $T_1^{\dagger} = T_1^{-1}$, $T_{m'} T_m = T_m T_{m'}$, $T_1 T_4 = I$
and $T_1 T_3 T_5= I$ (or $T_1 T_3 = T_2$)
where $T_{m} = R_0^{m-1} T_1 R_0^{1-m}$
as explained in section~\ref{sec:2Dorb}. 

From $T_1^{\dagger} (= T_1^{-1} = T_4) = R_0^3 T_1 R_0^{-3}$,
we derive the relation:
\begin{align}
\submatrix{M}{k-q}{k}{[-q]\dagger}=(-1)^{q}\submatrix{M}{k}{k-q}{[q]}.
\label{D-Z6-Mdagger}
\end{align}
Taking $q=0$ in eq.~\eqref{D-Z6-Mdagger}, 
we find that $\submatrix{M}{k}{k}{[0]\dagger}=\submatrix{M}{k}{k}{[0]}$
and $\submatrix{M}{k}{k}{[0]}$ is diagonalized 
as $(\submatrix{M}{k}{k}{[0]})_{ij}=a_k^i\delta_{ij}$ $(a_k^i\in {\mathbb R})$
by a unitary transformation,
without modifying $R_0$,
where $(\submatrix{M}{k}{k}{[0]})_{ij}$
are ($i,j$) elements of $\submatrix{M}{k}{k}{[0]}$.

As shown in appendix~\ref{app:MM}, 
from $T_{m'} T_{m} = T_{m} T_{m'}$, i.e., 
$T_1 R_0^{m-m'} T_1 = R_0^{m-m'} T_1 R_0^{m'-m} T_1 R_0^{m-m'}$, 
we derive the relation:
\begin{align}
\submatrix{M}{k}{k-q'}{[q']}\submatrix{M}{k-q'}{k-q}{[q-q']}
=\submatrix{M}{k}{k-q+q'}{[q-q']}\submatrix{M}{k-q+q'}{k-q}{[q']}.
\label{D-Z6-qq'}
\end{align}
Setting $q'=0$ in eq.~\eqref{D-Z6-qq'} 
and using $(\submatrix{M}{k}{k}{[0]})_{ij}=a_k^i\delta_{ij}$,
we obtain the relation:
\begin{align}
(a_k^i-a_{k-q}^j)(\submatrix{M}{k}{k-q}{[q]})_{ij}=0,
\label{D-Z6-(ai-aj)}
\end{align}
and rearrange $T_1$ to be a block-diagonal matrix, 
while keeping $R_0$ a diagonal one:
\begin{align}
R_0=
    \begin{pmatrix}
        R_0^{(1)} &          &        &         \\
                 & R_0^{(2)} &        &         \\
                 &          & \ddots &         \\
                 &          &        & R_0^{(M)}
    \end{pmatrix},\qquad
T_1=
    \begin{pmatrix}
        T_1^{(1)} &          &        &         \\
                 & T_1^{(2)} &        &         \\
                 &          & \ddots &         \\
                 &          &        & T_1^{(M)}
    \end{pmatrix},
\label{D-Z6-R0T1-block}
\end{align}
where $R_0^{(m)}$ and $T_1^{(m)}$ $(m = 1, 2, \cdots, M)$ are 
$n^{(m)} \times n^{(m)}$ matrices given by
\begin{align}
R_0^{(m)}=
    \begin{pmatrix}
    \eta I_{n_1^{(m)}}& & & & & \\
       &\eta^2 I_{n_2^{(m)}}& & & & \\
       & &-I_{n_3^{(m)}}& & & \\
       & & &-\eta I_{n_4^{(m)}} & & \\
      & & & & -\eta^2 I_{n_5^{(m)}} & \\
      & & & & & I_{n_6^{(m)}}
    \end{pmatrix},
\label{D-Z6-R0m}
\end{align}
and
\begin{align}
&T_1^{(m)}=
    \begin{pmatrix}
        a^{(m)}I_{n_1^{(m)}} & \submatrix{M}{1}{2}{(m)[-1]} 
& \submatrix{M}{1}{3}{(m)[-2]} & \submatrix{M}{1}{4}{(m)[-3]}
& \submatrix{M}{1}{5}{(m)[2]} & \submatrix{M}{1}{6}{(m)[1]} \\
       \submatrix{M}{2}{1}{(m)[1]} & a^{(m)}I_{n_2^{(m)}} 
& \submatrix{M}{2}{3}{(m)[-1]} & \submatrix{M}{2}{4}{(m)[-2]}  
& \submatrix{M}{2}{5}{(m)[-3]} & \submatrix{M}{2}{6}{(m)[2]}     \\
       \submatrix{M}{3}{1}{(m)[2]} & \submatrix{M}{3}{2}{(m)[1]} 
& a^{(m)}I_{n_3^{(m)}} & \submatrix{M}{3}{4}{(m)[-1]}
& \submatrix{M}{3}{5}{(m)[-2]} & \submatrix{M}{3}{6}{(m)[-3]} \\
       \submatrix{M}{4}{1}{(m)[3]} & \submatrix{M}{4}{2}{(m)[2]}
& \submatrix{M}{4}{3}{(m)[1]} & a^{(m)}I_{n_4^{(m)}}
& \submatrix{M}{4}{5}{(m)[-1]} & \submatrix{M}{4}{6}{(m)[-2]} \\
       \submatrix{M}{5}{1}{(m)[-2]} & \submatrix{M}{5}{2}{(m)[3]} 
& \submatrix{M}{5}{3}{(m)[2]} & \submatrix{M}{5}{4}{(m)[1]} &
a^{(m)}I_{n_5^{(m)}} & \submatrix{M}{5}{6}{(m)[-1]} \\
       \submatrix{M}{6}{1}{(m)[-1]} & \submatrix{M}{6}{2}{(m)[-2]} 
& \submatrix{M}{6}{3}{(m)[3]} & \submatrix{M}{6}{4}{(m)[2]} 
& \submatrix{M}{6}{5}{(m)[1]} & a^{(m)}I_{n_6^{(m)}} 
    \end{pmatrix},
\label{D-Z6-T1m}
\end{align}
respectively.
Here, $n^{(m)} = \sum_{k=1}^6 n_k^{(m)}$ and
submatrices in $T_1^{(m)}$ are denoted
as $(T_1^{(m)})_{(kl)} = \submatrix{M}{k}{l}{(m)[k-l]}$
for $k \ne l$ and $(T_1^{(m)})_{(kk)} = \submatrix{M}{k}{k}{(m)[0]}=a^{(m)} I_{n_k^{(m)}}$.
Real parameters $a^{(m)}$ satisfy $a^{(m)} \ne a^{(m')}$ for $m \ne m'$.
Submatrices obey the following relations 
that come from $T_1^{\dagger} = T_4$ and $T_{m'} T_{m} = T_{m} T_{m'}$,
\begin{align}
& \submatrix{M}{k-q}{k}{(m)[-q]\dagger}=(-1)^{q}\submatrix{M}{k}{k-q}{(m)[q]},
\label{D-Z6-Mdaggerm}\\
& \submatrix{M}{k}{k-q'}{(m)[q']}\submatrix{M}{k-q'}{k-q}{(m)[q-q']}
=\submatrix{M}{k}{k-q+q'}{(m)[q-q']}\submatrix{M}{k-q+q'}{k-q}{(m)[q']}.
\label{D-Z6-qq'm}
\end{align}
 
From $T_1 T_4 = T_1 R_0^3 T_1 R_0^{-3} = I$ 
and $T_2 = T_1 T_3$, i.e., $T_1 = R_0^{-1} T_1 R_0^2 T_1 R_0^{-1}$,
we derive the relations:
\begin{align}
\sum_{q'} (-1)^{q-q'}
\submatrix{M}{k}{k-q'}{(m)[q']}\submatrix{M}{k-q'}{k-q}{(m)[q-q']} 
= \delta_{k~\!\!k-q}I_{n_k^{(m)}},
\label{D-Z6-1MM}
\end{align}
and
\begin{align}
\submatrix{M}{k}{k-q}{(m)[q]}
=\sum_{q'} \eta^{q-2q'}
\submatrix{M}{k}{k-q'}{(m)[q']}\submatrix{M}{k-q'}{k-q}{(m)[q-q']},
\label{D-Z6-M=MM}
\end{align}
respectively.
Here the summation over $q'$ can be taken for any successive six integers.
Setting $q=0, 2$ in eq.~\eqref{D-Z6-1MM} and $q=0, 1, 2, 3$ in eq.~\eqref{D-Z6-M=MM}
and using eqs.~\eqref{D-Z6-Mdaggerm} and \eqref{D-Z6-qq'm},
we obtain the relations:
\begin{align}
& 2\submatrix{M}{k}{k-1}{(m)[1]}\submatrix{M}{k}{k-1}{(m)[1]\dagger}
+2\submatrix{M}{k}{k-2}{(m)[2]}\submatrix{M}{k}{k-2}{(m)[2]\dagger}
+ \submatrix{M}{k}{k-3}{(m)[3]}\submatrix{M}{k}{k-3}{(m)[3]\dagger}
=\left(1-a^{(m)2}\right)I_{n_k^{(m)}},
& \quad \mbox{for } q=0,
\label{D-Z6-MM0m}\\
& 2a^{(m)}\submatrix{M}{k}{k-2}{(m)[2]}
 + \submatrix{M}{k}{k+2}{(m)[-2]}\submatrix{M}{k+2}{k-2}{(m)[-2]}
=\submatrix{M}{k}{k-1}{(m)[1]}\submatrix{M}{k-1}{k-2}{(m)[1]}
+2\submatrix{M}{k}{k+1}{(m)[-1]}\submatrix{M}{k+1}{k-2}{(m)[3]},
& \quad \mbox{for } q=2,
\label{D-Z6-MM2m}
\end{align}
and
\begin{align}
& \submatrix{M}{k}{k-1}{(m)[1]}\submatrix{M}{k}{k-1}{(m)[1]\dagger}
 - \submatrix{M}{k}{k-2}{(m)[2]}\submatrix{M}{k}{k-2}{(m)[2]\dagger}
 - \submatrix{M}{k}{k-3}{(m)[3]}\submatrix{M}{k}{k-3}{(m)[3]\dagger}
 = \left(a^{(m)} - a^{(m)2}\right) I_{n_k^{(m)}},
& \quad \mbox{for } q=0,
\label{D-Z6-M=MM0m}\\
& \left(1-a^{(m)}\right)\submatrix{M}{k}{k-1}{(m)[1]}
= \submatrix{M}{k}{k-3}{(m)[3]}\submatrix{M}{k-3}{k-1}{(m)[-2]}
 - 2 \submatrix{M}{k}{k-2}{(m)[2]}\submatrix{M}{k-2}{k-1}{(m)[-1]},
& \quad \mbox{for } q=1,
\label{D-Z6-M=MM1m}\\
& \left(1+a^{(m)}\right)\submatrix{M}{k}{k-2}{(m)[2]}
= \submatrix{M}{k}{k-1}{(m)[1]}\submatrix{M}{k-1}{k-2}{(m)[1]}
 - \submatrix{M}{k}{k+1}{(m)[-1]}\submatrix{M}{k+1}{k-2}{(m)[3]} &
\nonumber \\
& ~~~~~~~~~~~~~~~~~~~~~~~~~~~~~~
 + \submatrix{M}{k}{k+2}{(m)[-2]}\submatrix{M}{k+2}{k-2}{(m)[-2]},
& \quad \mbox{for } q=2,
\label{D-Z6-M=MM2m}\\
& \left(1+2a^{(m)}\right)\submatrix{M}{k}{k-3}{(m)[3]}
= \submatrix{M}{k}{k-1}{(m)[1]}\submatrix{M}{k-1}{k-3}{(m)[2]}
 + \submatrix{M}{k}{k+2}{(m)[-2]}\submatrix{M}{k+2}{k-3}{(m)[-1]},
& \quad \mbox{for } q=3.
\label{D-Z6-M=MM3m}
\end{align}
The relations obtained by setting $q=1$, $q=3$ and $q=5$ in eq.~\eqref{D-Z6-1MM} 
are not independent of eq.~\eqref{D-Z6-qq'm}, because they are also derived from
linear combinations of eq.~\eqref{D-Z6-qq'm}.
The relation obtained by setting $q=4$ in eq.~\eqref{D-Z6-1MM} 
is the Hermitian conjugate of eq.~\eqref{D-Z6-MM2m}.
The relations obtained by setting $q=4$ and $q=5$ in eq.~\eqref{D-Z6-M=MM} 
are the Hermitian conjugates of eqs.~\eqref{D-Z6-M=MM2m} and \eqref{D-Z6-M=MM1m}, respectively.
From eq.~\eqref{D-Z6-MM0m}, we find that $0 \leq a^{(m)2} \leq 1$.
Note that $T_1^{(m)}$ is already diagonal for $a^{(m)}=\pm 1$,
and then $R_0$ and $T_1$ contain
$n^{(m)} \times n^{(m)}$ diagonal matrices whose submatrices
are given by $(R_0^{(m)})_{(kl)} = \eta^k \delta_{kl} I_{n_k^{(m)}}$
and $(T_1^{(m)})_{(kl)} = \pm \delta_{kl} I_{n_k^{(m)}}$
for $a^{(m)}=\pm 1$
(double sign in the same order)
as members of block-diagonal matrices.
Hereafter, we focus on $-1 < a^{(m)} < 1$.

\subsection{Restriction of $\submatrix{M}{k}{l}{(m)[k-l]}$}
\label{D2}

Next, let us restrict the form of $\submatrix{M}{k}{l}{(m)[k-l]}$,
using relations obtained in the previous subsection.

In the same way as on $T^2/\Z4$, 
we obtain the counterpart of eq.~\eqref{C-Z4-qq'm0Dk}, i.e., 
$\left(\submatrix{M}{k}{k-q'}{(m)[q']}\submatrix{M}{k}{k-q'}{(m)[q']\dagger}\right)'_{\rm D}
= \left(\submatrix{M}{k+q'}{k}{(m)[q']}\submatrix{M}{k+q'}{k}{(m)[q']\dagger}\right)'_{\rm D}$
($q' = 1, 2, 3$), and using them 
with $q'=1$ iteratively, we derive the relation:
\begin{align}
& \left(\submatrix{M}{k}{k-1}{(m)[1]}\submatrix{M}{k}{k-1}{(m)[1]\dagger}\right)'_{\rm D}
= \left(\submatrix{M}{k+1}{k}{(m)[1]}\submatrix{M}{k+1}{k}{(m)[1]\dagger}\right)'_{\rm D}
= \left(\submatrix{M}{k+2}{k+1}{(m)[1]}
\submatrix{M}{k+2}{k+1}{(m)[1]\dagger}\right)'_{\rm D}
\nonumber \\
& = \left(\submatrix{M}{k+3}{k+2}{(m)[1]}
\submatrix{M}{k+3}{k+2}{(m)[1]\dagger}\right)'_{\rm D}
= \left(\submatrix{M}{k+4}{k+3}{(m)[1]}
\submatrix{M}{k+4}{k+3}{(m)[1]\dagger}\right)'_{\rm D}
= \left(\submatrix{M}{k+5}{k+4}{(m)[1]}
\submatrix{M}{k+5}{k+4}{(m)[1]\dagger}\right)'_{\rm D}.
\label{D-Z6-qq'm0Dq=1}
\end{align}
From eq.~\eqref{D-Z6-qq'm0Dq=1},
we find that 
$\left(\submatrix{M}{k}{k-1}{(m)[1]}\submatrix{M}{k}{k-1}{(m)[1]\dagger}\right)'_{\rm D}$
are independent of $k$ and $\submatrix{M}{k}{k-1}{(m)[1]}$ have a common rank 
$r^{(m)} (= {\rm rank}(\submatrix{M}{k}{k-1}{(m)[1]}))$.
Then, based on the argument in appendix~\ref{ap:mmdag},
we choose the basis such that $\submatrix{M}{k}{k-1}{(m)[1]}$ have the form:
\begin{align}
\submatrix{M}{k}{k-1}{(m)[1]}&=\left(
\begin{array}{ccc|cc|cc}
 &           &                           &   &  &  \\
 & {\hat M}^{(m)[1]}\submatrix{U}{k}{k-1}{(m)} &  & \text{\large{0}} & & \text{\large{0}}\\
 &           &  &   & & \\
\hline
 & & & & & \\
 & \text{\large{0}} &  & 
 \text{\large{0}}
& & \text{\large{0}} \\
 &           &  &   & & \\
\hline
 & & & & & \\
 & \text{\large{0}} &  & \text{\large{0}}  &  & 
 \text{\large{0}}
\end{array}
\right),
\label{D-Z6-Mmq=1}
\end{align}
where ${\hat M}^{(m)[1]}$ are $r^{(m)} \times r^{(m)}$ diagonal matrices
with positive elements, i.e., $({\hat M}^{(m)[1]})_{ii} > 0$,
and $\submatrix{U}{k}{k-1}{(m)}$ are $r^{(m)} \times r^{(m)}$ unitary matrices.
From eqs.~\eqref{D-Z6-Mdaggerm} and \eqref{D-Z6-Mmq=1}, 
$\submatrix{M}{k-1}{k}{(m)[-1]}$ are given by
\begin{align}
\submatrix{M}{k-1}{k}{(m)[-1]}&=-\submatrix{M}{k}{k-1}{(m)[1]\dagger}=\left(
\begin{array}{ccc|cc|cc}
 &           &                           &   &  &  \\
 & {\hat M}^{(m)[1]}\submatrix{U}{k-1}{k}{(m)} &  & \text{\large{0}} & & \text{\large{0}}\\
 &           &  &   & & \\
\hline
 & & & & & \\
 & \text{\large{0}} &  & 
 \text{\large{0}}
& & \text{\large{0}} \\
 &           &  &   & & \\
\hline
 & & & & & \\
 & \text{\large{0}} &  & \text{\large{0}}  &  & 
 \text{\large{0}}
\end{array}
\right),
\label{D-Z6-Mmq=-1}
\end{align}
where $-\submatrix{U}{k}{k-1}{(m)\dagger}$ are denoted as $\submatrix{U}{k-1}{k}{(m)}$.

From the counterparts of
eqs.~\eqref{C-Z4-qq'm0-rank} and \eqref{C-Z4-qq'm0Dk} with $q' = 2$ and $q' = 3$, 
we derive the relations:
\begin{align}
& {\rm rank}(\submatrix{M}{k}{k-2}{(m)[2]})
= {\rm rank}(\submatrix{M}{k+2}{k}{(m)[2]}),
\label{D-Z6-qq'm0-rankq=2}\\
& \left(\submatrix{M}{k}{k-2}{(m)[2]}\submatrix{M}{k}{k-2}{(m)[2]\dagger}\right)'_{\rm D}
= \left(\submatrix{M}{k+2}{k}{(m)[2]}\submatrix{M}{k+2}{k}{(m)[2]\dagger}\right)'_{\rm D},
\label{D-Z6-qq'm0Dq=2}\\
& {\rm rank}(\submatrix{M}{k}{k-3}{(m)[3]}) 
= {\rm rank}(\submatrix{M}{k+3}{k}{(m)[3]}),
\label{D-Z6-qq'm0-rankq=3}\\
& \left(\submatrix{M}{k}{k-3}{(m)[3]}\submatrix{M}{k}{k-3}{(m)[3]\dagger}\right)'_{\rm D}
= \left(\submatrix{M}{k+3}{k}{(m)[3]}\submatrix{M}{k+3}{k}{(m)[3]\dagger}\right)'_{\rm D}.
\label{D-Z6-qq'm0Dq=3}
\end{align}
Using eqs.~\eqref{D-Z6-MM0m} and \eqref{D-Z6-M=MM0m} and the fact that 
$\left(\submatrix{M}{k}{k-1}{(m)[1]}\submatrix{M}{k}{k-1}{(m)[1]\dagger}\right)'_{\rm D}$
are independent of $k$ and 
$\submatrix{M}{k}{k-1}{(m)[1]}\submatrix{M}{k}{k-1}{(m)[1]\dagger}$ are diagonal 
in the present basis, i.e., 
$\submatrix{M}{k}{k-1}{(m)[1]}\submatrix{M}{k}{k-1}{(m)[1]\dagger} 
= ({\hat M}^{(m)[1]})^2 \oplus {\bm 0}$,
we find that the $r^{(m)} \times r^{(m)}$ submatrices in
$\left(\submatrix{M}{k}{k-2}{(m)[2]}\submatrix{M}{k}{k-2}{(m)[2]\dagger}\right)'_{\rm D}$
and
$\left(\submatrix{M}{k}{k-3}{(m)[3]}\submatrix{M}{k}{k-3}{(m)[3]\dagger}\right)'_{\rm D}$
corresponding to $({\hat M}^{(m)[1]})^2$
are also independent of $k$ and 
$\submatrix{M}{k}{k-2}{(m)[2]}\submatrix{M}{k}{k-2}{(m)[2]\dagger}$
and $\submatrix{M}{k}{k-3}{(m)[3]}\submatrix{M}{k}{k-3}{(m)[3]\dagger}$
are also diagonal.
Then, the forms of $\submatrix{M}{k}{k-2}{(m)[2]}$ and $\submatrix{M}{k}{k-3}{(m)[3]}$
can be written as $\submatrix{M}{k}{k-2}{(m)[2]} 
= {\hat M}^{(m)[2]}\submatrix{U}{k}{k-2}{(m)} \oplus \tilde{M}_{k~\!\!k-2}^{(m)}$
and $\submatrix{M}{k}{k-3}{(m)[3]} 
= {\hat M}^{(m)[3]}\submatrix{U}{k}{k-3}{(m)} \oplus \tilde{M}_{k~\!\!k-3}^{(m)}$
where both ${\hat M}^{(m)[2]}$ and ${\hat M}^{(m)[3]}$
are $r^{(m)} \times r^{(m)}$ diagonal matrices
with non-negative elements $({\hat M}^{(m)[2]})_{ii} \ge 0$
and $({\hat M}^{(m)[3]})_{ii} \ge 0$,
and $\submatrix{U}{k}{k-2}{(m)}$ and $\submatrix{U}{k}{k-3}{(m)}$ 
are $r^{(m)} \times r^{(m)}$ unitary matrices.
Using  eqs.~\eqref{D-Z6-MM0m} -- \eqref{D-Z6-M=MM3m}, 
$\submatrix{M}{k}{k-2}{(m)[2]}$ and $\submatrix{M}{k}{k-3}{(m)[3]}$ can be written as
\begin{align}
\submatrix{M}{k}{k-2}{(m)[2]}&=\left(
\begin{array}{ccc|cc|cc}
 &           &                           &   &  &  \\
 & {\hat M}^{(m)[2]}\submatrix{U}{k}{k-2}{(m)} &  & \text{\large{0}} & & \text{\large{0}}\\
 &           &  &   & & \\
\hline
 & & & & & \\
 & \text{\large{0}} &  & 
\frac{2}{3}~\! \tilde{U}_{k~\!\!k-2}^{(m)} 
& & \text{\large{0}} \\
 &           &  &   & & \\
\hline
 & & & & & \\
 & \text{\large{0}} &  & \text{\large{0}}  &  & \text{\large{0}}
\end{array}
\right),
\label{D-Z6-Mmq=2}\\
\submatrix{M}{k}{k-3}{(m)[3]}&=\left(
\begin{array}{ccc|cc|cc}
 &           &                           &   &  &  \\
 & {\hat M}^{(m)[3]}\submatrix{U}{k}{k-3}{(m)} &  & \text{\large{0}} & & \text{\large{0}}\\
 &           &  &   & & \\
\hline
 & & & & & \\
 & \text{\large{0}} &  & \text{\large{0}}
& & \text{\large{0}} \\
 &           &  &   & & \\
\hline
 & & & & & \\
 & \text{\large{0}} &  & \text{\large{0}}  &  & 
\pm \frac{\sqrt{3}}{2}~\! \tilde{U}_{k~\!\!k-3}^{(m)}
\end{array}
\right),
\label{D-Z6-Mmq=3}
\end{align}
where $\tilde{U}_{k~\!\!k-2}^{(m)}$ and $\tilde{U}_{k~\!\!k-3}^{(m)}$
satisfy the relations
$\tilde{U}_{k~\!\!k-2}^{(m)}\tilde{U}_{k~\!\!k-2}^{(m)\dagger}=I_{n_k^{(m)'}}$ 
and $\tilde{U}_{k~\!\!k-3}^{(m)}\tilde{U}_{k~\!\!k-3}^{(m)\dagger}=I_{n_k^{(m)''}}$
with $n_{k}^{(m)}=r^{(m)}+n_{k}^{(m)'}+n_{k}^{(m)''}$.
Here, $n_{k}^{(m)'} \ne 0$ and $n_{k}^{(m)''} \ne 0$
for $a^{(m)} = -1/3$ and $a^{(m)} = -1/2$, respectively.
From eqs.~\eqref{D-Z6-Mdaggerm} and \eqref{D-Z6-Mmq=2}, 
$\submatrix{M}{k-2}{k}{(m)[-2]}$ are given by
\begin{align}
\submatrix{M}{k-2}{k}{(m)[-2]}&=\submatrix{M}{k}{k-2}{(m)[2]\dagger}=\left(
\begin{array}{ccc|cc|cc}
 &           &                           &   &  &  \\
 & {\hat M}^{(m)[2]}\submatrix{U}{k-2}{k}{(m)} &  & \text{\large{0}} & & \text{\large{0}}\\
 &           &  &   & & \\
\hline
 & & & & & \\
 & \text{\large{0}} &  & 
\frac{2}{3}~\! \tilde{U}_{k-2~\!\!k}^{(m)} 
& & \text{\large{0}} \\
 &           &  &   & & \\
\hline
 & & & & & \\
 & \text{\large{0}} &  & \text{\large{0}}  &  & \text{\large{0}}
\end{array}
\right),
\label{D-Z6-Mmq=-2}
\end{align}
where $\submatrix{U}{k}{k-2}{(m)\dagger}$ and $\tilde{U}_{k~\!\!k-2}^{(m)\dagger}$
are denoted as $\submatrix{U}{k-2}{k}{(m)}$
and $\tilde{U}_{k-2~\!\!k}^{(m)}$, respectively.
In the same way, we often denote $-\submatrix{U}{k}{k-3}{(m)\dagger}$ 
and $-\tilde{U}_{k~\!\!k-3}^{(m)\dagger}$
as $\submatrix{U}{k-3}{k}{(m)}$ and $\tilde{U}_{k-3~\!\!k}^{(m)}$, respectively. 
Replacing $k$ with $k+2$ 
for $\tilde{U}_{k~\!\!k-2}^{(m)}\tilde{U}_{k~\!\!k-2}^{(m)\dagger}=I_{n_k^{(m)'}}$
and with $k+3$ 
for $\tilde{U}_{k~\!\!k-3}^{(m)}\tilde{U}_{k~\!\!k-3}^{(m)\dagger}=I_{n_k^{(m)''}}$,
we obtain 
$\tilde{U}_{k+2~\!\!k}^{(m)}\tilde{U}_{k+2~\!\!k}^{(m)\dagger}=I_{n_{k+2}^{(m)'}}$
and $\tilde{U}_{k+3~\!\!k}^{(m)}\tilde{U}_{k+3~\!\!k}^{(m)\dagger}=I_{n_{k+3}^{(m)''}}$,
respectively,
where $n_{k+2}^{(m)}=r^{(m)}+n_{k+2}^{(m)'}+n_{k+2}^{(m)''}$
and $n_{k+3}^{(m)}=r^{(m)}+n_{k+3}^{(m)'}+n_{k+3}^{(m)''}$.
Then, we obtain the relations:
\begin{align}
& {\rm rank}(\submatrix{M}{k}{k-2}{(m)[2]})={\rm rank}({\hat M}^{(m)[2]})+ n_k^{(m)'},\quad
{\rm rank}(\submatrix{M}{k+2}{k}{(m)[2]})={\rm rank}({\hat M}^{(m)[2]})+ n_{k+2}^{(m)'},
\label{D-Z6-rankMk+2}\\
& {\rm rank}(\submatrix{M}{k}{k-3}{(m)[3]})
={\rm rank}({\hat M}^{(m)[3]})+ n_k^{(m)''}, \quad
{\rm rank}(\submatrix{M}{k+3}{k}{(m)[3]})
={\rm rank}({\hat M}^{(m)[3]})+ n_{k+3}^{(m)''}.
\label{D-Z6-rankMk+3(3)}
\end{align}
From eqs.~\eqref{D-Z6-qq'm0-rankq=2}, \eqref{D-Z6-qq'm0-rankq=3}, 
\eqref{D-Z6-rankMk+2} and \eqref{D-Z6-rankMk+3(3)},
we find that $n_k^{(m)'} = n_{k+2}^{(m)'}$, i.e.,
$n_1^{(m)'} = n_3^{(m)'} = n_5^{(m)'}$ 
and $n_2^{(m)'} = n_4^{(m)'} = n_6^{(m)'}$, and
$n_k^{(m)''} = n_{k+3}^{(m)''}$, i.e.,
$n_1^{(m)''} = n_4^{(m)''}$, $n_2^{(m)''} = n_5^{(m)''}$ and $n_3^{(m)''} = n_6^{(m)''}$.
Thus $\tilde{U}_{k~\!\!k-2}^{(m)}$ and $\tilde{U}_{k~\!\!k-3}^{(m)}$ are 
$n_k^{(m)'} \times n_k^{(m)'}$ and $n_k^{(m)''} \times n_k^{(m)''}$ unitary matrices,
respectively.

Inserting eqs.~\eqref{D-Z6-Mmq=1}, \eqref{D-Z6-Mmq=2} and \eqref{D-Z6-Mmq=3}
into eqs.~\eqref{D-Z6-MM0m} and \eqref{D-Z6-M=MM0m},
we obtain the relations:
\begin{align}
& 2({\hat M}^{(m)[1]})^2 + 2({\hat M}^{(m)[2]})^2 + ({\hat M}^{(m)[3]})^2 
= (1-a^{(m)2})I_{r^{(m)}},
\label{D-Z6-MM0m-hat}\\
& ({\hat M}^{(m)[1]})^2 - ({\hat M}^{(m)[2]})^2 - ({\hat M}^{(m)[3]})^2 
= (a^{(m)} - a^{(m)2})I_{r^{(m)}}.
\label{D-Z6-M=MM0m-hat}
\end{align}
In a similar way as on $T^2/\Z4$,
the submatrices of ${\hat M}^{(m)[q]}\submatrix{U}{k}{k-q}{(m)}$ ($q=1, 2, 3$)
are described as $({\hat M}^{(m)[q]}\submatrix{U}{k}{k-q}{(m)})_{(k'l')}
= m^{(m)[q]}_{k'}\submatrix{u}{k}{k-q}{(m, k')} \delta_{k'l'}$
with non-negative numbers $m^{(m)[q]}_{k'}$ 
and unitary matrices $\submatrix{u}{k}{k-q}{(m, k')}$, and  
the commutation relation 
$[{\hat M}^{(m)[q]}, \submatrix{U}{k}{k-q'}{(m)}]=0$
holds for any integers $q$ and $q'$, with ${\hat M}^{(m)[0]} = a^{(m)}I_{r^{(m)}}$,
$\submatrix{U}{k}{k}{(m)}=I_{r^{(m)}}$, ${\hat M}^{(m)[-q]}={\hat M}^{(m)[q]}$
and $\submatrix{U}{k-q}{k}{(m)}=(-1)^q \submatrix{U}{k}{k-q}{(m)\dagger}$.

Inserting eqs.~\eqref{D-Z6-Mmq=1}, \eqref{D-Z6-Mmq=-1} 
and \eqref{D-Z6-Mmq=2} -- \eqref{D-Z6-Mmq=-2} into eq.~\eqref{D-Z6-qq'm},
we obtain the relation
$\submatrix{U}{k}{k-q'}{(m)}\submatrix{U}{k-q'}{k-q}{(m)}
= \submatrix{U}{k}{k-q+q'}{(m)}\submatrix{U}{k-q+q'}{k-q}{(m)}$
and write down the relation:
\begin{align}
\submatrix{U}{k-r}{k-r-q'}{(m)}\submatrix{U}{k-r-q'}{k-r-q}{(m)}
= \submatrix{U}{k-r}{k-r-q+q'}{(m)}\submatrix{U}{k-r-q+q'}{k-r-q}{(m)},
\label{D-Z6-U-rel-r}
\end{align}
where $k$ is replaced with $k-r$.
In particular, from eq.~\eqref{D-Z6-U-rel-r}, we obtain the relations:
\begin{align}
\submatrix{U}{k-q}{k-q-q'}{(m)}\submatrix{U}{k-q-q'}{k}{(m)}
= \submatrix{U}{k-q}{k+q'}{(m)}\submatrix{U}{k+q'}{k}{(m)}, \qquad
\submatrix{U}{k}{k-q'}{(m)}\submatrix{U}{k-q'}{k+q}{(m)}
= \submatrix{U}{k}{k+q+q'}{(m)}\submatrix{U}{k+q+q'}{k+q}{(m)},
\label{D-Z6-U-rel}
\end{align}
where $r$ and $q$ are replaced with $q$ ($0$) and $-q$ ($-q$)
in the first (second) relation.
Exactly speaking, eq.~\eqref{D-Z6-U-rel-r} is obtained
except for the parts involving $m^{(m)[2]}_{k'}=0$ and/or $m^{(m)[3]}_{k'}=0$. 
Because unitary submatrices for $m^{(m)[2]}_{k'}=0$ and $m^{(m)[3]}_{k'}=0$
are arbitrary and they can be taken to make the relations also hold for
the parts relating to $m^{(m)[2]}_{k'}=0$ and/or $m^{(m)[3]}_{k'}=0$,
we choose them in this way
and write down following relations in the same fashion that
unitary submatrices for $m^{(m)[2]}_{k'}=0$ 
and $m^{(m)[3]}_{k'}=0$ can be contained as in eq.~\eqref{D-Z6-U-rel-r}. 
Using eq.~\eqref{D-Z6-U-rel},  
we derive the relations:
\begin{align}
& J_{kk}^{(m)} \equiv 
\submatrix{U}{k}{k-1}{(m)}\submatrix{U}{k-1}{k-2}{(m)}\submatrix{U}{k-2}{k}{(m)}
= \submatrix{U}{k}{k-1}{(m)}\submatrix{U}{k-1}{k+1}{(m)}\submatrix{U}{k+1}{k}{(m)}
= \submatrix{U}{k}{k+2}{(m)}\submatrix{U}{k+2}{k+1}{(m)}\submatrix{U}{k+1}{k}{(m)},
\label{D-Z6-U-rel2}\\
& K_{kk}^{(m)} \equiv 
\submatrix{U}{k}{k+1}{(m)}\submatrix{U}{k+1}{k-2}{(m)}\submatrix{U}{k-2}{k}{(m)}
= \submatrix{U}{k}{k+1}{(m)}\submatrix{U}{k+1}{k+3}{(m)}\submatrix{U}{k+3}{k}{(m)}
= \submatrix{U}{k}{k+2}{(m)}\submatrix{U}{k+2}{k+3}{(m)}\submatrix{U}{k+3}{k}{(m)}
\nonumber \\
& ~~~~~~~\!
= \submatrix{U}{k}{k+2}{(m)}\submatrix{U}{k+2}{k-1}{(m)}\submatrix{U}{k-1}{k}{(m)}
= \submatrix{U}{k}{k+3}{(m)}\submatrix{U}{k+3}{k-1}{(m)}\submatrix{U}{k-1}{k}{(m)}
= \submatrix{U}{k}{k-3}{(m)}\submatrix{U}{k-3}{k-2}{(m)}\submatrix{U}{k-2}{k}{(m)}.
\label{D-Z6-U-rel3}
\end{align}
Using eqs.~\eqref{D-Z6-U-rel} -- \eqref{D-Z6-U-rel3},
it is shown that $J_{kk}^{(m)}$ and $K_{kk}^{(m)}$ commute with each other as follows,
\begin{align}
& J_{kk}^{(m)}K_{kk}^{(m)}
=\submatrix{U}{k}{k+2}{(m)}\submatrix{U}{k+2}{k+1}{(m)}\submatrix{U}{k+1}{k}{(m)}
\cdot \submatrix{U}{k}{k+1}{(m)}\submatrix{U}{k+1}{k-2}{(m)}\submatrix{U}{k-2}{k}{(m)}
\nonumber \\
& = -\submatrix{U}{k}{k+2}{(m)}\submatrix{U}{k+2}{k+1}{(m)}
\submatrix{U}{k+1}{k-2}{(m)}\submatrix{U}{k-2}{k}{(m)}
= -\submatrix{U}{k}{k+2}{(m)}\submatrix{U}{k+2}{k-1}{(m)}
\submatrix{U}{k-1}{k-2}{(m)}\submatrix{U}{k-2}{k}{(m)}
\nonumber \\
& = \submatrix{U}{k}{k+2}{(m)}\submatrix{U}{k+2}{k-1}{(m)}\submatrix{U}{k-1}{k}{(m)}
\cdot \submatrix{U}{k}{k-1}{(m)}\submatrix{U}{k-1}{k-2}{(m)}\submatrix{U}{k-2}{k}{(m)}
= K_{kk}^{(m)}J_{kk}^{(m)}.
\label{D-Z6-JK=KJ}
\end{align}
Then, $J_{kk}^{(m)}$ and $K_{kk}^{(m)}$ 
are diagonalized simultaneously by a suitable unitary transformation.
Furthermore, inserting $-\submatrix{U}{k+2}{k-3}{(m)} \submatrix{U}{k-3}{k}{(m)}
\submatrix{U}{k}{k-3}{(m)} \submatrix{U}{k-3}{k-2}{(m)} \submatrix{U}{k-2}{k}{(m)}
\submatrix{U}{k}{k-2}{(m)} \submatrix{U}{k-2}{k-3}{(m)} \submatrix{U}{k-3}{k+2}{(m)}
= I_{n_k^{(m)}}$ into 
$\submatrix{U}{k}{k+2}{(m)} \submatrix{U}{k+2}{k-2}{(m)} \submatrix{U}{k-2}{k}{(m)}$ and
using eqs.~\eqref{D-Z6-U-rel-r} -- \eqref{D-Z6-U-rel3},
$\submatrix{U}{k}{k+2}{(m)} \submatrix{U}{k+2}{k-2}{(m)} \submatrix{U}{k-2}{k}{(m)}
= -K_{kk}^{(m)} K_{kk}^{(m)} J_{kk}^{(m)}$ is derived as follows,
\begin{align}
& \submatrix{U}{k}{k+2}{(m)}\submatrix{U}{k+2}{k-2}{(m)}\submatrix{U}{k-2}{k}{(m)}
\nonumber \\
& = -\submatrix{U}{k}{k+2}{(m)}\cdot \submatrix{U}{k+2}{k-3}{(m)}\submatrix{U}{k-3}{k}{(m)}
\submatrix{U}{k}{k-3}{(m)}\submatrix{U}{k-3}{k-2}{(m)}\submatrix{U}{k-2}{k}{(m)}
\submatrix{U}{k}{k-2}{(m)}\submatrix{U}{k-2}{k-3}{(m)}\submatrix{U}{k-3}{k+2}{(m)}
\cdot \submatrix{U}{k+2}{k-2}{(m)}\submatrix{U}{k-2}{k}{(m)}
\nonumber \\
& = -K_{kk}^{(m)} K_{kk}^{(m)} 
\submatrix{U}{k}{k-2}{(m)}\submatrix{U}{k-2}{k-3}{(m)}\submatrix{U}{k-3}{k+2}{(m)}
\submatrix{U}{k+2}{k-2}{(m)}\submatrix{U}{k-2}{k}{(m)}
\nonumber \\
& = -K_{kk}^{(m)} K_{kk}^{(m)} 
\submatrix{U}{k}{k-2}{(m)}\submatrix{U}{k-2}{k-3}{(m)}\submatrix{U}{k-3}{k-1}{(m)}
\submatrix{U}{k-1}{k-2}{(m)}\submatrix{U}{k-2}{k}{(m)}
\nonumber \\
& = -K_{kk}^{(m)} K_{kk}^{(m)} 
\submatrix{U}{k}{k-2}{(m)}\submatrix{U}{k-2}{k}{(m)}\submatrix{U}{k}{k-1}{(m)}
\submatrix{U}{k-1}{k-2}{(m)}\submatrix{U}{k-2}{k}{(m)}
 = -K_{kk}^{(m)} K_{kk}^{(m)} 
\submatrix{U}{k}{k-1}{(m)}\submatrix{U}{k-1}{k-2}{(m)}\submatrix{U}{k-2}{k}{(m)}
\nonumber \\
& = -K_{kk}^{(m)} K_{kk}^{(m)} J_{kk}^{(m)}.
\label{D-Z6-KKJ}
\end{align}

Inserting eqs.~\eqref{D-Z6-Mmq=1}, \eqref{D-Z6-Mmq=-1} and \eqref{D-Z6-Mmq=2} -- \eqref{D-Z6-Mmq=-2}
into eqs.~\eqref{D-Z6-MM2m}, 
we obtain the relations:
\begin{align}
& 2a^{(m)}{\hat M}^{(m)[2]}
 - ({\hat M}^{(m)[2]})^2 K_{kk}^{(m)} K_{kk}^{(m)} J_{kk}^{(m)}
= ({\hat M}^{(m)[1]})^2 J_{kk}^{(m)}
+ 2{\hat M}^{(m)[1]}{\hat M}^{(m)[3]} K_{kk}^{(m)},
\label{D-Z6-MM2m-U}\\
& \tilde{U}_{k~\!\!k-2}^{(m)} - \tilde{U}_{k~\!\!k+2}^{(m)}\tilde{U}_{k+2~\!\!k-2}^{(m)} = 0.
\label{D-Z6-MM2m-Utilde}
\end{align}
In the same way, inserting eqs.~\eqref{D-Z6-Mmq=1}, \eqref{D-Z6-Mmq=-1} 
and \eqref{D-Z6-Mmq=2} -- \eqref{D-Z6-Mmq=-2}
into eqs.~\eqref{D-Z6-M=MM1m} -- \eqref{D-Z6-M=MM3m}, 
we obtain the relations:
\begin{align}
& (1-a^{(m)}){\hat M}^{(m)[1]}
= - {\hat M}^{(m)[3]}{\hat M}^{(m)[2]}K_{kk}^{(m)} 
+ 2{\hat M}^{(m)[2]}{\hat M}^{(m)[1]} J_{kk}^{(m)\dagger},
\label{D-Z6-M=MM1m-U}\\
& (1+a^{(m)}){\hat M}^{(m)[2]}
= ({\hat M}^{(m)[1]})^2 J_{kk}^{(m)} - {\hat M}^{(m)[1]}{\hat M}^{(m)[3]}K_{kk}^{(m)} 
- ({\hat M}^{(m)[2]})^2 K_{kk}^{(m)} K_{kk}^{(m)} J_{kk}^{(m)},
\label{D-Z6-M=MM2m-U}\\
& (1+2a^{(m)}){\hat M}^{(m)[3]}
= - {\hat M}^{(m)[1]}{\hat M}^{(m)[2]}(K_{kk}^{(m)\dagger} + K_{kk}^{(m)}),
\label{D-Z6-M=MM3m-U}
\end{align}
respectively, besides eq.~\eqref{D-Z6-MM2m-Utilde}.

In this way, the form of $\submatrix{M}{k}{l}{(m)[k-l]}$
is restricted as eqs.~\eqref{D-Z6-Mmq=1}, \eqref{D-Z6-Mmq=-1}
and \eqref{D-Z6-Mmq=2}--\eqref{D-Z6-Mmq=-2}
with ${\hat M}^{(m)[1]}$, ${\hat M}^{(m)[2]}$ and ${\hat M}^{(m)[3]}$ 
satisfying eqs.~\eqref{D-Z6-MM0m-hat}, \eqref{D-Z6-M=MM0m-hat},
\eqref{D-Z6-MM2m-U} and \eqref{D-Z6-M=MM1m-U}--\eqref{D-Z6-M=MM3m-U}, 
$\submatrix{U}{k}{l}{(m)}$ satisfying eqs.~\eqref{D-Z6-U-rel-r}, \eqref{D-Z6-MM2m-U}
and \eqref{D-Z6-M=MM1m-U}--\eqref{D-Z6-M=MM3m-U}, 
and $\tilde{U}_{k~\!\!k-2}^{(m)}$ satisfying eq.~\eqref{D-Z6-MM2m-Utilde}.

\subsection{Diagonalization of submatrices in $T_1^{(m)}$ and rearrangement}
\label{D3}

Based on eqs.~\eqref{D-Z6-Mmq=1}, \eqref{D-Z6-Mmq=-1} 
and \eqref{D-Z6-Mmq=2} -- \eqref{D-Z6-Mmq=-2},
we can rearrange $R_0^{(m)}$ and $T_1^{(m)}$ to be the form of 
block-diagonal ones such as
\begin{align}
&R_0^{(m)}=
	\begin{pmatrix}
		R_0^{(m)'}& & \\
		&R_0^{(m)''}& \\
		& & R_0^{(m)'''}
	\end{pmatrix}
=R_0^{(m)'}\oplus R_0^{(m)''}\oplus R_0^{(m)'''},
\label{Z6-R0m-de} \\
&R_0^{(m)'}=
  \begin{pmatrix}
      \eta I_{r^{(m)}}& & & & & \\
       & \eta^2 I_{r^{(m)}}& & & & \\
       & &-I_{r^{(m)}}& & & \\
       & & &-\eta I_{r^{(m)}} & & \\
      & & & & -\eta^2 I_{r^{(m)}} & \\
      & & & & & I_{r^{(m)}}
  \end{pmatrix},
\label{Z6-R0m'}\\
&R_0^{(m)''}=
  \begin{pmatrix}
      \eta I_{n_1^{(m)'}}& & & & & \\
       & \eta^2 I_{n_2^{(m)'}}& & & & \\
       & &-I_{n_1^{(m)'}}& & & \\
       & & &-\eta I_{n_2^{(m)'}} & & \\
      & & & & -\eta^2 I_{n_1^{(m)'}} & \\
      & & & & & I_{n_2^{(m)'}}
  \end{pmatrix},
\label{Z6-R0m''}\\
&R_0^{(m)'''}=
  \begin{pmatrix}
      \eta I_{n_1^{(m)''}}& & & & & \\
       & \eta^2 I_{n_2^{(m)''}}& & & & \\
       & &-I_{n_3^{(m)''}}& & & \\
       & & &-\eta I_{n_1^{(m)''}} & & \\
      & & & & -\eta^2 I_{n_2^{(m)''}} & \\
      & & & & & I_{n_3^{(m)''}}
  \end{pmatrix},
\label{Z6-R0m'''}
\end{align}
and
\begin{align}
&T_1^{(m)}=
	\begin{pmatrix}
		T_1^{(m)'}& \\
		& T_1^{(m)''}\\
		& & T_1^{(m)'''}
	\end{pmatrix}
=T_1^{(m)'}\oplus T_1^{(m)''}\oplus T_1^{(m)'''},
\label{Z6-T1m-de} \\
& T_1^{(m)'}=
    \begin{pmatrix}
        a^{(m)}I_{r^{(m)}} & {\hat M}^{(m)[1]}\submatrix{U}{1}{2}{(m)} 
& {\hat M}^{(m)[2]}\submatrix{U}{1}{3}{(m)} 
& {\hat M}^{(m)[3]}\submatrix{U}{1}{4}{(m)} 
& {\hat M}^{(m)[2]}\submatrix{U}{1}{5}{(m)} 
& {\hat M}^{(m)[1]}\submatrix{U}{1}{6}{(m)} \\
{\hat M}^{(m)[1]}\submatrix{U}{2}{1}{(m)} 
& a^{(m)}I_{r^{(m)}} 
& {\hat M}^{(m)[1]}\submatrix{U}{2}{3}{(m)} 
& {\hat M}^{(m)[2]}\submatrix{U}{2}{4}{(m)} 
& {\hat M}^{(m)[3]}\submatrix{U}{2}{5}{(m)} 
& {\hat M}^{(m)[2]}\submatrix{U}{2}{6}{(m)} \\
{\hat M}^{(m)[2]}\submatrix{U}{3}{1}{(m)} 
& {\hat M}^{(m)[1]}\submatrix{U}{3}{2}{(m)} 
& a^{(m)}I_{r^{(m)}} 
& {\hat M}^{(m)[1]}\submatrix{U}{3}{4}{(m)} 
& {\hat M}^{(m)[2]}\submatrix{U}{3}{5}{(m)} 
& {\hat M}^{(m)[3]}\submatrix{U}{3}{6}{(m)} \\
 {\hat M}^{(m)[3]}\submatrix{U}{4}{1}{(m)} 
& {\hat M}^{(m)[2]}\submatrix{U}{4}{2}{(m)} 
& {\hat M}^{(m)[1]}\submatrix{U}{4}{3}{(m)} & a^{(m)}I_{r^{(m)}}
& {\hat M}^{(m)[1]}\submatrix{U}{4}{5}{(m)} 
& {\hat M}^{(m)[2]}\submatrix{U}{4}{6}{(m)} \\
 {\hat M}^{(m)[2]}\submatrix{U}{5}{1}{(m)} 
& {\hat M}^{(m)[3]}\submatrix{U}{5}{2}{(m)} 
& {\hat M}^{(m)[2]}\submatrix{U}{5}{3}{(m)}
& {\hat M}^{(m)[1]}\submatrix{U}{5}{4}{(m)}  & a^{(m)}I_{r^{(m)}}
& {\hat M}^{(m)[1]}\submatrix{U}{5}{6}{(m)} \\
 {\hat M}^{(m)[1]}\submatrix{U}{6}{1}{(m)} 
& {\hat M}^{(m)[2]}\submatrix{U}{6}{2}{(m)} 
& {\hat M}^{(m)[3]}\submatrix{U}{6}{3}{(m)}
& {\hat M}^{(m)[2]}\submatrix{U}{6}{4}{(m)}
& {\hat M}^{(m)[1]}\submatrix{U}{6}{5}{(m)}   & a^{(m)}I_{r^{(m)}}
    \end{pmatrix},
\label{Z6-T1m'}\\
&T_1^{(m)''}=
    \begin{pmatrix}
        -\frac{1}{3}I_{n_1^{(m)'}} & 0 & \frac{2}{3}\submatrix{\tilde U}{1}{3}{(m)} 
& 0 & \frac{2}{3}\submatrix{\tilde U}{1}{5}{(m)} & 0\\
        0 & -\frac{1}{3}I_{n_2^{(m)'}} & 0 & \frac{2}{3}\submatrix{\tilde U}{2}{4}{(m)} 
& 0 & \frac{2}{3}\submatrix{\tilde U}{2}{6}{(m)}\\
        \frac{2}{3}\submatrix{\tilde U}{3}{1}{(m)} & 0 & -\frac{1}{3}I_{n_1^{(m)'}} 
& 0 & \frac{2}{3}\submatrix{\tilde U}{3}{5}{(m)} & 0\\
        0 & \frac{2}{3}\submatrix{\tilde U}{4}{2}{(m)} & 0 & -\frac{1}{3}I_{n_2^{(m)'}} 
& 0 & \frac{2}{3}\submatrix{\tilde U}{4}{6}{(m)}\\
	   \frac{2}{3}\submatrix{\tilde U}{5}{1}{(m)} & 0 
& \frac{2}{3}\submatrix{\tilde U}{5}{3}{(m)} & 0 & -\frac{1}{3}I_{n_1^{(m)'}} & 0\\
	   0 & \frac{2}{3}\submatrix{\tilde U}{6}{2}{(m)} & 0 
& \frac{2}{3}\submatrix{\tilde U}{6}{4}{(m)}& 0 & -\frac{1}{3}I_{n_2^{(m)'}}\\
    \end{pmatrix},
\label{Z6-T1m''}\\
&T_1^{(m)'''}=
    \begin{pmatrix}
        -\frac{1}{2}I_{n_1^{(m)''}} & 0 & 0 
& \pm\frac{\sqrt{3}}{2}\submatrix{\tilde U}{1}{4}{(m)} & 0 & 0\\
        0 & -\frac{1}{2}I_{n_2^{(m)''}} & 0 & 0 
& \pm\frac{\sqrt{3}}{2}\submatrix{\tilde U}{2}{5}{(m)} & 0\\
        0 & 0 & -\frac{1}{2}I_{n_3^{(m)''}} & 0 & 0 
& \pm\frac{\sqrt{3}}{2}\submatrix{\tilde U}{3}{6}{(m)}\\
        \pm\frac{\sqrt{3}}{2}\submatrix{\tilde U}{4}{1}{(m)} & 0 & 0 
& -\frac{1}{2}I_{n_1^{(m)''}} & 0 & 0\\
	   0 & \pm\frac{\sqrt{3}}{2}\submatrix{\tilde U}{5}{2}{(m)} & 0 & 0 
& -\frac{1}{2}I_{n_2^{(m)''}} & 0\\
	   0 & 0 & \pm\frac{\sqrt{3}}{2}\submatrix{\tilde U}{6}{3}{(m)} & 0 & 0 
& -\frac{1}{2}I_{n_3^{(m)''}}\\
    \end{pmatrix},
\nonumber \\
&~~~~~~~~~~~~~~~(\text{double sign in the same order}),
\label{Z6-T1m'''}
\end{align}
respectively.

For the convenience of a later calculation, 
we write down the following relations
obtained by setting $k=2$ in eqs.~\eqref{D-Z6-U-rel2}, \eqref{D-Z6-U-rel3} and \eqref{D-Z6-KKJ},
\begin{align}
& J_{22}^{(m)} =
\submatrix{U}{2}{1}{(m)}\submatrix{U}{1}{6}{(m)}\submatrix{U}{6}{2}{(m)}
= \submatrix{U}{2}{1}{(m)}\submatrix{U}{1}{3}{(m)}\submatrix{U}{3}{2}{(m)}
= \submatrix{U}{2}{4}{(m)}\submatrix{U}{4}{3}{(m)}\submatrix{U}{3}{2}{(m)},
\label{D-Z6-U-rel2k=2}\\
& K_{22}^{(m)} =
\submatrix{U}{2}{3}{(m)}\submatrix{U}{3}{6}{(m)}\submatrix{U}{6}{2}{(m)}
= \submatrix{U}{2}{3}{(m)}\submatrix{U}{3}{5}{(m)}\submatrix{U}{5}{2}{(m)}
= \submatrix{U}{2}{4}{(m)}\submatrix{U}{4}{5}{(m)}\submatrix{U}{5}{2}{(m)}
\nonumber \\
& ~~~~~~~\!
= \submatrix{U}{2}{4}{(m)}\submatrix{U}{4}{1}{(m)}\submatrix{U}{1}{2}{(m)}
= \submatrix{U}{2}{5}{(m)}\submatrix{U}{5}{1}{(m)}\submatrix{U}{1}{2}{(m)}
= \submatrix{U}{2}{5}{(m)}\submatrix{U}{5}{6}{(m)}\submatrix{U}{6}{2}{(m)},
\label{D-Z6-U-rel3k=2}\\
& \submatrix{U}{2}{4}{(m)}\submatrix{U}{4}{6}{(m)}\submatrix{U}{6}{2}{(m)}
= - K_{22}^{(m)}K_{22}^{(m)}J_{22}^{(m)}.
\label{D-Z6-U-rel6k=2}
\end{align}
From eq.~\eqref{D-Z6-MM2m-Utilde}, we obtain the relations:
\begin{align}
\tilde{U}_{3~\!\!5}^{(m)}\tilde{U}_{5~\!\!1}^{(m)}\tilde{U}_{1~\!\!3}^{(m)} = I_{n_1^{(m)'}},
\qquad
\tilde{U}_{4~\!\!6}^{(m)}\tilde{U}_{6~\!\!2}^{(m)}\tilde{U}_{2~\!\!4}^{(m)} = I_{n_2^{(m)'}}.
\label{D-Z6-Utilde-rel}
\end{align}

Let us perform a unitary transformation $V^{(m)}T_1^{(m)}V^{(m)\dagger}$
with $V^{(m)}=V^{(m)'} \oplus V^{(m)''} \oplus V^{(m)'''}$,
where $V^{(m)'}$, $V^{(m)''}$ and $V^{(m)'''}$ are given by
\begin{align}
&V^{(m)'}=
    \begin{pmatrix}
        \hat{\varTheta}^{(m)[1]\dagger} U^{(m)} \submatrix{U}{2}{1}{(m)} 
         &  &  &  &  & \\
         & U^{(m)} &  &  &  & \\
         &  & -\hat{\varTheta}^{(m)[1]} U^{(m)} \submatrix{U}{2}{3}{(m)}
         &  &  & \\
         &  &  & \hat{\varTheta}^{(m)[2]} U^{(m)} \submatrix{U}{2}{4}{(m)}
         &  & \\
         &  &  &  & i U^{(m)} \submatrix{U}{2}{5}{(m)} &  \\
         &  &  &  &  & \hat{\varTheta}^{(m)[2]\dagger} U^{(m)} \submatrix{U}{2}{6}{(m)} 
    \end{pmatrix},
\label{D-Z6-Vm'}\\
&V^{(m)''}=
    \begin{pmatrix}
        {\tilde U}_{3~\!\!1}^{(m)} &  &  &  &  & \\
         & {\tilde U}_{4~\!\!2}^{(m)} &  &  &  & \\
         &  & I_{n_1^{(m)'}} &  &  & \\
         &  &  & I_{n_2^{(m)'}}  &  & \\
         &  &  &  & {\tilde U}_{3~\!\!5}^{(m)} &  \\
         &  &  &  &  & {\tilde U}_{4~\!\!6}^{(m)} 
    \end{pmatrix},
\label{D-Z6-Vm''}
\end{align}
and
\begin{align}
&V^{(m)'''}=
    \begin{pmatrix}
        -i{\tilde U}_{4~\!\!1}^{(m)} &  &  &  & & \\
         & -i{\tilde U}_{5~\!\!2}^{(m)} &  &  & & \\
         &  & -i{\tilde U}_{6~\!\!3}^{(m)} &  &  & \\
         &  &  & I_{n_1^{(m)''}} &  & \\
         &  &  &  &  I_{n_2^{(m)''}}  & \\
         &  &  &  &  &  I_{n_3^{(m)''}}
    \end{pmatrix}.
\label{D-Z6-Vm'''}
\end{align}
Here $U^{(m)}$ are unitary matrices 
that make $J_{22}^{(m)}$ and $K_{22}^{(m)}$ diagonal ones 
$\hat{J}_{22}^{(m)}(=U^{(m)} J_{22}^{(m)} U^{(m)\dagger})$ 
and $\hat{K}_{22}^{(m)}(=U^{(m)} K_{22}^{(m)} U^{(m)\dagger})$ simultaneously and
commute with ${\hat M}^{(m)[q]}$,
$\hat{\varTheta}^{(m)[1]}$ are diagonal unitary matrices
whose cubes agree with $i\hat{K}_{22}^{(m)\dagger}\hat{J}_{22}^{(m)}$,
and $\hat{\varTheta}^{(m)[2]}$ are diagonal unitary matrices
whose cubes agree with $-\hat{J}_{22}^{(m)\dagger}\hat{K}_{22}^{(m)\dagger 2}$.
Then, we find that $R_0^{(m)'}$, $R_0^{(m)''}$ and $R_0^{(m)'''}$ remain unchanged
and, using $\submatrix{U}{k-q}{k}{(m)}=(-1)^q\submatrix{U}{k}{k-q}{(m)\dagger}$,
${\tilde U}_{k-q~\!\!k}^{(m)}=(-1)^q{\tilde U}_{k~\!\!k-q}^{(m)\dagger}$ and
eqs.~\eqref{D-Z6-U-rel2k=2} -- \eqref{D-Z6-Utilde-rel}, 
$T_1^{(m)'}$, $T_1^{(m)''}$ and $T_1^{(m)'''}$
are transformed into matrices 
\begin{align}
& T_1^{(m)'}=
    \begin{pmatrix}
        a^{(m)}I_{r^{(m)}} & -{\hat M}^{(m)[1]}\submatrix{\hat \Theta}{}{}{[1]\dagger} 
& {\hat M}^{(m)[2]}\submatrix{\hat \Theta}{}{}{[2]\dagger} 
& -i{\hat M}^{(m)[3]}I_{r^{(m)}} 
& {\hat M}^{(m)[2]}\submatrix{\hat \Theta}{}{}{[2]} 
& {\hat M}^{(m)[1]}\submatrix{\hat \Theta}{}{}{[1]} \\
{\hat M}^{(m)[1]}\submatrix{\hat \Theta}{}{}{[1]} 
& a^{(m)}I_{r^{(m)}} 
& -{\hat M}^{(m)[1]}\submatrix{\hat \Theta}{}{}{[1]\dagger} 
& {\hat M}^{(m)[2]}\submatrix{\hat \Theta}{}{}{[2]\dagger} 
& -i{\hat M}^{(m)[3]}I_{r^{(m)}} 
& {\hat M}^{(m)[2]}\submatrix{\hat \Theta}{}{}{[2]} \\
{\hat M}^{(m)[2]}\submatrix{\hat \Theta}{}{}{[2]} 
& {\hat M}^{(m)[1]}\submatrix{\hat \Theta}{}{}{[1]} 
& a^{(m)}I_{r^{(m)}} 
& -{\hat M}^{(m)[1]}\submatrix{\hat \Theta}{}{}{[1]\dagger} 
& {\hat M}^{(m)[2]}\submatrix{\hat \Theta}{}{}{[2]\dagger}
& -i{\hat M}^{(m)[3]}I_{r^{(m)}}  \\
 -i{\hat M}^{(m)[3]}I_{r^{(m)}} 
& {\hat M}^{(m)[2]}\submatrix{\hat \Theta}{}{}{[2]} 
& {\hat M}^{(m)[1]}\submatrix{\hat \Theta}{}{}{[1]} & a^{(m)}I_{r^{(m)}}
& -{\hat M}^{(m)[1]}\submatrix{\hat \Theta}{}{}{[1]\dagger} 
& {\hat M}^{(m)[2]}\submatrix{\hat \Theta}{}{}{[2]\dagger} \\
 {\hat M}^{(m)[2]}\submatrix{\hat \Theta}{}{}{[2]\dagger} 
&  -i{\hat M}^{(m)[3]}I_{r^{(m)}}
& {\hat M}^{(m)[2]}\submatrix{\hat \Theta}{}{}{[2]}
& {\hat M}^{(m)[1]}\submatrix{\hat \Theta}{}{}{[1]}  & a^{(m)}I_{r^{(m)}}
& -{\hat M}^{(m)[1]}\submatrix{\hat \Theta}{}{}{[1]\dagger} \\
 -{\hat M}^{(m)[1]}\submatrix{\hat \Theta}{}{}{[1]\dagger} 
& {\hat M}^{(m)[2]}\submatrix{\hat \Theta}{}{}{[2]\dagger} 
& -i{\hat M}^{(m)[3]}I_{r^{(m)}}
& {\hat M}^{(m)[2]}\submatrix{\hat \Theta}{}{}{[2]}
& {\hat M}^{(m)[1]}\submatrix{\hat \Theta}{}{}{[1]}   & a^{(m)}I_{r^{(m)}}
    \end{pmatrix},
\label{Z6-T1m'-tr}\\
&T_1^{(m)''}=
    \begin{pmatrix}
        -\frac{1}{3}I_{n_1^{(m)'}} & 0 & \frac{2}{3}I_{n_1^{(m)'}} & 0 
& \frac{2}{3}I_{n_1^{(m)'}} & 0\\
        0 & -\frac{1}{3}I_{n_2^{(m)'}} & 0 & \frac{2}{3}I_{n_2^{(m)'}} & 0 
& \frac{2}{3}I_{n_2^{(m)'}}\\
        \frac{2}{3}I_{n_1^{(m)'}} & 0 & -\frac{1}{3}I_{n_1^{(m)'}} & 0 
& \frac{2}{3}I_{n_1^{(m)'}} & 0\\
        0 & \frac{2}{3}I_{n_2^{(m)'}} & 0 & -\frac{1}{3}I_{n_2^{(m)'}} & 0 
& \frac{2}{3}I_{n_2^{(m)'}}\\
	   \frac{2}{3}I_{n_1^{(m)'}} & 0 & \frac{2}{3}I_{n_1^{(m)'}} & 0 
& -\frac{1}{3}I_{n_1^{(m)'}} & 0\\
	   0 & \frac{2}{3}I_{n_2^{(m)'}} & 0 & \frac{2}{3}I_{n_2^{(m)'}}& 0 
& -\frac{1}{3}I_{n_2^{(m)'}}\\
    \end{pmatrix},
\label{Z6-T1m''-tr}\\
&T_1^{(m)'''}=
    \begin{pmatrix}
        -\frac{1}{2}I_{n_1^{(m)''}} & 0 & 0 & \pm\frac{\sqrt{3}}{2}i I_{n_1^{(m)''}} & 0 & 0\\
        0 & -\frac{1}{2}I_{n_2^{(m)''}} & 0 & 0 & \pm\frac{\sqrt{3}}{2}i I_{n_2^{(m)''}} & 0\\
        0 & 0 & -\frac{1}{2}I_{n_3^{(m)''}} & 0 & 0 & \pm\frac{\sqrt{3}}{2}i I_{n_3^{(m)''}}\\
        \pm\frac{\sqrt{3}}{2}i I_{n_1^{(m)''}} & 0 & 0 & -\frac{1}{2}I_{n_1^{(m)''}} & 0 & 0\\
	   0 & \pm\frac{\sqrt{3}}{2}i I_{n_2^{(m)''}} & 0 & 0 & -\frac{1}{2}I_{n_2^{(m)''}} & 0\\
	   0 & 0 & \pm\frac{\sqrt{3}}{2}i I_{n_3^{(m)''}} & 0 & 0 & -\frac{1}{2}I_{n_3^{(m)''}}\\
    \end{pmatrix},
\nonumber \\
&~~~~~~~~~~~~~~~(\text{double sign in the same order}),
\label{Z6-T1m'''-tr}
\end{align}
respectively.
Those submatrices are compactly expressed as
\begin{align}
& (T_1^{(m)'})_{(k k-q)}
={\hat M}^{(m)[q]}\hat{\varTheta}^{(m)[q]}I_{r^{(m)}},\qquad
(T_1^{(m)''})_{(k k-q)}
= \left(-\frac{1}{3}\delta_{q~\!\! 0} 
+ \frac{2}{3}\delta_{q~\!\!\pm 2}\right)I_{n_k^{(m)'}},
\nonumber \\
& (T_1^{(m)'''})_{(k k-q)}
= \left(-\frac{1}{2}\delta_{q~\!\! 0}
\pm\frac{\sqrt{3}}{2}i \delta_{q~\!\!3}\right)I_{n_k^{(m)''}},
\label{D-Z6-submatrices-tr}
\end{align}
where $\hat{\varTheta}^{(m)[-q]}=(-1)^q \hat{\varTheta}^{(m)[q]\dagger}$,
$\hat{\varTheta}^{(m)[0]}=I_{r^{(m)}}$, 
$\hat{\varTheta}^{(m)[3]}= -i I_{r^{(m)}}$, $n_k^{(m)'}=n_{k+2}^{(m)'}$
and $n_k^{(m)''}=n_{k+3}^{(m)''}$.
Using $\hat{\varTheta}^{(m)[1]}$, $\hat{\varTheta}^{(m)[2]}$
and those hermitian conjugates, 
eqs.~\eqref{D-Z6-MM2m-U} and \eqref{D-Z6-M=MM1m-U} -- \eqref{D-Z6-M=MM3m-U} are rewritten as
\begin{align}
& 2a^{(m)}{\hat M}^{(m)[2]}\hat{\varTheta}^{(m)[2]}
 + ({\hat M}^{(m)[2]}\hat{\varTheta}^{(m)[2]\dagger})^2
 = ({\hat M}^{(m)[1]}\hat{\varTheta}^{(m)[1]})^2
- 2{\hat M}^{(m)[1]}\hat{\varTheta}^{(m)[1]\dagger}{\hat M}^{(m)[3]}
\hat{\varTheta}^{(m)[3]},
\label{D-Z6-MM2m-U1-M}\\
& (1 - a^{(m)}){\hat M}^{(m)[1]}\hat{\varTheta}^{(m)[1]}
 = {\hat M}^{(m)[3]}\hat{\varTheta}^{(m)[3]}{\hat M}^{(m)[2]}\hat{\varTheta}^{(m)[2]\dagger}
   + 2{\hat M}^{(m)[2]}\hat{\varTheta}^{(m)[2]}{\hat M}^{(m)[1]}\hat{\varTheta}^{(m)[1]\dagger},
\label{D-Z6-M=MM1m-U1-M}\\
& (1 + a^{(m)}){\hat M}^{(m)[2]}\hat{\varTheta}^{(m)[2]}
 = ({\hat M}^{(m)[1]}\hat{\varTheta}^{(m)[1]})^2
 + {\hat M}^{(m)[1]}\hat{\varTheta}^{(m)[1]\dagger}{\hat M}^{(m)[3]}\hat{\varTheta}^{(m)[3]}
 + ({\hat M}^{(m)[2]}\hat{\varTheta}^{(m)[2]\dagger})^2,
\label{D-Z6-M=MM2m-U1-M}\\
& (1 + 2 a^{(m)}){\hat M}^{(m)[3]}\hat{\varTheta}^{(m)[3]}
 = {\hat M}^{(m)[1]}\hat{\varTheta}^{(m)[1]}{\hat M}^{(m)[2]}\hat{\varTheta}^{(m)[2]}
   - {\hat M}^{(m)[2]}\hat{\varTheta}^{(m)[2]\dagger}
  {\hat M}^{(m)[1]}\hat{\varTheta}^{(m)[1]\dagger},
\label{D-Z6-M=MM3m-U1-M}
\end{align}
respectively.

\section{Possible forms of a matrix $A$ under the condition
  $AA^\dag$ and $A^\dag A$ are diagonal}
\label{ap:mmdag}

Let $A$ be an $l_1\times l_2$ matrix with rank $r$.  In general, $A$
can be expressed by using $l_i\times l_i$ unitary matrices $U_i$
$(i=1,2)$ and a matrix $\hat A$ as $ A=U_1^\dag \hat A U_2$,
where $\hat A$ is an $l_1\times l_2$ matrix and written by using a
diagonal matrix $\hat A_r$ with rank $r$ as
\begin{align}
  \hat A=
  \left(
  \begin{array}{c|c}
    \hat A_r& 0\\\hline
    0&0
  \end{array}
       \right).
\label{A-hatA}
\end{align}
We can take $\hat A_r$ as 
\begin{align}
\hat A_r=
       \begin{pmatrix}
           a_1I_{n_1}&&&\\&a_2I_{n_2}&&\\&&\ddots&\\&&&a_qI_{n_q}
       \end{pmatrix},
\qquad a_k>0 \qquad {\rm for} \quad k=1,\dots, q,
\label{A-hatAr}
\end{align}
where $a_k\neq a_{k'}$ for $k\neq k'$, ${n_k}$ is a positive integer
satisfying $\sum_{k=1}^qn_k=r$, and $I_{n_k}$ is the $n_k\times n_k$
unit matrix. Then, we find
\begin{align}
U_1AA^\dag U_1^\dag=  \hat A\hat A^\dag =
  \begin{pmatrix}
      \hat A_r^2&0\\0&0
  \end{pmatrix},\qquad 
U_2A^\dag AU_2^\dag=  \hat A^\dag \hat A=
  \begin{pmatrix}
      \hat A_r^2&0\\0&0
  \end{pmatrix}. 
\end{align}
Note that the former (latter) is an $l_1\times l_1$ ($l_2\times l_2$)
matrix.

Let us take a basis where $AA^\dag$ and $A^\dag A$
are already diagonal and satisfy $AA^\dag=\hat A\hat A^\dag $ and
$A^\dag A=\hat A^\dag\hat A $. Then, we find
$U_1AA^\dag U_1^\dag=AA^\dag $ and $U_2A^\dag A U_2^\dag=A^\dag A $,
which implies $[U_1,AA^\dag]=0$ and $[U_2,A^\dag A]=0$. In this case,
a possible form of $U_i$ is restricted as
\begin{align}
  U_i=\left(
  \begin{array}{ccccc}
    u_i^{(1)}&&&&\\
    &u_i^{(2)}&&&\\
    &&\ddots&&\\
    &&&u_i^{(q)}&\\
    &&&&\tilde u_i
  \end{array}\right),
\end{align}
where $u_i^{(k)}$ and $\tilde u_i$ are $n_k\times n_k$
and $(l_i-r)\times (l_i-r)$ 
unitary matrices, respectively. Therefore, we find that 
a general form of $A$ is given by 
\begin{align}
  A=U_1^\dag \hat A U_2=
    \left(
  \begin{array}{c|c}
    \tilde A_r&0\\\hline 0&0
  \end{array}
                            \right), \qquad
\tilde A_r=                            
  \left(
  \begin{array}{cccc}
    a_1\tilde u^{(1)}&&&\\
    &a_2\tilde u^{(2)}&&\\
    &&\ddots&\\
    &&&a_q\tilde u^{(q)}
  \end{array}\right) =\hat{A}_r U,
\label{A-ArU}
\end{align}
where $\hat{A}_r$ is a diagonal matrix with submatrices 
$(\hat{A}_r)_{(kl)} = a_k I_{n_k} \delta_{kl}$ ($k, l = 1, \cdots, q$)
as seen from eq.~\eqref{A-hatAr},
and $U$ is a unitary matrix with a diagonal submatrices 
$(U)_{(kl)} = \tilde{u}^{(k)} \delta_{kl}$.
Here $\tilde{u}^{(k)}$ is an $n_k\times n_k$ unitary matrix 
defined by $\tilde u^{(k)}=u_1^{(k)\dag}u_2^{(k)}$.
Then, it is easy to see that $U$ commutes with $\hat{A}_r$.

%
\section{Derivation of eqs.~\eqref{C-Z4-qq'} and \eqref{D-Z6-qq'}}
\label{app:MM}
%

The unitary matrices $R_0$ and $T_1$ on $T^2/\Z N$ ($N = 3, 4, 6$)
contain submatrices denoted as $(R_0)_{(kl)}=\tau^k \delta_{kl} I_{n_{k}}$
with $\tau = e^{2\pi i/N}$
and $(T_1)_{(kl)} = \submatrix{M}{k}{l}{[k-l]}$.
The upper index $k-l=q$ represents the charge of the $\Z N$ symmetry 
generated by 
$R_0$: $(R_0T_1R_0^{-1})_{(k\,k-q)}=\tau^{q}\submatrix{M}{k}{k-q}{[q]}$. 
We use a notation of 
$\submatrix{M}{k}{l}{[k-l]}=\submatrix{M}{k}{l}{[k'-l']}=\submatrix{M}{k'}{l'}{[k-l]}$
with $k'=k$ (mod $N$) and $l'=l$ (mod $N$). 

Eqs.~\eqref{C-Z4-qq'} and \eqref{D-Z6-qq'}, i.e.,
$\submatrix{M}{k}{k-q'}{[q']}\submatrix{M}{k-q'}{k-q}{[q-q']}
=\submatrix{M}{k}{k-q+q'}{[q-q']}\submatrix{M}{k-q+q'}{k-q}{[q']}$, are derived as follows.
We define $T_{m}$ as $T_{m} \equiv R_0^{m-1} T_1 R_0^{1-m}$.
Then the relation $T_{m'} T_m = T_m T_{m'}$ leads to 
$T_1 R_0^{m-m'} T_1 = R_0^{m-m'} T_1 R_0^{m'-m} T_1 R_0^{m-m'}$ 
and, using it, we derive the relation:
\begin{align}
\sum_{q'} \tau^{l(k-q')} \submatrix{M}{k}{k-q'}{[q']}\submatrix{M}{k-q'}{k-q}{[q-q']}
= \sum_{q'} \tau^{l(k+q'-q)}\submatrix{M}{k}{k-q'}{[q']}\submatrix{M}{k-q'}{k-q}{[q-q']}.
\label{ZN-tauMM}
\end{align}
where $l=m-m'$ takes an integer 
and the summation over $q'$ can be taken for any successive $N$ integers.
Changing $q'$ into $q-q'$ in the right-hand side of eq.~\eqref{ZN-tauMM},
we obtain the relation:
\begin{align}
\sum_{q'} \tau^{l(k-q')} \submatrix{M}{k}{k-q'}{[q']}\submatrix{M}{k-q'}{k-q}{[q-q']}
= \sum_{q'} \tau^{l(k-q')}\submatrix{M}{k}{k-q+q'}{[q-q']}\submatrix{M}{k-q+q'}{k-q}{[q']}.
\label{ZN-tauMM-q'}
\end{align}
Multiplying $\frac{1}{N}\sum_{l=1}^N\tau^{l(q''-k)}$ 
by eq.~\eqref{ZN-tauMM-q'}
and using the relation $\frac{1}{N} \sum_{l=1}^N \tau^{l(q''-q')} = \delta_{q''~\!\!q'}$,
we obtain the relation:
\begin{align}
\submatrix{M}{k}{k-q'}{[q']}\submatrix{M}{k-q'}{k-q}{[q-q']}
=\submatrix{M}{k}{k-q+q'}{[q-q']}\submatrix{M}{k-q+q'}{k-q}{[q']},
\label{ZN-qq'}
\end{align}
where $q''$ is replaced with $q'$.
For reference, using eq.~\eqref{ZN-qq'}, we obtain the relation:
\begin{align}
& \submatrix{M}{k}{k-q''}{[q'']}\submatrix{M}{k-q''}{k-q''-q'}{[q']}
\submatrix{M}{k-q''-q'}{k-q''-q}{[q-q']}
=\submatrix{M}{k}{k-q'}{[q']}\submatrix{M}{k-q'}{k-q''-q'}{[q'']}
\submatrix{M}{k-q''-q'}{k-q''-q}{[q-q']}
\nonumber \\
& \qquad =\submatrix{M}{k}{k-q'}{[q']}\submatrix{M}{k-q'}{k-q}{[q-q']}
\submatrix{M}{k-q}{k-q-q''}{[q'']}.
\label{ZN-qq'q''}
\end{align}
Taking $q=0$ and $q''=0$ in eq.~\eqref{ZN-qq'q''},
we obtain the relation:
\begin{align}
[\submatrix{M}{k}{k}{[0]}, \submatrix{M}{k}{k-q'}{[q']}
\submatrix{M}{k-q'}{k}{[-q']}] = 0.
\label{ZN-M0-comm}
\end{align}

%
\section{Derivation of eqs.~\eqref{Gtr.|alpha|} -- \eqref{Gtr.t1Y}}
\label{app:ap}
%

The specific relations for $N=3$ are given by
\beqn
a_1^3+a_2^3+a_3^3-3a_1 a_2 a_3=1, \quad
\abs{a_1}^2+\abs{a_2}^2+\abs{a_3}^2=1,\quad
\overline{a_1} a_3 + \overline{a_3} a_2 + \overline{a_2} a_1 =0.
\label{C-rN=3}
\eeqn
From the first relation and the combination of the second and third ones 
in eq.~\eqref{C-rN=3},
we obtain the relations:
\beqn
 &&(\omega a_1 + \omega^2 a_2 + a_3)(\omega^2 a_1 + \omega a_2 + a_3)(a_1+a_2+a_3)=1,
\label{C-r1N=3}\\
 &&\abs{\omega a_1 + \omega^2 a_2 + a_3}^2
= \abs{\omega^2 a_1 + \omega a_2 + a_3}^2 = \abs{a_1+a_2+a_3}^2=1,
\label{C-r2N=3}
\eeqn
respectively.
Using $\alpha_j \equiv \sum_{p=1}^{3} a_j \omega^{jp}$,
eqs.~\eqref{C-r1N=3} and \eqref{C-r2N=3} are rewritten compactly as
\beqn
\alpha_1 \alpha_2 \alpha_3 = 1, \qquad \abs{\alpha_j}^2 = 1, \quad (j=1, 2, 3),
\eeqn
respectively.

In the same way, 
eqs.~\eqref{Gtr.|alpha|}, \eqref{Gtr.alpha-N=4} and \eqref{Gtr.alpha-N=6} are obtained,
using the specific relations such that 
$2|a_1|^2 + a_2^2 + a_4^2 = 1$ and $2a_2 a_4 = a_1^2 + \overline{a_1}^2$ 
with $a_2 = \overline{a_2}$, $a_3 = -\overline{a_1}$ and $a_4 = \overline{a_4}$ for $N=4$
and $2|a_1|^2 + 2|a_2|^2 + |a_3|^2 + a_6^2 = 1$,
$|a_1|^2 - |a_2|^2 - |a_3|^2 + a_6^2 = a_6$,
$2a_2 a_6 + \overline{a_2}^2 = a_1^2 - 2 \overline{a_1}a_3$,
$a_1 a_6 + a_3 \overline{a_2} + 2 a_2 \overline{a_1} = a_1$,
$-a_2 a_6 + a_1^2 + \overline{a_1} a_3 + \overline{a_2}^2 = a_2$
and $-2a_3 a_6 + a_1 a_2 - \overline{a_1} \overline{a_2} = a_3$
with $a_3 = -\overline{a_3}$, $a_4 = \overline{a_2}$, $a_5 = -\overline{a_1}$
and $a_6 = \overline{a_6}$ for $N=6$.

Here, we explain the reason why the relations 
\eqref{Gtr.|alpha|} -- \eqref{Gtr.alpha-N=6} hold.
From $t_1 = \sum_{p=1}^N a_p Y^p$ and $t_{m} \equiv r_0^{m-1} t_1 r_0^{1-m}$, 
we obtain the relation:
\bequ
t_m = X^{m-1} (\sum_{p=1}^N a_p Y^p) X^{1-m}
= \sum_{p=1}^N a_p \tau^{(m-1)p} Y^p,
\label{C-tm}
\eequ
using the relation $X^m Y^{m'} = \tau^{m m'} Y^{m'} X^m$.
The unitary matrices $t_m$ are diagonalized simultaneously by the unitary transformation:
\bequ
U t_m U^{\dagger} = \sum_{p=1}^N a_p \tau^{(m-1)p} (UYU^{\dagger})^p
= \sum_{p=1}^N a_p \tau^{(m-1)p} X^p,
\label{C-tm-diag}
\eequ
where the $(j, j')$ elements of $X$, $Y$ and $U$ are given by
\bequ
\left(X\right)_{jj'} = \tau^{j} \delta_{j~\!\!j'},
\qquad \left(Y\right)_{jj'} = \delta_{j~\!\!j'+1},
\qquad \left(U\right)_{jj'} = \frac{1}{\sqrt{N}} \tau^{j(j'+1)},
\label{Gtr.U-comp}
\eequ
respectively.
Because $U t_m U^{\dagger}$ obey the same constraints as $t_m$,
those eigenvalues should do.
The $(1,1)$ element of $U t_m U^{\dagger}$
is given by $(U t_m U^{\dagger})_{11} = \sum_{p=1}^N a_p \tau^{mp}$,
and we denote it as $\alpha_m = \sum_{p=1}^N a_p \tau^{mp}$.
Thus, we find that $\alpha_m$ satisfy the relations
\eqref{Gtr.|alpha|} -- \eqref{Gtr.alpha-N=6}
corresponding to $t_m^{\dagger} t_m = I$
and peculiar constraints in Table~\ref{T-gauge}.
The same relations are obtained for other diagonal elements of $U t_m U^{\dagger}$.  

Finally, we derive eq.~\eqref{Gtr.t1Y}, i.e.,
$t_1 = \sum_{p=1}^N a_{p} Y^{p} = e^{i\left(\theta Y + \bar{\theta} Y^{N-1}\right)}$.
From eqs.~\eqref{Gtr.|alpha|} -- \eqref{Gtr.alpha-N=6}, 
$\alpha_j= \sum_{p=1}^N a_p \tau^{jp}$ 
are parametrized by a complex number $\theta$
as follows,
\bequ
 \alpha_j = \sum_{p=1}^N a_{p} \tau^{jp} 
= e^{i\left(\theta \tau^j + \bar{\theta} \bar{\tau}^j\right)},
\label{Gtr.t1-theta}
\eequ
because the number of independent parameters on $\alpha_j$ is two 
for the $N=3, 4, 6$ cases.
Here, we derive eq.~\eqref{Gtr.t1-theta} for $N=6$.
From eq.~\eqref{Gtr.|alpha|}, $\alpha_j$ are written by $\alpha_j = e^{i\varphi_j}$
using real parameters $\varphi_j$.
From eq.~\eqref{Gtr.alpha-N=6}, $\varphi_j$ obey 
$\varphi_1 + \varphi_4 = \varphi_2 + \varphi_5 = \varphi_3 + \varphi_6 
=\varphi_1 + \varphi_3 + \varphi_5 = \varphi_2 + \varphi_4 + \varphi_6 = 0$ (mod $2\pi$).
Taking $\varphi_2$ and $\varphi_6$ as independent ones, others are
determined as $\varphi_1 = \varphi_2 + \varphi_6$, $\varphi_3 = -\varphi_6$,
$\varphi_4 = -\varphi_2 - \varphi_6$ and $\varphi_5 = - \varphi_2$ (mod $2\pi$).
Then, using a complex parameter 
$\theta = \frac{1}{2}\varphi_6 - \frac{i}{2\sqrt{3}}(2\varphi_2 + \varphi_6)$
made of $\varphi_2$ and $\varphi_6$,
$\varphi_j$ are parametrized as $\varphi_j = \theta \eta^j + \bar{\theta}\bar{\eta}^j$
with $\eta = e^{2\pi i/6}$.
In the same way, $\alpha_j$ are expressed as eq.~\eqref{Gtr.t1-theta}
for $N=3,4$.
Using eq.~\eqref{Gtr.t1-theta} and $\left(X\right)_{jj'} = \tau^{j} \delta_{j~\!\!j'}$,
we obtain the relation:
\bequ
 \sum_{p=1}^N a_{p} X^{p} = e^{i\left(\theta X + \bar{\theta} \bar{X}\right)}
= e^{i\left(\theta X + \bar{\theta} X^{N-1}\right)}.
\label{Gtr.t1X}
\eequ
Performing the unitary transformation such as $U^{\dagger}X U = Y$
for eq.~\eqref{Gtr.t1X}, we arrive at eq.~\eqref{Gtr.t1Y}:
\bequ
t_1 = \sum_{p=1}^N a_{p} Y^{p} = e^{i\left(\theta Y + \bar{\theta} Y^{N-1}\right)}.
\label{C.t1Y}
\eequ
For reference, 
multiplying $\frac{1}{N}\sum_{j=1}^N\tau^{-jp'}$ by eq.~\eqref{Gtr.t1-theta}
and using $\frac{1}{N} \sum_{j=1}^N \tau^{j(p-p')} = \delta_{p~\!\!p'}$,
we obtain the relation:
\bequ
a_p = \frac{1}{N} \sum_{j=1}^{N} \alpha_j \tau^{-jp}
= \frac{1}{N} \sum_{j=1}^{N} \tau^{-jp} 
e^{i\left(\theta \tau^j + \bar{\theta}\bar{\tau}^j\right)},
\label{Gtr.ap}
\eequ
where $p'$ is replaced with $p$.   

\section*{Acknowledgments}
This work was supported in part by scientific grants 
from the Ministry of Education, Culture,
Sports, Science and Technology under Grant No.~22K03632 (YK).



\begin{thebibliography}{99}
\bibitem{M}
N.~Manton, 
{\it A new six-dimensional approach to the Weinberg-Salam model}, 
{\it Nucl.\ Phys.} B {\bf 158} (1979), 141.

\bibitem{GUT}
  H.~Georgi and S.~L.~Glashow,
{\it Unity of All Elementary Particle Forces},
{\it Phys.\ Rev.\ Lett.} {\bf 32} (1974) 438.

\bibitem{SUSYGUT-DG}
S.~Dimopoulos and H.~Georgi, {\it Softly broken supersymmetry and SU(5)},
{\it Nucl.\ Phys.} B {\bf 193} (1981) 150.

\bibitem{SUSYGUT-S}
N.~Sakai, {\it Naturalness in supersymmetric GUTS},
{\it Z.\ Phys.} C {\bf 11} (1981) 153.

\bibitem{K1}
Y.~Kawamura, {\it Gauge Symmetry Reduction from the Extra Space $S^1/Z_2$},
{\it Prog.\ Theor.\ Phys.} {\bf 103} (2000) 613 [hep-ph/9902423].

\bibitem{K2}
Y.~Kawamura, {\it Triplet-doublet Splitting, Proton Stability and an Extra Dimension},
{\it Prog.\ Theor.\ Phys.} {\bf 105} (2001) 999 [hep-ph/0012125].

\bibitem{H&N}
L.~Hall and Y.~Nomura, {\it Gauge Unification in Higher Dimensions},
{\it Phys.\ Rev.} D {\bf 64} (2001) 055003 [hep-ph/0103125].

\bibitem{GHU-KL&Y}
M.~Kubo, C.~S.~Lim and H.~Yamashita,
{\it The Hosotani mechanism in bulk gauge theories with an orbifold
extra space $S^1 / Z_2$}, {\it Mod.\ Phys.\ Lett.} A {\bf 17} (2002) 2249 [hep-ph/0111327].

\bibitem{GHU-CG&M}
C.~Csaki, C.~Grojean and H.~Murayama,
{\it Standard model Higgs from higher dimensional gauge fields},
{\it Phys.\ Rev.} D {\bf 67} (2003) 085012 [hep-ph/0210133].

\bibitem{GHU-SS&S}
C.~A.~Scrucca, M.~Serone and L.~Silvestrini,
{\it Electroweak symmetry breaking and fermion masses from extra dimensions},
{\it Nucl.\ Phys.} B {\bf 669} (2003) 128 [hep-ph/0304220].

\bibitem{finiteness-K}
N.~V.~Krasnikov,
{\it Ultraviolet Fixed Point Behavior Of The Five-Dimensional Yang-Mills 
Theory, The Gauge Hierarchy Problem And A Possible New Dimension At The Tev Scale},
{\it Phys.\ Lett.} B {\bf 273} 246 (1991).

\bibitem{finiteness-HI&L}
H.~Hatanaka, T.~Inami and C.~S.~Lim,
{\it The gauge hierarchy problem and higher dimensional gauge theories},
{\it Mod.\ Phys.\ Lett.} A {\bf 13} 2601 (1998) [hep-th/9805067].

\bibitem{finiteness-AC&G}
N.~Arkani-Hamed, A.~G.~Cohen and H.~Georgi,
{\it Electroweak symmetry breaking from dimensional deconstruction},
{\it Phys.\ Lett.} B {\bf 513} (2001) 232 [hep-ph/0105239].

\bibitem{finiteness-M&Y}
N.~Maru and T.~Yamashita,
{\it Two-loop calculation of Higgs mass in gauge-Higgs unification: 5D  massless
QED compactified on $S^1$},
{\it Nucl.\ Phys.} B {\bf 754} 127 (2006) [hep-ph/0603237].

\bibitem{finiteness-HMT&Y}
Y.~Hosotani, N.~Maru, K.~Takenaga and T.~Yamashita,
{\it Two loop finiteness of Higgs mass and potential in the gauge-Higgs unification},
{\it Prog.\ Theor.\ Phys.}  {\bf 118} 1053 (2007) [arXiv:0709.2844 [hep-ph]].

\bibitem{divcont}
J.~Hisano, Y.~Shoji and A.~Yamada,
{\it To be, or not to be finite? The Higgs potential in Gauge Higgs Unification},
{\it JHEP} \textbf{02} (2020), 193 [arXiv:1908.09158 [hep-ph]].

\bibitem{GHU-HHK&Y}
N.~Haba, Y.~Hosotani, Y.~Kawamura and T.~Yamashita,
{\it Dynamical symmetry breaking in gauge Higgs unification on orbifold},
{\it Phys.\ Rev.} D {\bf 70} (2004) 015010 [hep-ph/0401183].

\bibitem{Lim:2007jv}
C.~S.~Lim and N.~Maru,
{\it Towards a realistic grand gauge-Higgs unification},
{\it Phys.\ Lett.} B \textbf{653} (2007) 320-324 
[arXiv:0706.1397 [hep-ph]].

\bibitem{Hosotani:2015hoa}
Y.~Hosotani and N.~Yamatsu,
{\it Gauge\textendash{}Higgs grand unification},
{\it Prog.\ Theor.\ Exp.\ Phys.} \textbf{2015} (2015) 111B01 
[arXiv:1504.03817 [hep-ph]].

\bibitem{gGHU-E6}
K.~Kojima, K.~Takenaga and T.~Yamashita,
{\it The Standard Model Gauge Symmetry from Higher-Rank Unified Groups 
in Grand Gauge-Higgs Unification Models},
{\it JHEP} \textbf{06} (2017) 018 [arXiv:1704.04840 [hep-ph]].

\bibitem{gGHU}
  K.~Kojima, K.~Takenaga and T.~Yamashita,
  {\it Grand Gauge-Higgs Unification},
  {\it Phys.\ Rev.} D {\bf 84} (2011) 051701 [arXiv:1103.1234 [hep-ph]].
%
\bibitem{gGHU-KT&Y}
K.~Kojima, K.~Takenaga and T.~Yamashita,
{\it Gauge symmetry breaking patterns in an SU(5) grand gauge-Higgs unification model},
{\it Phys.\ Rev.} D {\bf 95} (2017) 015021 [arXiv:1608.05496 [hep-ph]].

\bibitem{gGHU-Y}
T.~Yamashita,
{\it Doublet-Triplet Splitting in an SU(5) Grand Unification},
{\it Phys.\ Rev.} D {\bf 84} (2011) 115016 [arXiv:1106.3229 [hep-ph]].
 
\bibitem{gGHU-pheno}
M.~Kakizaki, S.~Kanemura, H.~Taniguchi and T.~Yamashita,
{\it Higgs sector as a probe of supersymmetric grand unification with the Hosotani mechanism},
{\it Phys.\ Rev.} D {\bf 89} (2014) 075013 [arXiv:1312.7575 [hep-ph]].

\bibitem{gGHU-pheno-NSS&Y}
H.~Nakano, M.~Sato, O.~Seto and T.~Yamashita,
{\it Dirac gaugino from grand gauge-Higgs unification},
{\it Prog.\ Theor.\ Exp.\ Phys.} \textbf{2022} (2022) 033B06 [arXiv:2201.04428 [hep-ph]].

\bibitem{H1}
Y.~Hosotani, {\it Dynamical mass generation by compact extra dimensions},
{\it Phys.\ Lett.} B {\bf 126} (1983) 309.

\bibitem{H2}
Y.~Hosotani, {\it Dynamics of Nonintegrable Phases and Gauge Symmetry Breaking},
{\it Ann.\ of Phys} {\bf 190} (1989) 233.  

\bibitem{HHH&K}
N.~Haba, M.~Harada, Y.~Hosotani and Y.~Kawamura,
{\it Dynamical rearrangement of gauge symmetry on the Orbifold $S^1/Z_2$},
{\it Nucl.\ Phys.} B {\bf 657} (2003) 169
[{\it Errata ibid} B {\bf 669} (2003) 381] [hep-ph/0212035].

\bibitem{HH&K}
N.~Haba, Y.~Hosotani and Y.~Kawamura, 
{\it Classification and Dynamics of Equivalence Classes in $SU(N)$ gauge theory 
on the orbifold $S^1/Z_2$},
{\it Prog.\ Theor.\ Phys.} {\bf 111} (2004) 265
[hep-ph/0309088].

\bibitem{generalFormula}
N.~Haba and T.~Yamashita,
{\it A General formula of the effective potential in 5-D SU(N) gauge theory on orbifold},
{\it JHEP} \textbf{02} (2004) 059 [hep-ph/0401185].

\bibitem{KK&M}
Y.~Kawamura, T.~Kinami and T.~Miura, 
{\it Equivalence Classes of Boundary Conditions in Gauge Theory on $Z_3$ Orbifold},
{\it Prog.\ Theor.\ Phys.} {\bf 120} (2008) 815 [arXiv:0808.2333].

\bibitem{K&M}
Y.~Kawamura and T.~Miura, 
{\it Equivalence Classes of Boundary Conditions 
in $SU(N)$ Gauge Theory on 2-Dimensional Orbifolds},
{\it Prog.\ Theor.\ Phys.} {\bf 122} (2009) 847 [arXiv:0905.4123].

\bibitem{G&K}
Y.~Goto and Y.~Kawamura, 
{\it Orbifold family unification using vectorlike representation on six dimensions},
{\it Phys.\ Rev.} D {\bf 98} (2018) 035039 [arXiv:1712.06444].

\bibitem{HN&T}
Y.~Hosotani, S.~Noda and K.~Takenaga, 
{\it Dynamical gauge symmetry breaking and mass generation on the Orbifold $T^2/Z_2$},
{\it Phys.\ Rev.} D {\bf 69} (2004) 125014 [hep-ph/0403106].

\bibitem{K&N}
Y.~Kawamura and Y.~Nishikawa,
{\it On diagonal representatives in boundary condition matrices on orbifolds},
{\it Int. J. Mod. Phys.} A \textbf{35} (2020) 2050206
[arXiv:2009.10958].

\bibitem{tHooft}
G.~'t Hooft,
{\it A Property of Electric and Magnetic Flux in Nonabelian Gauge Theories},
{\it Nucl. Phys.} B {\bf 153} (1979) 141.

\bibitem{VG}
G.~von Gersdorff,
{\it A New Class of Rank Breaking Orbifolds},
{\it Nucl. Phys.} B {\bf 793} (2008) 192
[arXiv:0705.2410].

\bibitem{Bachas}
C.~Bachas, {\it A way to break supersymmetry}, [hep-ph/9503030].

\bibitem{S&S}
C.~A.~Scrucca and M.~Serone,
{\it Anomalies in field theories with extra dimensions},
{\it Int. J. Mod. Phys.} A {\bf 19} (2004) 2579 [hep-th/0403163].

\bibitem{treepot6D}
C.~A.~Scrucca, M.~Serone, L.~Silvestrini and A.~Wulzer,
{\it Gauge Higgs unification in orbifold models},
{\it JHEP} \textbf{02} (2004) 049 [hep-th/0312267].

\bibitem{FN&W}
S.~F\"{o}rste, H.~P.~Nilles and A.~Wingerter, 
{\it Geometry of Rank Reduction}, {\it Phys.\ Rev.} D {\bf 72} (2005) 026001 
[hep-ph/0504117].

\end{thebibliography}
\end{document}